\documentclass[fleqn,usenatbib]{mnras}

\usepackage{newtxtext,newtxmath}
\usepackage{deluxetable}
\usepackage{rotating}
\usepackage{graphicx}
\usepackage{adjustbox}
\usepackage{array, makecell} 

\usepackage[T1]{fontenc}

\DeclareRobustCommand{\VAN}[3]{#2}
\let\VANthebibliography\thebibliography
\def\thebibliography{\DeclareRobustCommand{\VAN}[3]{##3}\VANthebibliography}

\usepackage{graphicx}	
\usepackage{amsmath}	
\usepackage{amssymb}	
\usepackage{lscape}
\newcommand{\fermi}{{\it Fermi}-LAT }
\newcommand{\gray}{$\gamma$-ray }
\newcommand{\grays}{$\gamma$-rays }

\title[High redshift blazars]{Multiwavelength Study of High-Redshift Blazars}

\author[N. Sahakyan et al.]{
N. Sahakyan,$^{1,2}$\thanks{E-mail: narek@icra.it}
D. Israyelyan,$^{1}$
G. Harutyunyan$^{1}$
M. Khachatryan$^{1}$
and S. Gasparyan$^{1}$
\\
$^{1}$ICRANet Armenia, Marshall Baghramian Avenue 24a, Yerevan 0019, Republic of Armenia\\
$^{2}$ICRANet, Piazza della Repubblica 10, I-65122 Pescara, Italy \\
}

\date{Accepted XXX. Received YYY; in original form ZZZ}

\pubyear{2020}

\begin{document}
\label{firstpage}
\pagerange{\pageref{firstpage}--\pageref{lastpage}}
\maketitle

\begin{abstract}
High-redshift blazars are among the most powerful objects in the Universe. The spectral and temporal properties of thirty-three distant blazars ($z>2.5$) detected in the high energy $\gamma$-ray band are investigated by analyzing the {\it Fermi}-LAT and {\it Swift} UVOT/ XRT data. The considered sources have soft time averaged $\gamma$-ray spectra ($\Gamma_{\rm \gamma}\geq2.2$) whereas those that have been observed in the X-ray band have hard X-ray spectra ($\Gamma_{\rm X}=1.01-1.86$). The $\gamma$-ray flux of high-redshift blazars ranges from $4.84\times10^{-10}$ to $1.50\times10^{-7}\:{\rm photon\:cm^{-2}\:s^{-1}}$ and the luminosity is within $(0.10-5.54)\times10^{48}\:{\rm erg\:s^{-1}}$ which during the \gray flares increases up to $(0.1-1)\times10^{50}\:{\rm erg\:s^{-1}}$. In the X-ray band, only the emission of PKS 0438-43, B2 0743+25 and TXS 0222+185 is found to vary in different Swift XRT observations whereas in the \gray band, the emission is variable for fourteen sources: the flux of B3 1343+451 and PKS 0537-286 changes in sub-day scales, that of PKS 0347-211 and PKS 0451-28 in day scales, while the \gray variability of the others is in week or month scales. The properties of distant blazar jets are derived by modeling the multiwavelength spectral energy distributions within a one-zone leptonic scenario assuming that the X-ray and $\gamma$-ray emissions are produced from inverse Compton scattering of synchrotron and dusty torus photons. From the fitting, the emission region size is found to be $\leq0.05$ pc and the magnetic field and the Doppler factor are correspondingly within $0.10-1.74$ G and $10.0-27.4$. By modeling the optical-UV excess, we found that the central black hole masses and accretion disk luminosities are within $L_{\rm d}\simeq(1.09-10.94)\times10^{46}\:{\rm erg \: s^{-1}}$ and $(1.69-5.35)\times10^{9}\:M_\odot$, respectively.
\end{abstract}

\begin{keywords}
Galaxies: active --Gamma rays: galaxies--radiation mechanisms: non-thermal
\end{keywords}



\section{Introduction}
Blazars are one of the most luminous objects in the Universe. In the unification scheme of radio-loud active galactic nuclei (AGNs) blazars are a subclass with a relativistic jet making a small angle (a few degrees) with the observer's line of sight \citep[]{urry}. The jets are sources of non-thermal emission which cover the entire electromagnetic spectrum, from radio to high energy (HE; $>100$ MeV) or very high energy (VHE; $>100$ GeV) \gray bands \citep{2017A&ARv..25....2P}. This nonthermal emission is varying in short time scales \citep[e.g., in minute scales in the \gray band][]{2016ApJ,foschini11,foschini13, nalewajko, brown13, rani13,saito,hayashida15} with a substantial increase in luminosity. The flux variation as well as the observed superluminal motion, the high degree of polarization and other observed features can be explained by the relativistic beaming effects.\\
\begin{table*}
\caption{List of \gray emitting blazars at $z>2.5$. The results of observation of distant blazars from August 4, 2008, to August 4, 2018 by \fermi are presented in the right part of the table.}\label{sources}
\centering\renewcommand\cellalign{lc}
\begin{tabular}{llccc|cccccc} \hline \hline
Object & 4FGL name & Class &  RA & Dec &  $F_{\rm \gamma}$ $^1$ & Photon index & $TS$ & $E_{\rm max}$ $^2$ & Probability & $z$  \\
\hline
GB 1508+5714 & J1510.1+5702 & FSRQ &227.54& 57.04&  $0.76 \pm 0.15$ & $2.95 \pm 0.15$&  63.50 & $8.76$& 0.8268 & 4.31\\
PKS 1351-018 & J1354.3-0206 & FSRQ  &208.50&-2.11& $1.11\pm 0.15$ & $2.69\pm0.09$& 98.49 & $1.30$ & 0.7010 & 3.72\\
MG3 J163554+3629 & J1635.6+3628 & BCU&248.92&36.48& $1.46 \pm 0.27$ & $2.84 \pm 0.12$&  123.95 & $3.74$ &0.8413 & 3.65\\
\makecell{NVSS J121915 +365718} & J1219.0+3653&BCU&184.77&36.89& 	$0.22 \pm 0.08$ & $2.20\pm  0.14$&  40.56 & $2.67$& 0.8273 &3.53\\
PKS 0335-122 & J0337.8-1157& FSRQ  &    54.47  &   -11.96  & 		$0.72 \pm 0.16$ & $2.69 \pm 0.13$&  49.47 & $10.19$ & 0.9245& 3.44\\
PKS 0537-286 & J0539.9-2839& FSRQ  &    84.99  &   -28.66  & 		$4.38 \pm 0.18$ & $2.72 \pm 0.03$&  1694.80 & $7.60$& 0.9619 & 3.10\\
TXS 0800+618 & J0805.4+6147& FSRQ  &   121.36  &    61.80  & 		$2.23 \pm 0.14$ & $2.82 \pm 0.05$&  475.17 & $4.83$& 0.8441 & 3.03\\
S4 1427+543 & J1428.9+5406& FSRQ  &   217.23  &    54.11  & 		$0.59\pm 0.15$ & $2.62\pm  0.14$&  67.10 & $10.26$ & 0.7831& 3.01\\
GB6 J0733+0456 & J0733.8+0455& FSRQ  &   113.47  &     4.93  & $1.14 \pm 0.15$ & $2.39 \pm 0.07$&  197.26 & $17.65$ & 0.9854& 3.01\\
B2 0743+25 & J0746.4+2546&FSRQ  &   116.60  &    25.77  & 		$2.06 \pm 0.19$ & $2.87 \pm 0.07$&  280.99 & $1.52$ & 0.7449& 2.99\\
PKS 0347-211 & J0349.8-2103 & FSRQ  &    57.47  &   -21.06  & 		$3.25 \pm 0.15$ & $2.47 \pm 0.03$&  1372.81 & $5.83$ & 0.9690& 2.94\\
S4 1124+57 & J1127.4+5648& FSRQ  &   171.87  &    56.80  & 		$0.96 \pm 0.13$ & $2.75 \pm 0.09$&  147.30 & $7.09$ & 0.8495& 2.89\\
MRSS 291-081526 & J2313.9-4501& BCU  &348.49 &-45.02  &  $0.89 \pm 0.20$ & $2.80 \pm 0.13$&  81.12	 & $2.25$ & 0.7550& 2.88\\
PKS 0438-43 & J0440.3-4333& FSRQ  &    70.09  &   -43.55  &  $2.24 \pm 0.20$ & $2.60 \pm 0.05$&  574.84 & $2.00$ & 0.9225& 2.85\\
S4 2015+65 & J2015.4+6556 &FSRQ  &   303.86  &    65.95  & 		$0.39 \pm 0.15$ & $2.37 \pm 0.15$&  23.77 & $4.34$ & 0.7257& 2.84\\
\makecell{87GB 214302.1+095227} & J2145.5+1006& BL Lac &   326.38  & 10.12 & 	$0.048\pm0.014$ & $1.71 \pm 0.19$&  40.46	 & $67.15$ & 0.9874& 2.83\\
MG2 J174803+3403 &J1748.0+3403 & FSRQ  &   267.01  &    34.06  & 	$0.97 \pm 0.13$ & $2.31 \pm 0.07$&  284.21 & $31.63$ & 0.9934& 2.76\\
PKS 0834-20 & J0836.5-2026 & FSRQ  &   129.13  &   -20.45  & 	$1.93 \pm 0.19$ & $2.94 \pm 0.08$&  171.15 & $1.15$ & 0.5796& 2.75\\
TXS 0222+185 & J0224.9+1843 & FSRQ  &    36.24  &    18.72  & 		$1.72 \pm 0.22$ & $3.05 \pm 0.12$&  101.08 & $2.59$ & 0.5327& 2.69\\
OD 166 & J0242.3+1102& FSRQ  &    40.59  &    11.05  & 		$1.94 \pm 0.20$ & $2.59 \pm 0.06$&  252.83 & $6.91$ & 0.8651& 2.68\\
\makecell{CRATES J233930\\+024420} & J2339.6+0242& BCU   &  354.90  &     2.71  &  $1.00 \pm 0.21$ & $2.58 \pm 0.11$ &  94.50	 & $6.17$ & 0.8537& 2.66\\
TXS 0907+230 & J0910.6+2247& FSRQ  &   137.67  &    22.80  & 		$1.17 \pm 0.14$ & $2.37 \pm 0.06$&  262.58 & $5.38$ & 0.9578& 2.66\\
PMN J1441-1523 & J1441.6-1522 & FSRQ  &   220.41  &   -15.38  & 	$0.65 \pm 0.23$ & $2.32 \pm 0.13$&  72.02 & $11.77$ & 0.9535& 2.64\\
\makecell{CRATES J105433\\+392803} & J1054.2+3926 & BCU   &  163.56  &    39.44  &  $0.27\pm 0.09$ & $2.30 \pm 0.16$&  34.05 & $4.79$ & 0.9000& 2.63\\
MG1 J154930+1708 & J1549.6+1710& BL Lac &  237.41  &    17.18  & 	$0.17 \pm 0.08$ & $2.01 \pm 0.16$&  44.75 & $17.03$ & 0.9759& 2.62\\
TXS 1448+093 & J1450.4+0910& FSRQ  &   222.62  &     9.18  & 		$0.76 \pm 0.13$ & $2.35 \pm 0.08$&  130.24 & $10.58$ & 0.9599& 2.61\\
PMN J0226+0937 & J0226.5+0938 & FSRQ  &    36.63  &     9.63  & 	$0.48 \pm 0.12$ & $2.18 \pm 0.10$&  96.49 & $56.42$ & 0.9523& 2.61\\
PKS 0451-28 & J0453.1-2806 & FSRQ  &    73.29  &   -28.11  & 		$5.83 \pm 0.17$ & $2.66 \pm 0.02$&  3118.20 & $10.77$ & 0.9815& 2.56\\
B3 0908+416B &J0912.2+4127 & FSRQ  &   138.06  &    41.46  & 		$1.51 \pm 0.14$ & $2.42 \pm 0.05$&  539.57 & $12.10$ & 0.9903& 2.56\\
TXS 1616+517 &J1618.0+5139 & FSRQ  &   244.52  &    51.67  & 		$0.69\pm 0.13$ & $2.68 \pm 0.12$&  72.65	 & $12.38$ & 0.8408& 2.56\\
B3 1624+414 & J1625.7+4134 & FSRQ  &   246.45  &    41.57  & 		$1.38 \pm 0.14$ & $2.49 \pm 0.06$&  395.12& $9.69$ & 0.9787& 2.55\\
B3 1343+451 & J1345.5+4453c & FSRQ  &   206.39  &    44.88  & 		$15.01 \pm 0.16$ & $2.25\pm 0.008$&  34652.79 & $24.25$ & 0.9994&2.53\\
PKS 2107-105 & J2110.2-1021c & FSRQ  &   317.56  &   -10.36  & 		$0.88\pm 0.18$ & $2.66 \pm 0.13$&  53.29  & $5.46$ & 0.6500& 2.50\\
\hline
\multicolumn{8}{l}{%
 \begin{minipage}{6.5cm}%
    $^1$ Integrated \gray flux in units of ${\rm \times 10^{-8} photon\:cm^{-2}\:s^{-1}}$.\\
    $^2$ Photon energy in GeV.
  \end{minipage}%
}
\end{tabular}
\end{table*}
Blazars are usually grouped into two large classes based on the absence or presence of emission lines in their optical spectra, i.e., BL Lacertae objects (BL Lacs) have weak or no emission lines; the equivalent width (EW) of the emission line $< 5$ {\AA}  in the rest frame, while the flat-spectrum radio quasars (FSRQs) have stronger emission lines (EW $> 5$ {\AA} ) \citep{urry}. Based on the position of the synchrotron peak in the rest frame ($\nu_{\rm p}^{\rm syn}$), blazars are further classified as low-synchrotron-peaked (LSP for $\nu_{\rm p}^{\rm syn}<10^{14}$ Hz), intermediate-synchrotron-peaked (ISP for $10^{14}<\nu_{\rm p}^{\rm syn}<10^{15}$ Hz), and high-synchrotron-peaked (HSP for $\nu_{\rm p}^{\rm syn}>10^{15}$ Hz) objects \citep{1995ApJ...444..567P, 2010ApJ...716...30A}.\\
The broadband spectral energy distribution (SED) of blazars shows a typical double-peaked structure. The low-energy peak (extending from radio to UV/soft X-rays) is produced by synchrotron emission from the relativistic electrons in the jet. The HE component (above  X-ray band) is often attributed to Inverse Compton (IC) scattering of photons produced either inside \citep[synchrotron-self Compton, (SSC);][]{ghisellini,bloom, maraschi} or outside of the jet \citep[external inverse Compton, (EIC);][]{blazejowski, sikora, ghisellini009}. The external photons can be from the accretion disk, broad-line region and dusty torus surrounding the disc. In alternative models, the HE component is due to the interaction of protons accelerated along with electrons in the jet \citep[e.g.,][]{2013ApJ...768...54B}. In this case, the blazars also can emit VHE neutrinos \citep[e.g.,][]{2018ApJ...863L..10A, 2019NatAs...3...88G, 2019MNRAS.483L..12C, 2018ApJ...864...84K, 2018arXiv180705210L, 2018ApJ...866..109S, 2019A&A...622A.144S, 2018arXiv180900601W, 2019MNRAS.483L.127R}.\\
\begin{table}
\caption{Spectral parameters of the sources modeled with logparabola.}\label{sourcesLP}
\centering
\begin{tabular}{lccc}
\hline \hline
Object  &$F_{\rm \gamma}$ $^1$ & $\alpha$ & $\beta$ \\
\hline
 PKS 1351-018 & $( 6.86\pm1.66)\times10^{-9}$ & $2.20\pm0.23 $& $0.63\pm 0.23$   \\
 PKS 0537-286 & $(4.15 \pm 0.20)\times10^{-8}$ & $2.66 \pm 0.04$& $0.10\pm 0.03$  \\
 TXS 0800+618 & $(2.04\pm 0.16)\times10^{-8}$ & $2.67 \pm 0.08$& $0.17\pm0.06$  \\
 B2 0743+25 & $(1.72 \pm 0.23)\times10^{-8}$ & $2.53 \pm 0.14$& $0.40\pm0.13$  \\
 PKS 0347-211 & $(2.69 \pm 0.17)\times10^{-8}$ & $2.32 \pm 0.05$& $0.19\pm0.03$  \\
 PKS 0438-43 & $(1.15 \pm 0.23)\times10^{-8}$ & $2.35 \pm 0.12$& $0.48\pm0.11$  \\
 S4 2015+65 & $(1.66 \pm 1.00)\times10^{-9}$ & $2.42 \pm 0.35$& $0.42\pm0.31$  \\
 PKS 0834-20 & $(1.76 \pm 0.20)\times10^{-8}$ & $2.62 \pm 0.16$& $0.29\pm0.13$  \\
 OD 166 & $(1.45 \pm 0.24)\times10^{-8}$ & $2.43 \pm 0.11$& $0.24\pm0.08$ \\
TXS 0907+230 & $(8.68 \pm 1.77)\times10^{-8}$ & $2.28 \pm 0.09$&  $0.12\pm 0.06$  \\
PKS 0451-28 &$(5.56 \pm 0.18)\times10^{-8}$ & $2.56 \pm 0.04$& $0.09\pm0.02$  \\
B3 1343+451 & $(1.41 \pm 0.02)\times10^{-7}$ & $2.18\pm 0.01$& $0.07\pm0.006$  \\
\hline
\multicolumn{4}{l}{%
 \begin{minipage}{6.5cm}%
    $^1$ Integrated \gray flux in units of ${\rm photon\:cm^{-2}\:s^{-1}}$.
  \end{minipage}%
}
\end{tabular}
\end{table}
Blazars are the dominant sources in the extragalactic \gray sky. Among the total 5000 sources in the Fermi Large Area Telescope (\fermi) fourth source catalog of \gray sources \citep[4FGL;][]{2020ApJS..247...33A}, $\sim2800$ are blazars, 45 are radio galaxies, and 19 are other AGNs.  Low to high redshift blazars are observed showing different redshift distributions for FSRQs and BL Lacs; the distribution of FSRQs has a peak around $z=1$ with a median of $1.14\pm0.62$ and there is a high number of FSRQs at the redshift of $\simeq0.5-2.0$, while the peak of BL Lacs is at $z=0.3$ with a mean of $0.34\pm0.42$. There are only 105 sources detected with $z>2$ ($3.75$ \% of total sources) and only 33 with $z>2.5$ ($1.18$ \%).\\
Blazars harboring supermassive black holes are valuable sources for studying the relativistic outflows and formation and propagation of relativistic jets. In this context, the high redshift blazars ($z>2.5$) are of particular interest; they are  the most powerful non-explosive astrophysical sources having ever been detected in the \gray band. 
Their study can shed light on the further understanding of the cosmological evolution of blazars and supermassive black holes and also on the evolution of relativistic jets across different cosmic epochs \citep{2010A&ARv..18..279V}. Moreover, the \gray observations of distant blazars are important since a limit on the density of the Extragalactic Background Light (EBL) can be derived. The \grays, as they propagate from their sources to the Earth, are subject to absorption through two-photon pair production when interacting with EBL photons \citep[e.g.,][]{2006ApJ...648..774S, 2011MNRAS.410.2556D}. This absorption feature is visible in the spectrum of the nearby sources only at VHEs while for distant sources ($z>2.0$), it can be significant already at tens of GeV. Thus, the EBL density in a redshift-dependence way can be constrained or measured by analyzing the \gray data \citep[e.g.,][]{2010ApJ...723.1082A, 2012Sci...338.1190A, 2018Sci...362.1031F, 2019ApJ...874L...7D}. However, if the \gray spectrum does not extend above 10 GeV to constrain the EBL density, by the theoretical interpretation of the data, the intrinsic source processes can be investigated and separated from the propagation effects which can be a help in the observations of distant blazars by future telescopes (e.g., CTA).\\
\begin{table*}
\caption{Swift XRT data analyses results. For the sources, when several observations were available, they were merged to estimate the averaged X-ray spectra. The sources for which the number of counts was enough to constrain the flux and the photon index in a single observation are marked with an asterisk ($^*$).}\label{sourcesXray}
\centering
\begin{tabular}{lcccc}
\hline \hline
Object & $n_{H}\times10^{20}\:{\rm cm^{-2}}$ &  $\Gamma_{\rm X}$ & ${\rm Log(F_{(0.3-10) keV})/erg\:cm^{-2}\:s^{-1}}$ & C-stat./dof  \\
\hline
GB 1508+5714 & $1.56$ & $1.38\pm0.52$ & $-12.21\pm0.21$ & 0.58(20)\\
PKS 0537-286$^*$ & $2.20$ & $1.17\pm0.04$ & $-11.42\pm0.02$ & 1.28(679)\\
TXS 0800+618$^*$ & $4.67$ & $1.13\pm0.10$ & $-11.53\pm0.04$ & 1.08(355)\\
S4 1427+543 & $1.17$ & $1.41\pm0.23$ & $-12.24\pm0.09$ & 1.1(102)\\
GB6 J0733+0456 & $7.72$ & $1.65\pm0.65$ & $-12.80\pm0.23$ & 1.33(20)\\
B2 0743+25$^*$ & $4.50$ & $1.15\pm0.03$ & $-11.30\pm0.01$ & 1.39(773)\\
PKS 0347-211 &$4.23$ & $1.32\pm0.27$ & $-12.26\pm0.12$ & 1.24(75)\\
PKS 0438-43$^*$ & $1.41$ & $1.25\pm0.09$ & $-11.52\pm0.04$ & 1.4(345)\\
S4 2015+65 & $10.6$ & $1.79\pm0.69$ & $-12.28\pm0.23$ & 0.91(17)\\
MG2 J174803+3403 & $3.22$ & $1.36\pm0.47$ & $-12.35\pm0.18$ & 0.61(33)\\
PKS 0834-20$^*$ & $6.07$ & $1.07\pm0.09$ & $-11.67\pm0.04$ & 1.12(399)\\
TXS 0222+185$^*$ & $9.40$ & $1.11\pm0.04$ & $-10.93\pm0.02$ & 1.89(684)\\
OD 166 & $8.97$ & $1.86\pm0.40$ & $-12.37\pm0.15$ & 0.65(40)\\
TXS 0907+230 & $4.68$ & $1.14\pm0.40$ & $-12.49\pm0.18$ & 1.26(40)\\
PMN J1441-1523 & $7.71$ & $1.86\pm0.63$ & $-13.10\pm0.22$ & 0.51(25)\\
TXS 1448+093 & $2.11$ & $2.33\pm0.62$ & $-13.30\pm0.16$ & 1.04(26)\\
PMN J0226+0937 & $6.57$ & $1.49\pm0.37$ & $-12.83\pm0.15$ & 0.93(49) \\
PKS 0451-28$^*$ & $2.07$ & $1.55\pm0.10$ & $-11.42\pm0.04$ & 0.99(297) \\
B3 0908+416B & $1.42$ & $1.01\pm0.46$ & $-12.47\pm0.19$ & 1.11(30)\\
TXS 1616+517 & $1.98$ & $1.25\pm0.33$ & $-12.77\pm0.14$ & 1.17(59)\\   
B3 1343+451$^*$ & $1.78$ & $1.21\pm0.17$ & $-12.22\pm0.07$ & 0.85(160)\\
PKS 2107-105 & $6.23$ & $1.24\pm0.29$ & $-12.49\pm0.12$ & 1.67(72) \\
\hline
\end{tabular}
\end{table*}
The vast majority of high redshift blazars are LSPs so the HE peak in their SEDs is at MeV below the \fermi band. They are sometimes also called 'MeV' blazars \citep{1995A&A...293L...1B}, being bright and strong X-ray emitters. In the X-ray band, these sources usually have a hard spectrum which corresponds to the rising part of the HE component, so these data are crucial for investigation of the origin of nonthermal emission \citep{2010MNRAS.405..387G}. Yet, due to the shift of SED peaks, for some blazars the direct thermal emission from the accretion disc is visible in the optical-UV band (the big blue bump) which can be modeled to derive the accretion disc luminosity \citep{1973A&A....24..337S} and the black hole mass and so to understand the properties of the central source. So, the Neil Gehrels Swift Observatory (\cite{2004ApJ...611.1005G}, hereafter Swift), taking the data in Optical/UV and X-ray bands, is an ideal instrument (within its sensitivity limit) for observing distant blazars.\\
Even though in the \gray band the distant blazars are relatively faint as compared with the X-ray band, \fermi observations are still crucial. The \gray band corresponds to the falling part of the HE component which combined with the X-ray data will fully constrain the second peak in the SED. This is fundamental allowing to derive the physical parameters of the jets. The multiwavelength observations and theoretical interpretations are a regular approach and a unique way of probing the physical condition of the plasma in the jet. From the theoretical modeling point of view, distant blazars are excellent sources for studying accretion disc-jet connection in the early epoch of quasar formation as well as for probing the environments around supermassive black holes.\\
Identification of distant blazars and their investigation has always been one of the actively discussed topics in the blazar research \citep[e.g.,][]{2016ApJ...825...74P, 2015ApJ...804...74P, 2011MNRAS.411..901G, 2018ApJ...853..159L, 2016ApJ...826...76A, 2017ApJ...834...41K, 2017ApJ...837L...5A, 2018ApJ...859...80K, 2017ApJ...839...96M, 2019ApJ...871..211P}. In contrast to nearby blazars, good quality multiwavelength data are missing for the high redshift blazars, which significantly complicates their detailed study. However, due to improved sensitivity of the instruments (e.g., X-ray and \gray observatories) and wide-field surveys in the low energy bands (e.g., Sloan Digital Sky Survey \citep{2000AJ....120.1579Y} or WISE \citet[][]{2010AJ....140.1868W}), the number of high redshift blazars with sufficient multiwavelength data has been significantly increased. Along with continuous \gray observations of some distant blazars since 2008 this opens new perspectives for exploring the physics of distant blazars.\\ 
Motivated by the large number of detected high redshift \gray emitting blazars, with the aim to characterize their multiwavelength emission properties, we perform an intense broadband study of all the thirty-three known \gray blazars beyond redshift $2.5$. Using the improved Pass 8 dataset which is more suitable for studying weak sources, we perform a detailed spectral and temporal analysis of \fermi \gray data accumulated during 2008-2018.  The \gray flux variation, not well explored for distant blazars, is investigated for the considered ten years, using an improved adaptive binning method. To characterize the physical properties of the considered sources in the X-ray and optical/UV bands, the data from their observation with both Swift X-ray Telescope (XRT) and Ultraviolet and Optical Telescope (UVOT) in the previous fifteen years are analyzed. This allows to collect unprecedented data in the optical/UV, X-ray and \gray bands, which is used to constrain the multiwavelength SEDs. Then, through theoretical modeling of these SEDs, the physical parameters characterizing the sources (disc luminosity, black hole mass, etc.) and their jets (e.g., the distribution of underlying electrons, magnetic field, power, etc.) are derived, allowing a quantitative discussion and investigation of the state of plasma in these powerful jets. Taking into account the large number of the considered sources (thirty-three) and the amount of analyzed data, this is one of the most comprehensive studies of most distant and powerful blazars.\\
The paper is organized as follows. In Section \ref{sec1}, the list of high redshift blazars is presented. The \fermi and Swift data extraction and analyses are described in Sections \ref{sec2} and \ref{sec3}, respectively. The data analysis results are presented in Section \ref{sec4} and the origin of the multiwavelength emission is discussed in Section \ref{sec5}. The discussion and conclusion are given in Section \ref{sec6} and a summary in Section \ref{sec7}.
\section{The Sample}\label{sec1}
The fourth catalog of AGNs detected by \fermi contains more than 2863 AGNs detected above the $5\sigma$ limit \citep{2020ApJS..247...33A}. A small fraction of them ($1.18$\%) are distant blazars $z>2.5$ ($\sim20.7$ Gpc), including twenty-six FSRQs, two BL Lacs and five blazars of uncertain type (BCU). The coordinates, redshift and synchrotron peak classification of these sources are given in Table \ref{sources}. The most distant source at $z=4.31$ is the FSRQ GB 1508+5714, whereas the BL Lacs 87 GB 214302.1+095227 and MG1 J154930+1708 are at $z=2.83$ and $z=2.62$, respectively, which is interesting, since due to low \gray luminosity BL Lacs are rarely observed at these distances. In this paper, the data collected by \fermi and Swift are analyzed to study the multiwavelength emission from high redshift blazars ($z>2.5$) selected from the fourth catalog of AGNs detected by \fermi which are presented in Table \ref{sources}. Throughout this paper, we assume the following standard cosmological parameters of $H_{0}=71\:{\rm km\:s^{-1}\:Mpc^{-1}}$ and $\Omega_{\Lambda}=0.730$ \citep{2001ApJ...553...47F}.
\begin{table*}
\caption{Summary of Swift-UVOT observations of the considered sources. For the sources, when several observations were available, the fluxes in each band are computed from the summed images. Averaged flux in each band is in units of erg cm$^{-2}$ s$^{-1}$.} \label{tabeswift}
\begin{adjustbox}{angle=90}
\begin{tabular}{lcccccc} 
\hline
\hline      
Object & $V$ & $B$ & $U$ &  $W1$  & $M2$ & $W2$ \\       
\hline
PKS 1351-108 -- & -- & -- & $(4.54\pm2.53)\times10^{-14}$ & -- & --  \\
MG3 J163554+3629 & -- & -- & $(6.93\pm2.72)\times10^{-14}$ & -- & -- & $(7.50\pm4.11)\times10^{-14}$  \\
PKS 0537-286 & $(3.78\pm5.86)\times10^{-15}$& $(4.53\pm1.56)\times10^{-14}$ & $(3.54\pm0.42)\times10^{-13}$ & $(4.54\pm2.54)\times10^{-14}$ & $(6.84\pm0.75)\times10^{-14}$ & $(4.61\pm0.55)\times10^{-13}$ \\
TXS 0800+618 & $(7.69\pm0.12)\times10^{-15}$ & -- & $(2.39\pm0.6)\times10^{-13}$ & $(5.68\pm0.18)\times10^{-15}$ & -- & $(6.58\pm0.13)\times10^{-14}$\\
S4 1427+543 &  --& $(2.52\pm2.03)\times10^{-14}$ &  $(2.53\pm0.92)\times10^{-13}$ & -- & $(3.88\pm1.4)\times10^{-14}$ & $(1.50\pm0.02)\times10^{-15}$\\
GB6 J0733+0456 &-- & $(2.97\pm2.55)\times10^{-14}$ & -- & $(9.40\pm0.25)\times10^{-15}$ & -- & -- \\
B2 0743+25 & -- & $(4.92\pm0.82)\times10^{-14}$ & $(3.39\pm0.59)\times10^{-13}$ & -- & $(1.12\pm0.22)\times10^{-13}$ & $(4.76\pm0.76)\times10^{-13}$\\
PKS 0347-211 & -- & -- & $(2.80\pm0.52)\times10^{-13}$ & -- & $(1.69\pm0.5)\times10^{-13}$ & $(2.03\pm1.09)\times10^{-13}$\\
S4 1124+57 & $(1.16\pm0.85)\times10^{-14}$& -- & -- & -- & -- & -- \\
PKS 0438-43 & $(7.90\pm0.1)\times10^{-15}$& -- & $(3.45\pm0.38)\times10^{-13}$ & -- & -- & $(8.02\pm0.69)\times10^{-13}$ \\
S4 2015+65 & $(1.63\pm0.02)\times10^{-15}$ & -- & -- & $(6.91\pm0.33)\times10^{-14}$ & -- & -- \\
87 GB 214302.1+095227 & $(1.01\pm0.48)\times10^{-13}$ & $(2.49\pm0.36)\times10^{-13}$  & -- & $(2.78\pm0.72)\times10^{-13}$ & $(3.49\pm0.63)\times10^{-13}$ & --  \\
MG2 J174803+3403 & $(1.92\pm0.32)\times10^{-13}$ &  $(2.53\pm0.25)\times10^{-13}$ & $(4.68\pm0.87)\times10^{-13}$ & $(2.40\pm0.33)\times10^{-13}$ & $(3.73\pm0.61)\times10^{-13}$ & $(3.20\pm1.52)\times10^{-13}$ \\
PKS 0834-20 & -- & -- & $(7.82\pm0.54)\times10^{-13}$ & $(3.84\pm1.72)\times10^{-14}$ & $(2.39\pm0.36)\times10^{-13}$ & $(7.29\pm0.91)\times10^{-13}$\\
TXS 0222+185 & -- & -- & $(9.36\pm0.89)\times10^{-13}$ & -- & $(8.18\pm1.11)\times10^{-13}$ & $(1.12\pm0.21)\times10^{-12}$\\
OD 166 & -- & -- & -- & -- & -- & $(9.19\pm0.36)\times10^{-14}$\\
TXS 0907+230 & $(9.53\pm4.27)\times10^{-14}$ & $(9.66\pm3.17)\times10^{-14}$& $(5.28\pm2.66)\times10^{-13}$ & $(8.77\pm3.31)\times10^{-14}$ & $(2.09\pm1.55)\times10^{-13}$ & -- \\
PMN J1441-1523 & --& -- & $(6.33\pm4.81)\times10^{-14}$ & -- & -- & --\\
CRATES J05433+392803 & --& -- & -- & $(1.90\pm0.88)\times10^{-16}$ & -- & -- \\
TXS 1448+093 & -- & -- & $(6.53\pm2.36)\times10^{-14}$ & -- & -- & $(1.35\pm0.35)\times10^{-13}$\\
PMN J0226+0937 & $(9.95\pm4.17)\times10^{-14}$& $(5.13\pm0.9)\times10^{-13}$ & $(1.22\pm0.06)\times10^{-12}$ & $(4.39\pm1.53)\times10^{-13}$ & $(1.30\pm0.06)\times10^{-12}$ & $(1.26\pm0.08)\times10^{-12}$ \\
PKS 0451-28 & -- & $(9.52\pm1.21)\times10^{-14}$ & -- & -- & -- & --\\
B3 0908+416B & $(3.45\pm0.11)\times10^{-15}$ & $(3.46\pm2.22)\times10^{-14}$ & $(2.08\pm0.66)\times10^{-13}$ & $(4.94\pm1.95)\times10^{-14}$ & $(6.87\pm1.91)\times10^{-14}$ & $(2.67\pm1.2)\times10^{-13}$ \\
TXS 1616+517 & -- & -- &$(1.96\pm0.27)\times10^{-13}$ & -- & -- & $(1.25\pm0.34)\times10^{-13}$\\
B3 1343+451 & -- & $(3.12\pm2.77)\times10^{-14}$ & $(4.54\pm0.55)\times10^{-13}$ & $(7.41\pm0.23)\times10^{-15}$ & $(2.52\pm0.29)\times10^{-13}$ & $(4.47\pm0.65)\times10^{-13}$\\
PKS 2107-105 & -- & -- & $(1.74\pm0.1)\times10^{-12}$ & -- & -- & $(1.44\pm0.13)\times10^{-12}$\\
\hline
\end{tabular}
\end{adjustbox}
\end{table*}
%
\section{Fermi-LAT observations}\label{sec2}
The \fermi data used in the current study had been accumulated during the first ten years of operation, from 4 August 2008 to 4 August 2018. The PASS 8 events from a circular region with a radius of $12^{\circ}$ around each source in the energy range from 100 MeV to 500 GeV were downloaded and analyzed with the Fermi Tools (1.2.1) using P8R3 instrument response functions. Good data and time intervals were selected using {\it gtselect} and {\it gtmktime} tools (with selection cuts {\it "Event class=128"} and {\it "evtype=3"}) using maximum zenith angle value of $90^{\circ}$ to avoid the \gray detection from the earth's limb. Using {\it gtbin} tool, the events are binned into $16.9^{\circ}\times16.9^{\circ}$ a square region of interest (ROI) with pixels of $0.1^{\circ}\times0.1^{\circ}$ and into 37 equal logarithmically spaced energy bins. The model file of each source (point-like sources included in the ROI and background) is generated using the 4FGL-DR2 version of the 4FGL which is based on 10 years of survey. The model file includes all 4FGL sources falling within the ROI $+ 5^{\circ}$ region. The {\it gll\_iem\_v07} and {\it iso\_P8R3\_SOURCE\_V2\_v1} models are used to parameterize the Galactic and extragalactic diffuse emission components. The spectral parameters set for the sources located within the ROI are allowed to be free in the analysis. The normalization parameters for the two diffuse components were also kept free.\\
The spectral analysis was performed using the binned maximum-likelihood method implemented in the {\it gtlike} tool. The source detection significance is estimated using the test statistic ($TS$) \citep{1996ApJ...461..396M} defined as $TS = 2\times(log L_{1} -log L_0)$ where $L_{1}$ is the likelihood of the data with a point source at the given position and $L_{0}$ without the source. The \gray spectral models of each considered source were assumed to be the same as in the 4FGL, and for those with a log-parabola an additional fit with the power-law model was performed. The spectra were calculated by separately running {\it gtlike} for smaller energy intervals equal in logarithmic space. Then, using the {\it gtsrcprob} tool and the model file obtained from the likelihood fitting, the energy of the highest-energy photon detected from the direction of each source is computed.
\subsection{\gray variability}
The \gray light curves are calculated using the unbinned likelihood analysis method implemented in the {\it gtlike} tool. The spectra of the considered sources were modeled by a power-law which can provide a good representation of the data over the small bins of time. The normalization and photon index of the sample sources are allowed to vary while the photon indexes of all background sources within the ROI are fixed to their best values obtained from the fit of the entire 10-year data set. As the diffuse background emission should not be variable, the parameters of the background models are fixed as well. During the fitting, the events within 0.1-300 GeV with the appropriate quality cuts mentioned above are considered.\\
Initially, in order to study the variability, the \gray light curves with 30-day binning are calculated for all sources, considering the 0.1-300 GeV  range. Additional light curves with short periods (days or a week) were computed for the statistically significant ($\geq 5\sigma$) \gray emitting periods identified in the monthly light curves. Next, using the adaptively binned method of \citet{lott}, the \gray variability is studied further. At fixed time binning, when long time intervals are used, a possible fast flux variation will be smoothed out, while using short bins might result in many upper limits during the low-activity periods. In the adaptive binning approach, the time bin widths are adjusted to produce bins with constant flux uncertainty above the optimal energy ($E_{0}$ ) and this appraoch is proved to be a powerful method to search for \gray flux variation \citep[e.g.,][]{2013A&A...557A..71R, 2017MNRAS.470.2861S, 2018A&A...614A...6S, 2017ApJ...848..111B, 2018ApJ...863..114G, 2017ApJ...835..182D, 2017A&A...608A..37Z}. $E_{0}$ depends on the flux and photon index of each source and can be computed following \citet{lott}. For the considered sources, adaptive binning light curves with 15\% uncertainty are generated.
\section{Swift observations} \label{sec3}
Swift is a multi-frequency space observatory, designed as a rapid response mission for follow-up observation of gamma-ray bursts (GRBs) \citep{2004ApJ...611.1005G}. Although its primary scientific goal is the observation of GRBs, due to wide frequency coverage, it is suitable for blazar studies. The data from two of the instruments onboard Swift UVOT \citep[170-600 nm;][]{2005SSRv..120...95R} and XRT \citep[0.3-10.0 keV;][]{2005SSRv..120..165B} have been analyzed in the current paper. Twenty-nine sources from Table \ref{sources} (except MRSS 291-081526, CRATES J233930+024420, MG1 J154930+1708 and B3 1624+414) were at least once observed by Swift. There are available data from multiple observations of some sources; e.g., B2 0743+25, PKS 0438-43, TXS 0222+185, TXS 1448+093, PMN J0226+0937, TXS 1616+517 and PKS 2107-105 had been observed more than ten times. The Swift XRT data analysis was possible to apply only for twenty two sources (see Table \ref{sourcesXray}) and twenty six were detected in at least one of the optical-UV Swift-UVOT filters (see Table \ref{tabeswift}).\\
Swift UVOT data from all six bands are considered when available: UVW2 (188 nm), UVM2 (217 nm), UVW1 (251 nm), U (345 nm), B (439 nm) and V (544 nm). The photometry analysis of all our sources was performed using the standard UVOT software distributed within the HEAsoft package (v6.25) and the calibration included in the CALDB (v.20170922). The source counts for each filter are extracted from a circular region with a 5$''$ radius around the sources, while the background ones from a region with 20$''$ radius not being contaminated with any signal from the nearby sources. {\it uvotsource} tool was used to convert the source counts using the conversion factors provided by \cite{poole}. The fluxes were corrected for extinction using the reddening coefficient $E(B-V)$ from the Infrared Science Archive \footnote{\url{http://irsa.ipac.caltech.edu/applications/DUST/}}. In the case of several observations for the same source, the analysis is performed using the same background region but validated that it is not contaminated by nearby objects in any filter. Also, light curves have been generated for investigation of the flux variation in each band. Then, if no significant variation is found, the spectral points are computed from the summed images, resulting in the flux estimation with reduced uncertainties.\\
The XRT data, taken both in photon counting and window timing modes were analyzed with standard XRTDAS tool distributed within the HEAsoft package (v6.25), applying standard procedures, filtering and screening criteria. The source counts were selected from a 20-pixel (47$''$) circle centered on the coordinates of each source, while those for the background- from an annulus with the same center and inner and outer radii of 51 (120$''$) and 85 pixels (200$''$), respectively. The Cash statistics \citep{1979ApJ...228..939C} on ungrouped data was used, as for many observations the number of counts was low and did not contain the minimum number of counts required for Gaussian statistics. The 0.3-10 keV X-ray spectrum of each source is fitted with XSPEC12.10.1 adopting an absorbed power-law model with a neutral hydrogen column density fixed to its Galactic value in each direction.\\
Initially, for the considered sources, the X-ray spectral analysis was performed for each observation. However, for most of the sources, the count rate was below $20$, preventing spectral fitting, and the photon index and flux could be estimated only for a few bright sources. Then, when several observations of the same source were available, they were merged to increase the photon statistics, and the averaged X-ray spectra were obtained. The merging was done using the tool available from the UK Swift Science Data Centre \footnote{\url{https://www.swift.ac.uk/user_objects/}} \citep{2009MNRAS.397.1177E}. Again XSPEC was used to fit the 0.3-10 keV spectrum, testing both the absorbed power-law and log-parabola models. We note that the spectra of many sources could be constrained only by merging the observations.
\section{Results of data analyses}\label{sec4}
\begin{figure*}
   \includegraphics[width=0.49 \textwidth]{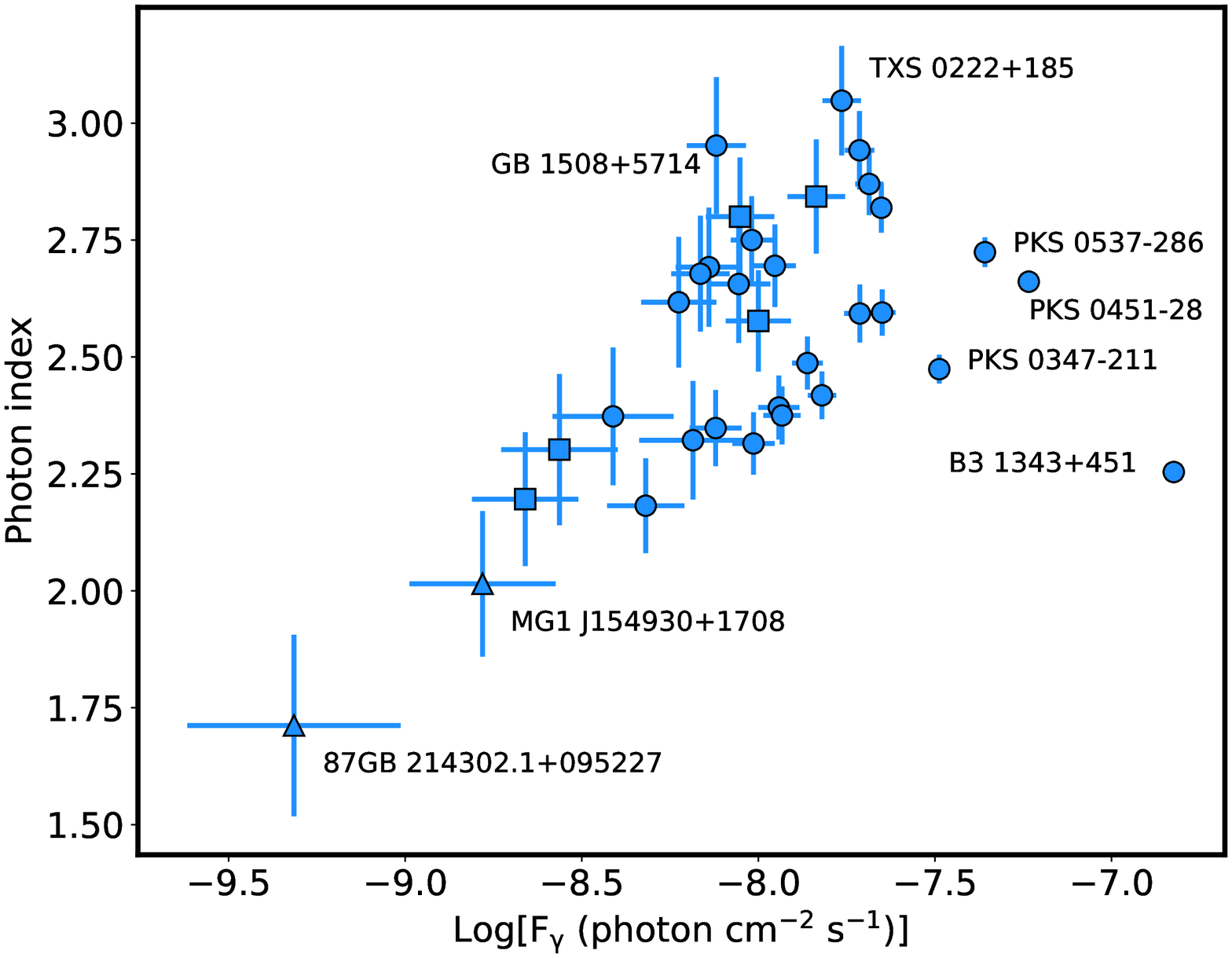}
    \includegraphics[width=0.49 \textwidth]{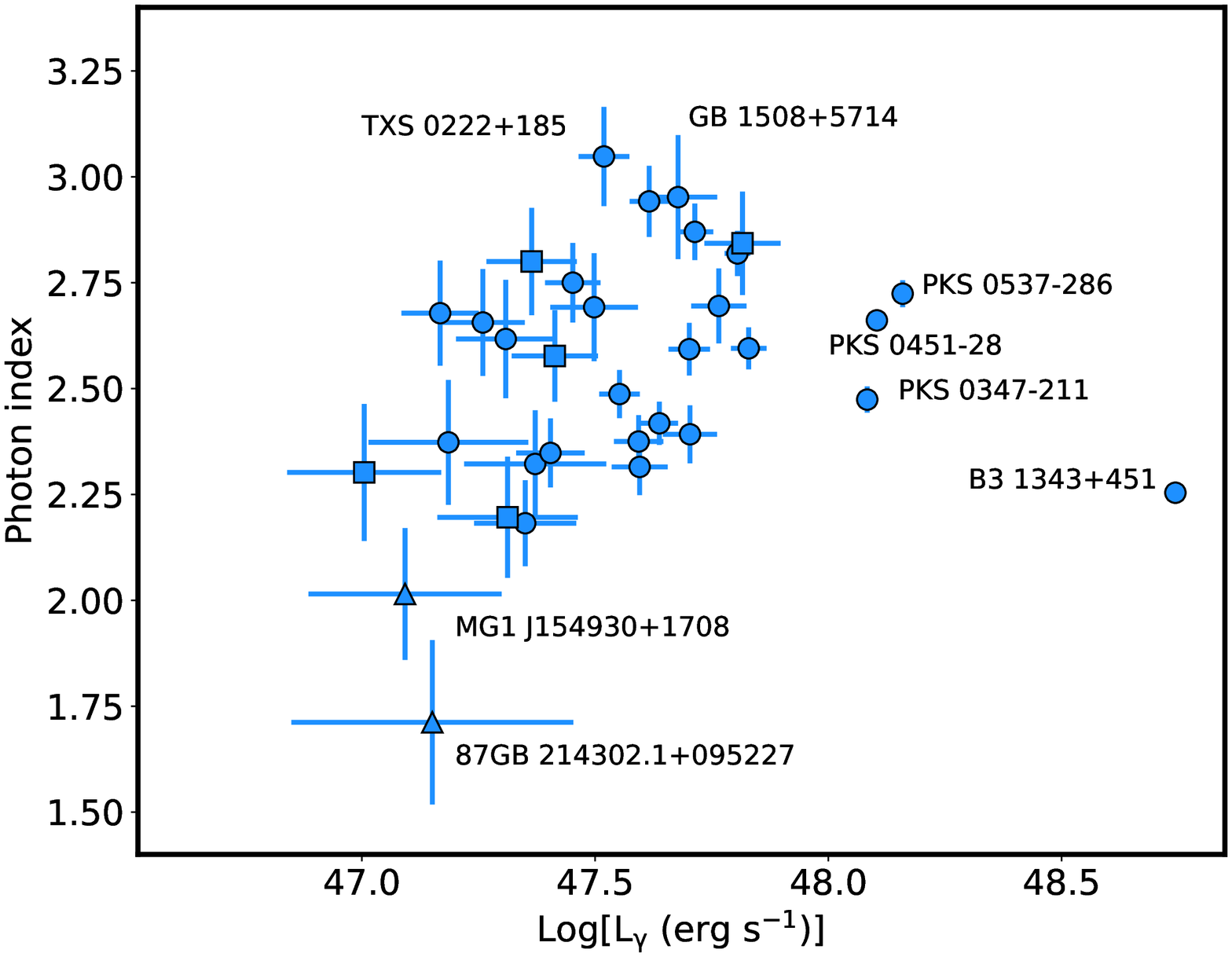}\\
     \includegraphics[width=0.49 \textwidth]{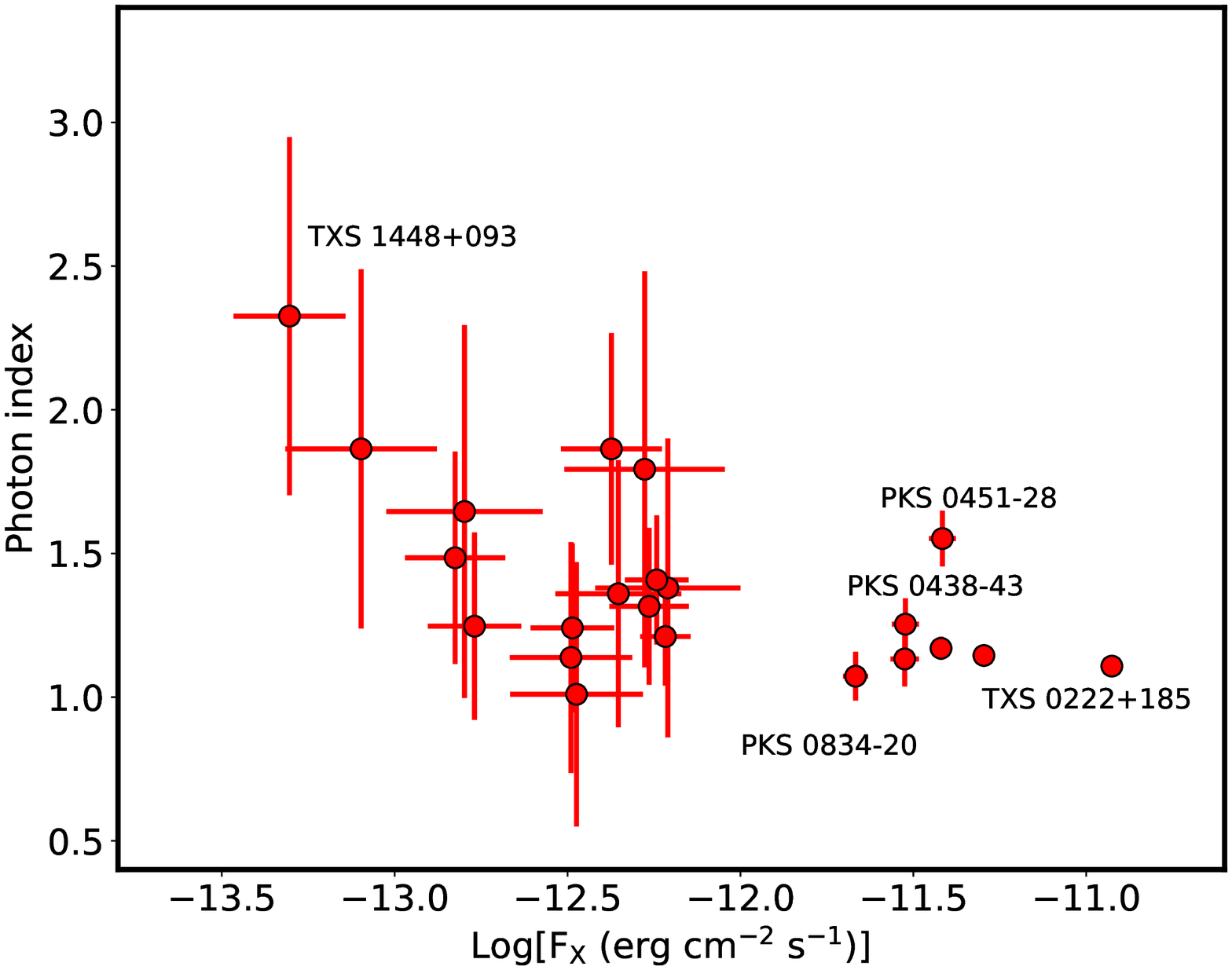}
     \includegraphics[width=0.49 \textwidth]{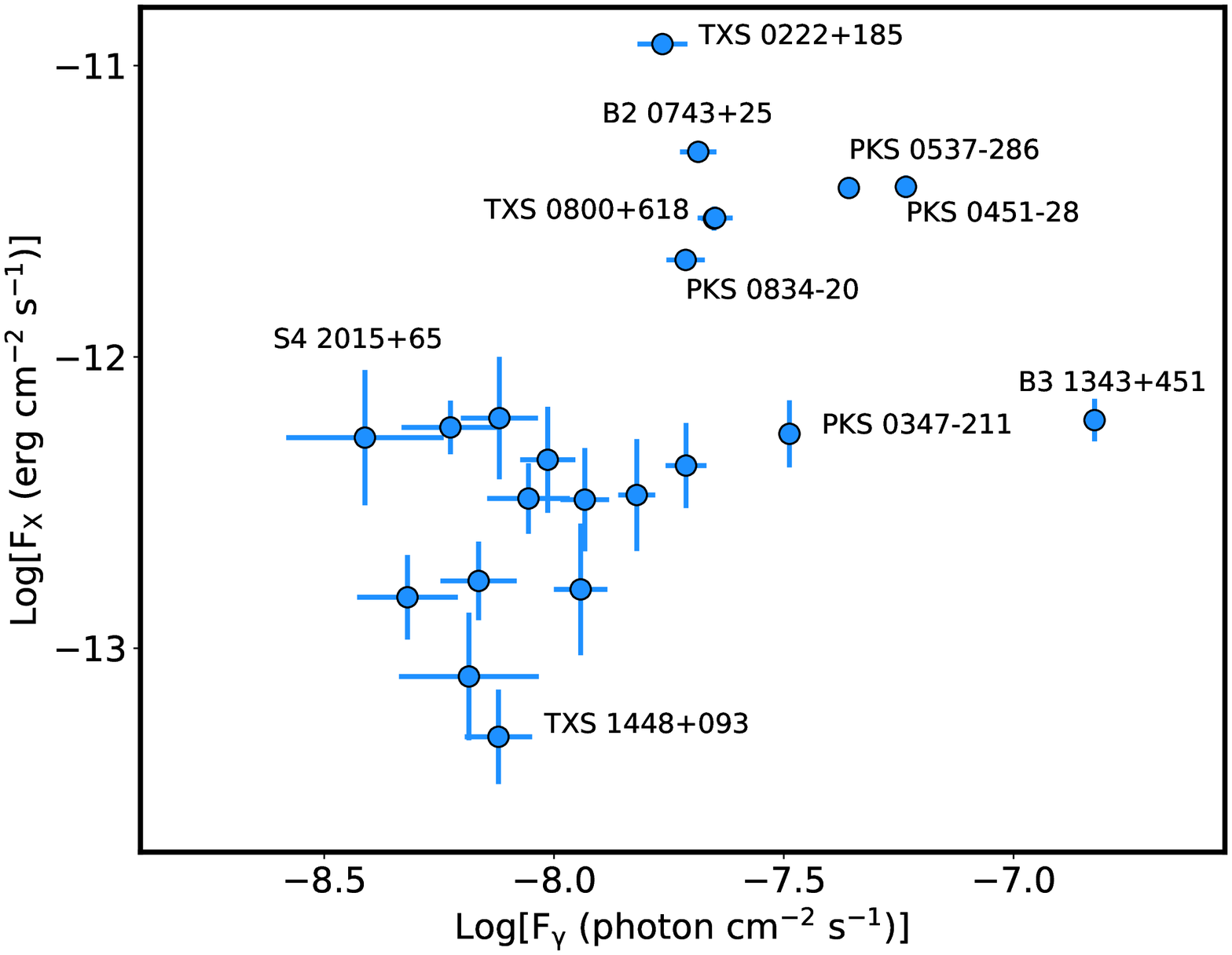}
      \caption{ The \gray flux ($>100$ MeV) and luminosity of considered sources versus the photon index are shown in the upper panels. BL lacs, FSRQs and BCUs are shown with triangles, circles, and squares, respectively. {\it Lower left panel:} The X-ray flux ($0.3-10$ keV) versus the photon index. {\it Lower right panel:} Comparison of \gray and X-ray (if available) fluxes.} 
         \label{gammaflux}
   \end{figure*}

The \gray data analysis results obtained from the power-law fit in the range from 100 MeV to 500 GeV are presented in Table \ref{sources} for each source, providing the \gray flux ($F_{\rm \gamma}$), photon index, the detection significance (TS), the energy of the highest-energy events ($E_{\rm max}$) detected from each object with the probability of its association with the target. The sources are detected with $TS>34.05$ significance, 
except for S4 2015+65, which appeared with $TS=23.77$. B3 1343+451 is detected with the highest significance of $TS=34652.79\:(186.15\sigma)$, allowing to perform a detailed spectral and variability analysis. In addition, for the sources with the spectra modeled by a log-parabola in 4FGL, an additional fit with a log-parabola model was performed, the results of which are presented in Table \ref{sourcesLP}. The curvature of the spectra of PKS 1351-018, B2 0743+25, PKS 0438-43, S4 2015+65, PKS 0834-20, and OD 166 is substantial ($\beta=0.24-0.63$), so their \gray spectrum quickly declines. The flux estimated from the power-law fitting which yielded soft \gray spectra as well does not substantially differ from that obtained with a log-parabola. The only noticeable difference is found for PKS 1351-018; the flux estimated from the log-parabola fitting is $(6.86\pm1.66)\times10^{-9}\:{\rm photon\:cm^{-2}\:s^{-1}}$ as compared with $(1.11\pm0.15)\times10^{-8}\:{\rm photon\:cm^{-2}\:s^{-1}}$ when the spectrum was modeled by a power-law. However, for this source $\beta=0.63\pm0.23$ was estimated implying its spectrum is curved significantly. Interestingly, the log-parabola fitting of S4 2015+65 spectrum resulted in detection of the source with $TS=27.96$.\\
The results presented in Table \ref{sources} are shown in Fig. \ref{gammaflux}. The \gray photon index versus flux (estimated from power-law fitting) is shown in the upper left panel, where the FSRQs are circles, BL LACs triangles and BCUs squares. Even though the number of sources is not enough for population studies, some difference in various blazar types can be seen. The photon index ($\sim E^{-\Gamma_{\rm \gamma}}$) estimated in the 0.1-500 GeV range ranges from $1.71$ to $3.05$ with a mean of $2.54$. The soft \gray spectra of the considered sources (except 87 GB 214302.1+095227 and MG1 J154930+1708) indicate that the peak of the HE component in their SED is in the MeV range. The flux of considered sources ranges from $4.84\times10^{-10}$ to $1.50\times10^{-7}\:{\rm photon\:cm^{-2}\:s^{-1}}$. The two BL Lacs detected beyond $z=2.5$, 87 GB 214302.1+095227 and MG1 J154930+1708, have the lowest flux, $(4.84\pm1.37)\times10^{-10}\:{\rm photon\:cm^{-2}\:s^{-1}}$ and $(1.66\pm0.79)\times10^{-9}\:{\rm photon\:cm^{-2}\:s^{-1}}$, respectively, but they have a harder \gray spectrum. The \gray photon index of 87GB 214302.1+095227, which is the only ISP object among the selected sources, is $1.71\pm0.19$ and that of MG1 J154930+1708 is $\Gamma_{\rm \gamma}=2.01\pm0.16$. The \gray flux of BCUs included in the sample ranges from $2.19\times10^{-9}\:{\rm photon\:cm^{-2}\:s^{-1}}$ to $1.46\times10^{-8}\:{\rm photon\:cm^{-2}\:s^{-1}}$ with $\Gamma_{\rm \gamma}$ within $2.20-2.84$. The FSRQs occupy the region of high flux $\geq3.89\times10^{-9}\:{\rm photon\:cm^{-2}\:s^{-1}}$ and $\Gamma_{\rm \gamma}\geq2.2$ with a mean $F_{\rm \gamma}\simeq 2.11\times10^{-8}\:{\rm photon\:cm^{-2}\:s^{-1}}$. The highest \gray flux of $(1.50\pm0.02)\times10^{-7}\:{\rm photon\:cm^{-2}\:s^{-1}}$ was observed from the bright FSRQ  B3 1343+451.\\
The highest energy events ($E_{\rm max}$) along with the probability of being associated with the sources are given in Table \ref{sources}. As the sources mostly have a soft \gray spectrum or the \gray data are better modeled by a log-parabola, their \gray spectra do not extend to HEs and the photon energies are below $20$ GeV; except for 87GB 214302.1+095227, MG2 J174803+3403, PMN J0226+0937 and B3 1343+451 from which photons with $67.15$, $31.63$, $56.42$ and $24.25$ GeV have been detected.\\
The flux, which is a function of the distance, is compared for each source in Fig. \ref{gammaflux} (upper left panel). Even though it is informative, the total energy released from each source cannot be investigated. Next, using the observed flux ($F_{\gamma}$) and photon index ($\Gamma_{\rm \gamma}$), the luminosity of each source is computed as:
\begin{equation}
L_{\gamma}=4\:\pi\:d_{\rm L}^2\:\frac{\Gamma_{\rm \gamma}-1}{\Gamma_{\rm \gamma}-2}\:\frac{{\rm E_{max}^{2-\Gamma_{\rm \gamma}}-E_{min}^{2-\Gamma_{\rm \gamma}}}}{{\rm E_{max}^{1-\Gamma_{\rm \gamma}}-E_{min}^{1-\Gamma_{\rm \gamma}}}}\: F_{\gamma}
\end{equation}
where ${\rm E_{min}=100}$ MeV and ${\rm E_{max}=500}$ GeV. Fig. \ref{gammaflux} (upper right panel) shows the distribution of the considered sources in the $\Gamma_{\rm \gamma}-L_{\gamma}$ plane. The \gray luminosity of considered sources ranges from $1.01\times10^{47}\:{\rm erg\:s^{-1}}$ to $5.54 \times 10^{48}\:{\rm erg\:s^{-1}}$. The lowest luminosity of $(1.01\pm0.38)\times10^{47}\:{\rm erg\:s^{-1}}$ has been estimated for CRATES J105433+392803 which is of the same order with that of the two BL Lacs included in the sample; $(1.42\pm0.98)\times10^{47}\:{\rm erg\:s^{-1}}$ for 87GB 214302.1+095227 and $(1.24\pm0.59)\times10^{47}\:{\rm erg\:s^{-1}}$ for MG1 J154930+1708. The luminosity of these BL Lacs corresponds to the highest end of the luminosity distribution of BL Lacs included in the fourth catalog of AGNs detected by \fermi \citep[Fig. 10 in][]{2020ApJ...892..105A}. The \gray luminosity of only PKS 0347-211, PKS 0451-28, PKS 0537-286 and B3 1343+451 exceeds $10^{48}\:{\rm erg\:s^{-1}}$ with the highest \gray luminosity of $(5.54\pm0.06)\times10^{48}\:{\rm erg\:s^{-1}}$,  estimated for B3 1343+451. Naturally, as compared to the distribution of all \gray emitting BL Lacs and FSRQs in the $\Gamma_{\rm \gamma}-L_{\gamma}$ plane \citep{2020ApJ...892..105A}, the blazars considered here occupy the highest luminosity range. We note that the luminosities shown in Fig. \ref{gammaflux} (upper right panel) have been computed based on the time-averaged \gray flux, and even higher luminosities are expected during \gray flares.\\
Table \ref{sourcesXray} shows the X-ray data analysis results, for each source presenting the neutral hydrogen column density, X-ray photon index ($\Gamma_{\rm X}$), flux, and C-stat/dof. The flux ranges from $\simeq5\times10^{-14}\:{\rm erg\:cm^{-2}\:s^{-1}}$ to $\simeq10^{-11}\:{\rm erg\:cm^{-2}\:s^{-1}}$, the highest flux of $(1.19\pm0.04)\times10^{-11}\:{\rm erg\:cm^{-2}\:s^{-1}}$ being observed for TXS 0222+185 ($z=2.69$). Interestingly, from the sources considered here, only FSRQs are detected in the X-ray band; among BL Lacs, there are no observations for MG1 J154930+1708, while for 87GB 214302.1+095227, even after merging six observations, only 10 counts are detected. Also, BCUs included in Table \ref{sources} have not been detected in the X-ray band.\\
The X-ray flux is plotted versus the photon index in the lower left panel of Fig. \ref{gammaflux}. For several sources, the number of detected counts was not high enough, so the flux and photon index were estimated with large uncertainties. The X-ray photon index of considered sources is $<2.0$, implying the X-ray spectra have a rising shape in the $\nu F\nu\sim\nu^{2-\Gamma_{\rm X}}$ representation, which is natural as LSP blazars are considered. The only exception is TXS 1448+093 with $\Gamma_{\rm X}=2.33\pm0.62$, but even when merging its all 26 observations, the observed counts were only 25. This source is relatively faint in the X-ray band with a flux of $(5.01\pm1.85)\times10^{-14}\:{\rm erg\:cm^{-2}\:s^{-1}}$, so even an exposure of $2.39\times10^{4}$ sec is not enough to detect a reasonable number of counts. B3 0908+416B has the hardest X-ray spectrum with $\Gamma_{\rm X}=1.01 \pm 0.46$. 
In the $F_{\rm X}-\Gamma_{\rm X}$ plane, PKS 0451-28, TXS 0222+185, PKS 0834-20, PKS 0537-286, TXS 0800+618, B2 0743+25 and PKS 0438-43 are detached from the other sources because they have a comparably high X-ray flux, $F_{\rm X-ray} \geq2.13\times10^{-12}\:{\rm erg\:cm^{-2}\:s^{-1}}$. In the lower right panel of Fig. \ref{gammaflux}, the \gray and X-ray (if available) fluxes of the considered sources are compared. Interestingly, the bright \gray sources PKS 0537-286 and PKS 0451-28 appear to be also bright X-ray emitters. The other bright \gray blazars (e.g., B3 1343+451, PKS 0451-28) do not have any distinguishable feature in the X-ray band, having a flux and photon index similar to those of the other considered sources. The bright X-ray sources TXS 0222+185, B2 0743+25, TXS 0800+618, and PKS 0834-20 appear with a similar flux in the \gray band, $F_{\rm \gamma}=(1.72-2.23)\times10^{-8}\:{\rm photon\:cm^{-2}\:s^{-1}}$. In Fig. \ref{sed1} the X-ray spectra of all sources included in Table \ref{sourcesXray} are shown in red.\\
The Swift UVOT data analysis was performed in all available filters. The results are consistent when different background regions are selected. Initially, all the single observations of the sources were analyzed to search for variability. However, the sources are relatively faint in the optical/UV bands and the large uncertainties in the flux estimation do not allow to investigate flux variation in time. Table \ref{tabeswift} summarizes the results of the UVOT data analysis after merging the observations for each source, presenting the fluxes in the six UVOT filters (if available) with errors. In Fig. \ref{sed1} these data are shown in light blue. In the SED of GB 1508+5714, PKS 1351-108, PKS 0537-286, TXS 0800+618, S4 1427+543, GB6 J0733+0456, B2 0743+25, PKS 0347-211, S41124+57, PKS 0438-43, S4 2015+65, PKS 0834-20, TXS 0222+185, TXS 0907+230, PMN J0226+0937, PKS 0451-28 and PKS 2107-105, a thermal blue-bump component can be seen, which may represent the emission directly from the disc \citep{2009MNRAS.399.2041G}.
\begin{figure*}
    \includegraphics[width=0.99 \textwidth]{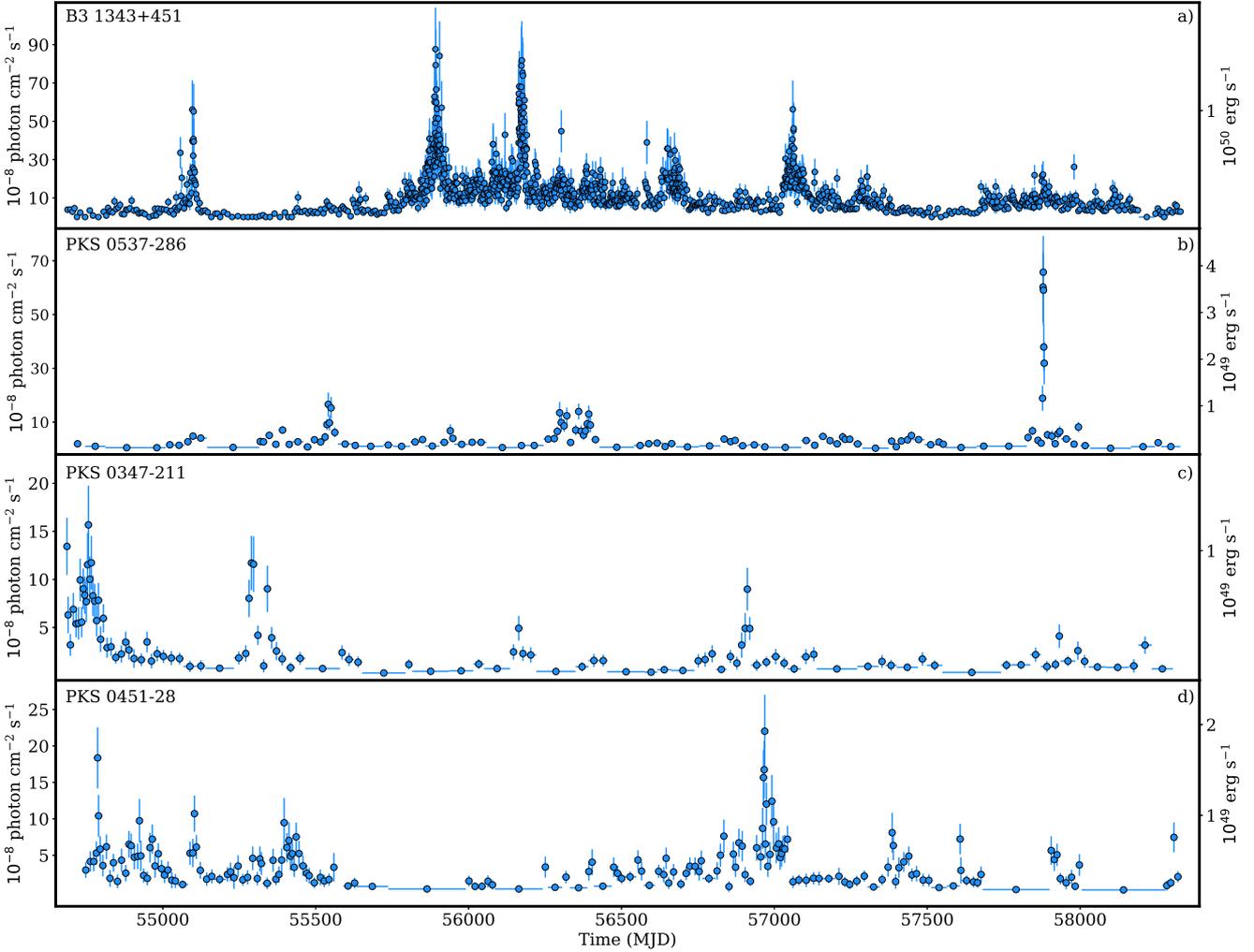}
      \caption{\gray light curve of B3 1343+451, PKS 0537-286, PKS 0347-211, and PKS  0451-28 for the period from August 2008  to August 2018 calculated using adaptively-binned timescales.
              }
         \label{lightcurve}
\end{figure*}

\begin{figure*}
   \includegraphics[width=0.49 \textwidth]{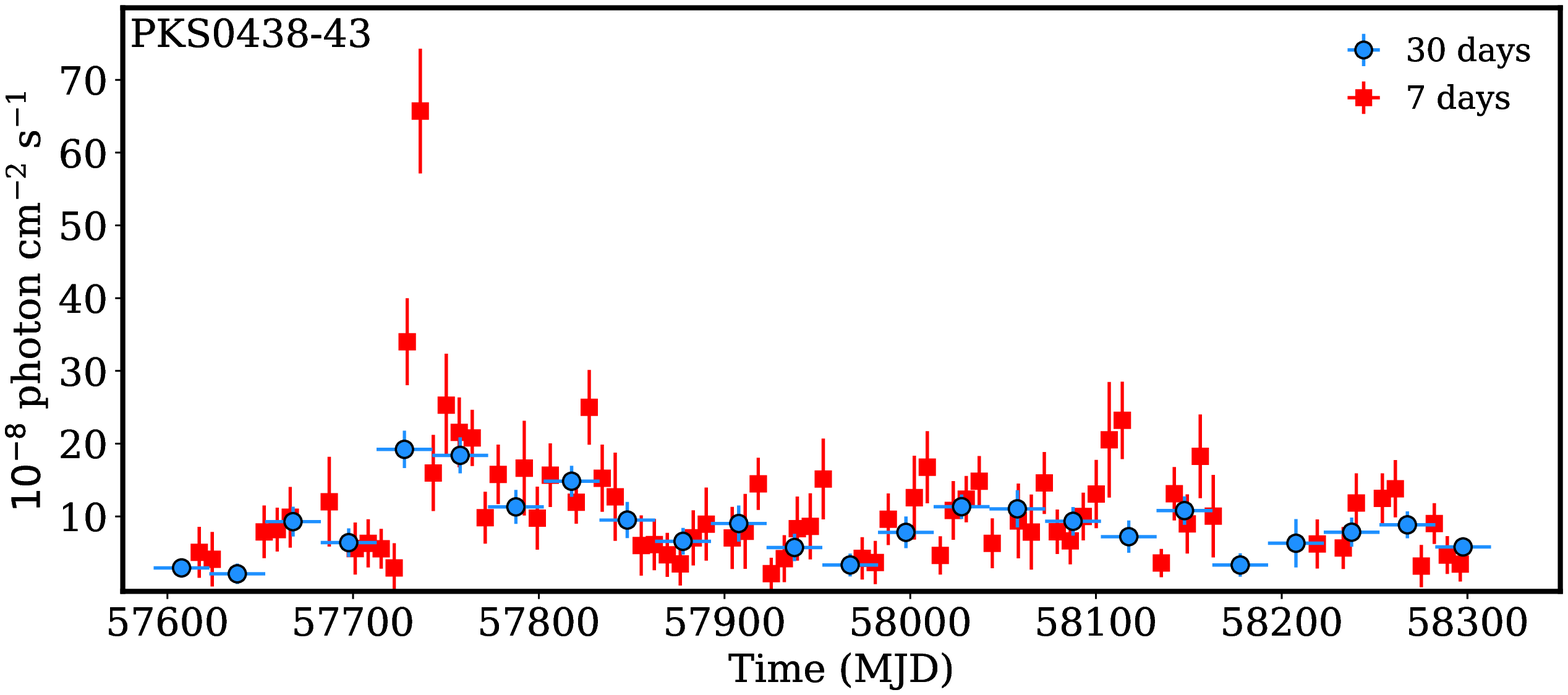}
   \includegraphics[width=0.49 \textwidth]{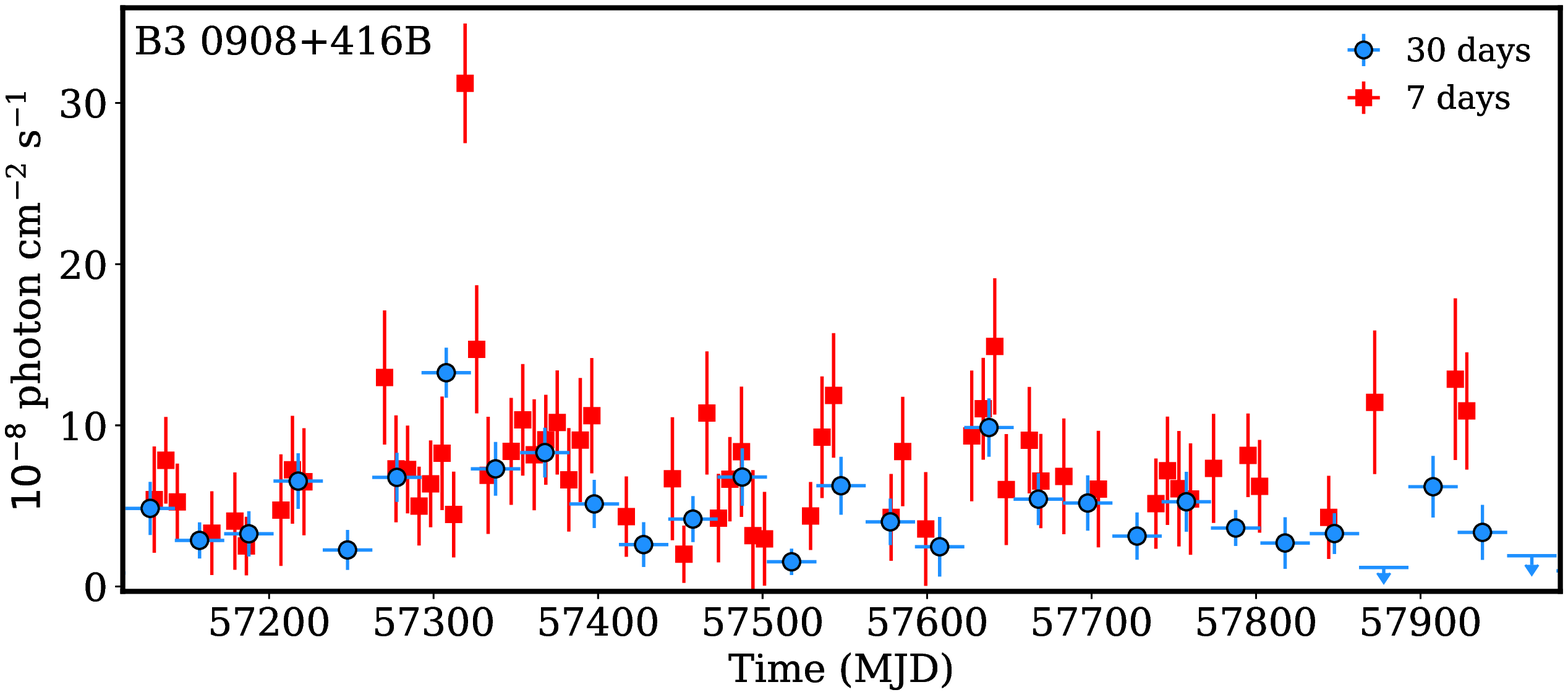}\\
    \includegraphics[width=0.49 \textwidth]{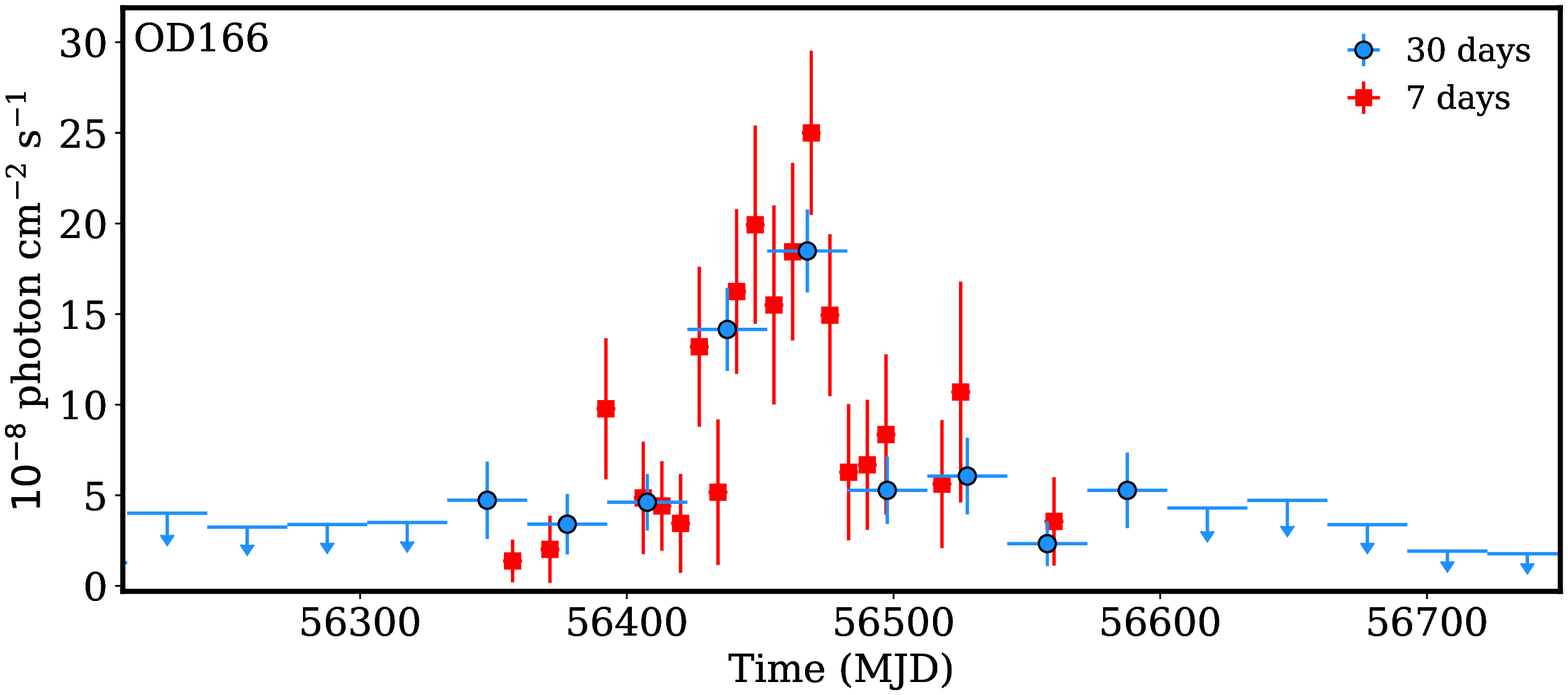}
   \includegraphics[width=0.49 \textwidth]{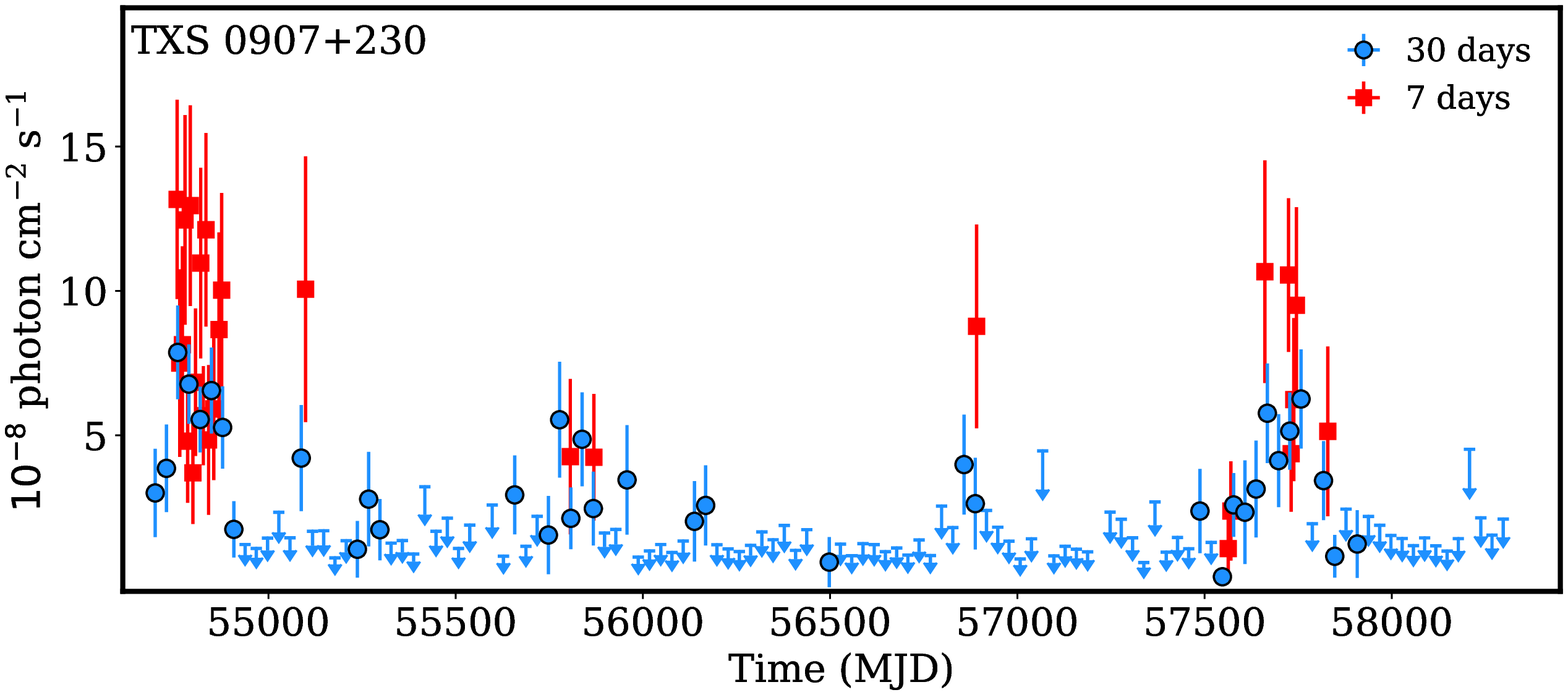}\\
    \includegraphics[width=0.49 \textwidth]{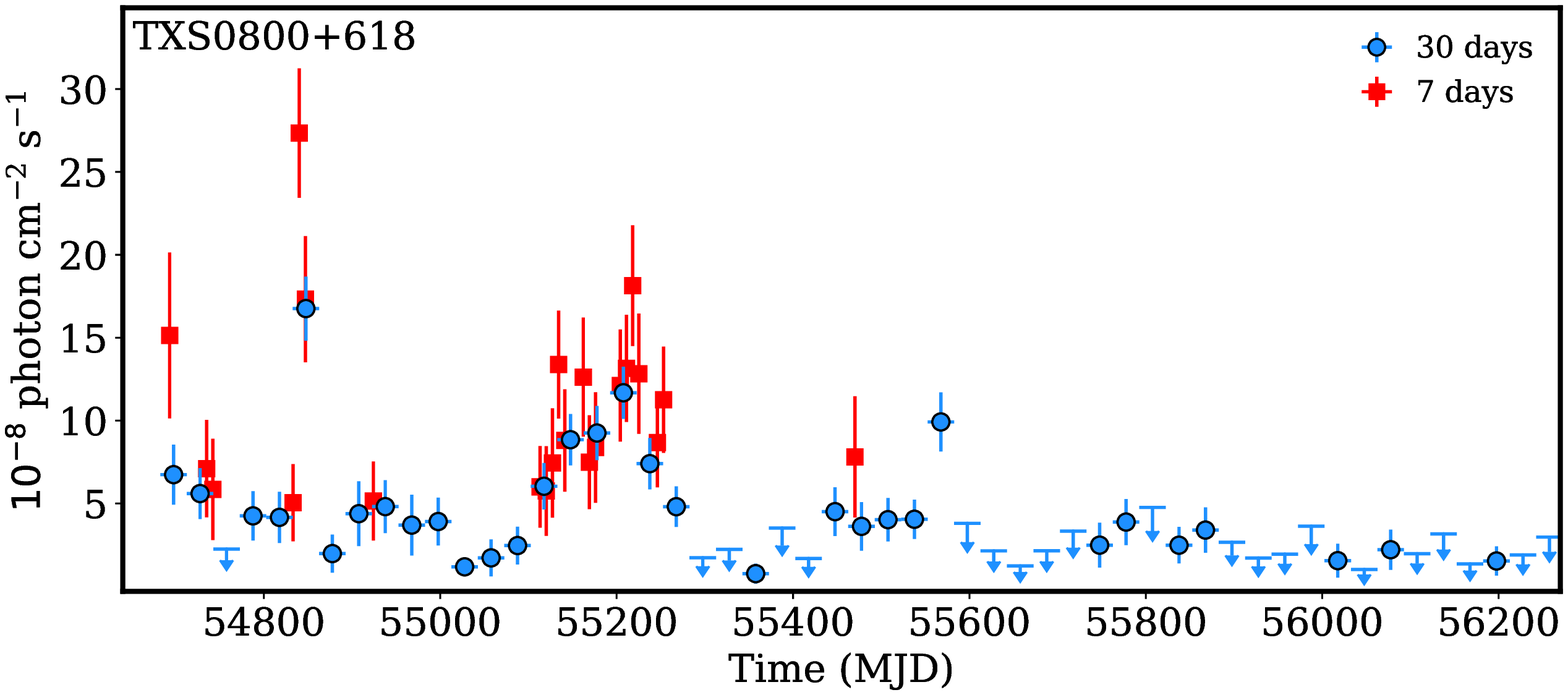}
   \includegraphics[width=0.49 \textwidth]{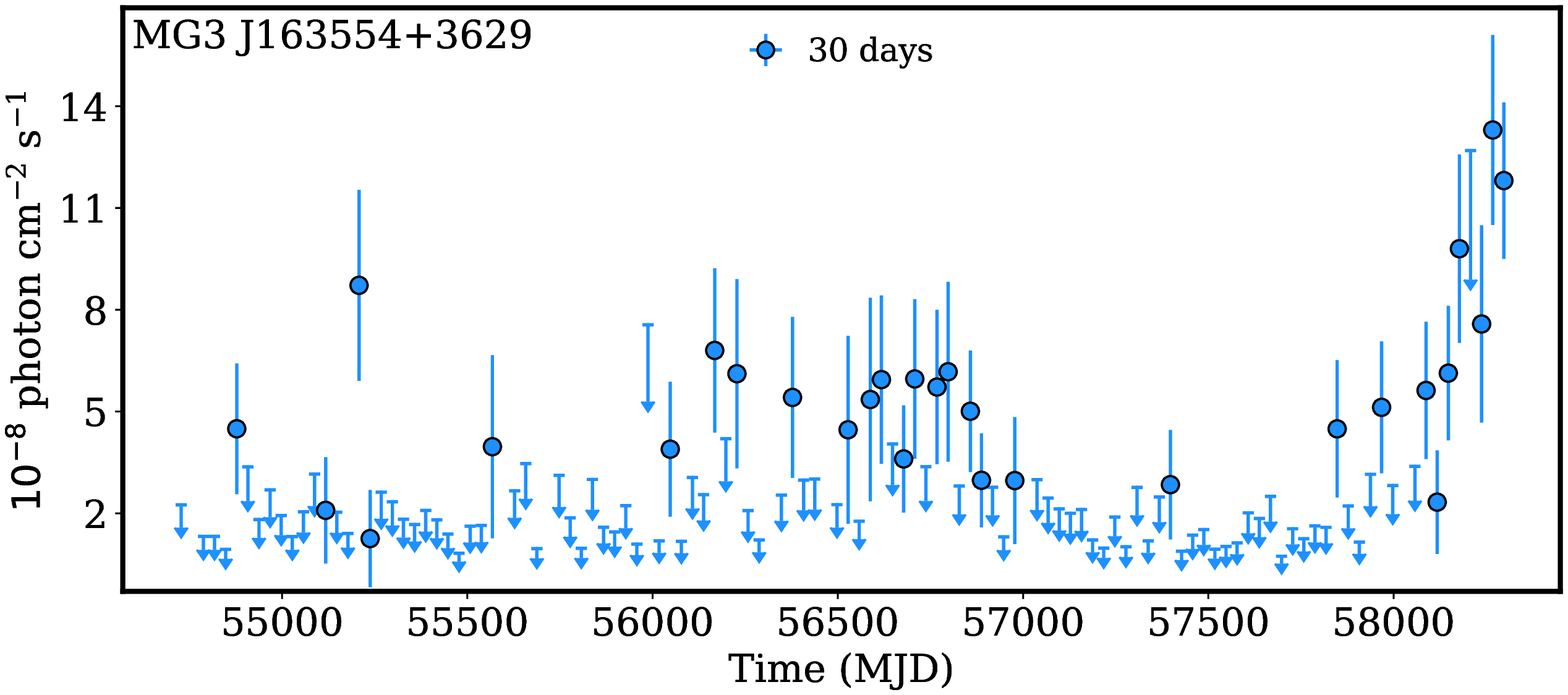}\\
   \includegraphics[width=0.49 \textwidth]{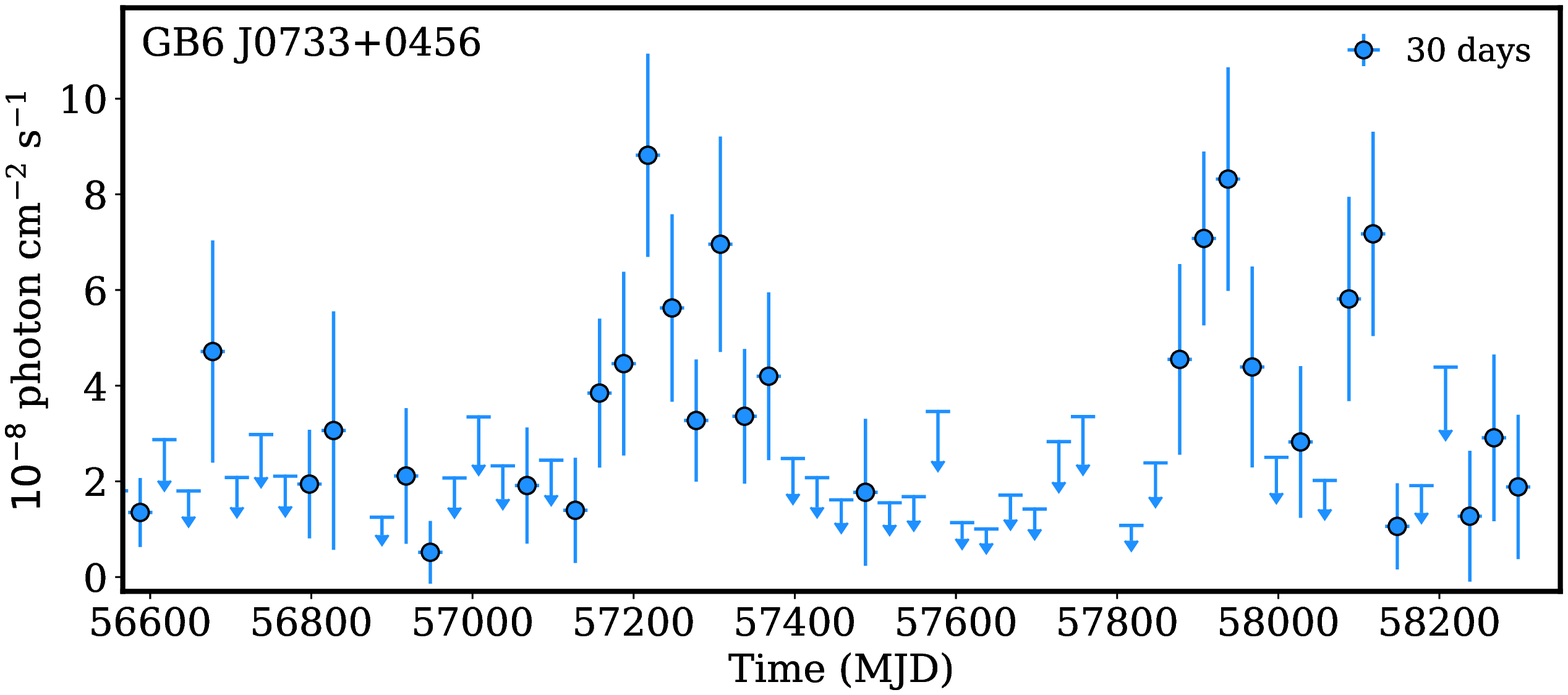}
   \includegraphics[width=0.49 \textwidth]{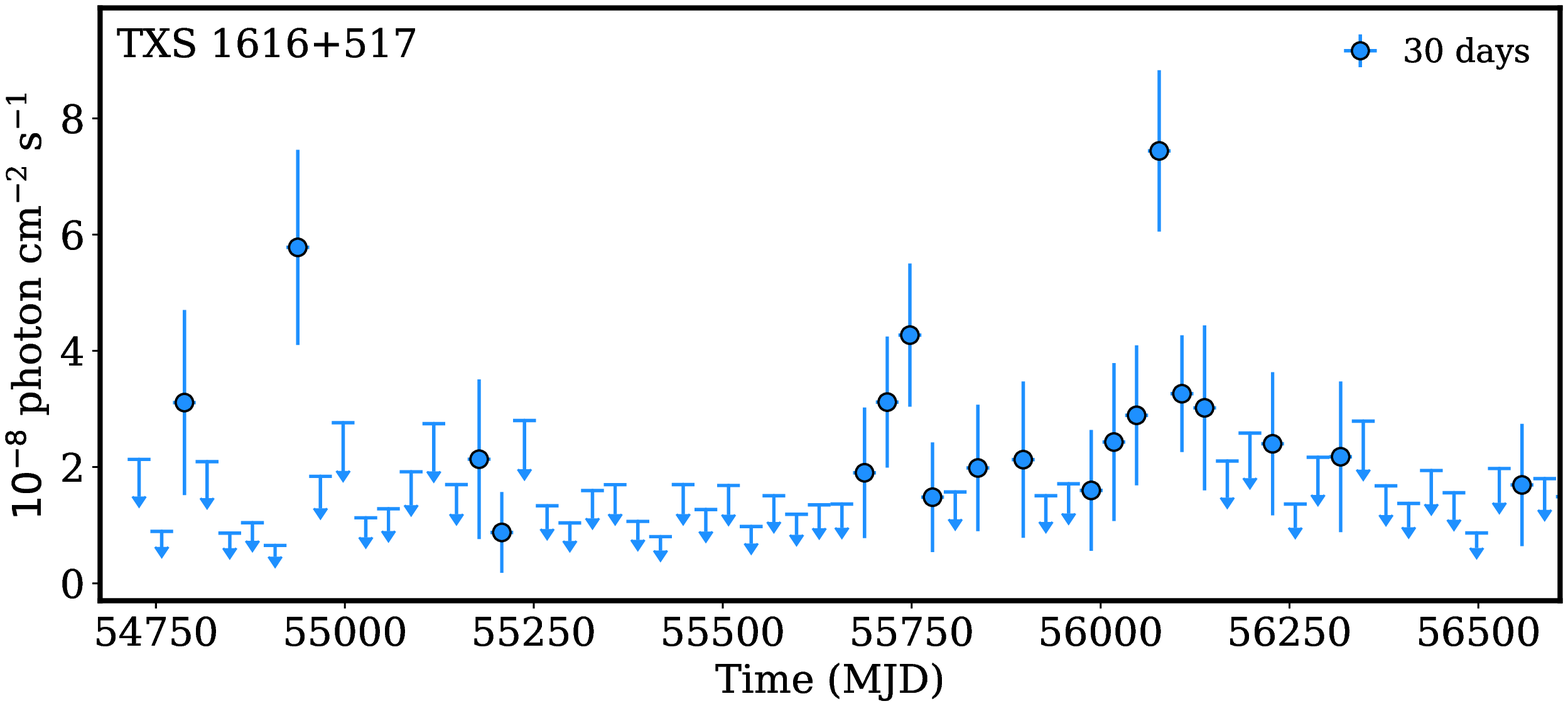}\\
  \includegraphics[width=0.49 \textwidth]{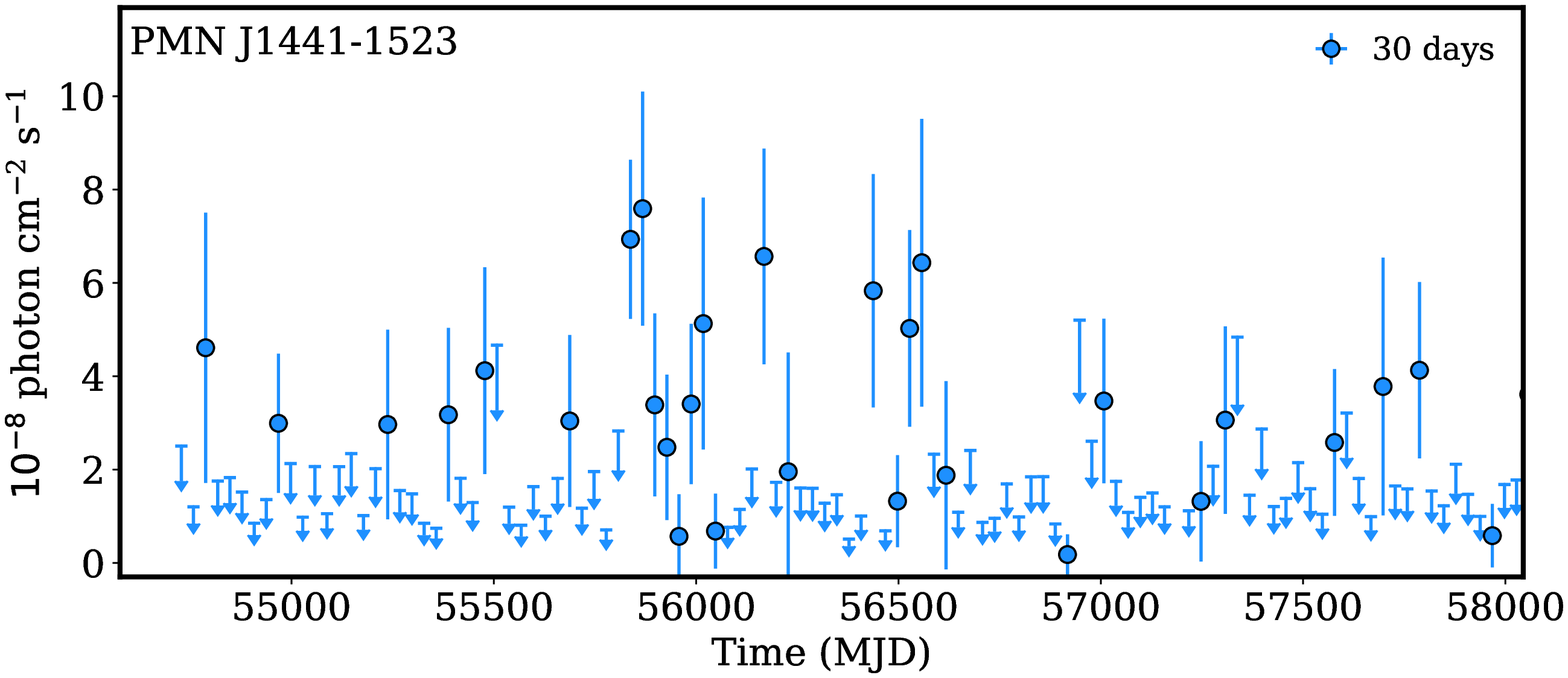}
	\includegraphics[width=0.49 \textwidth]{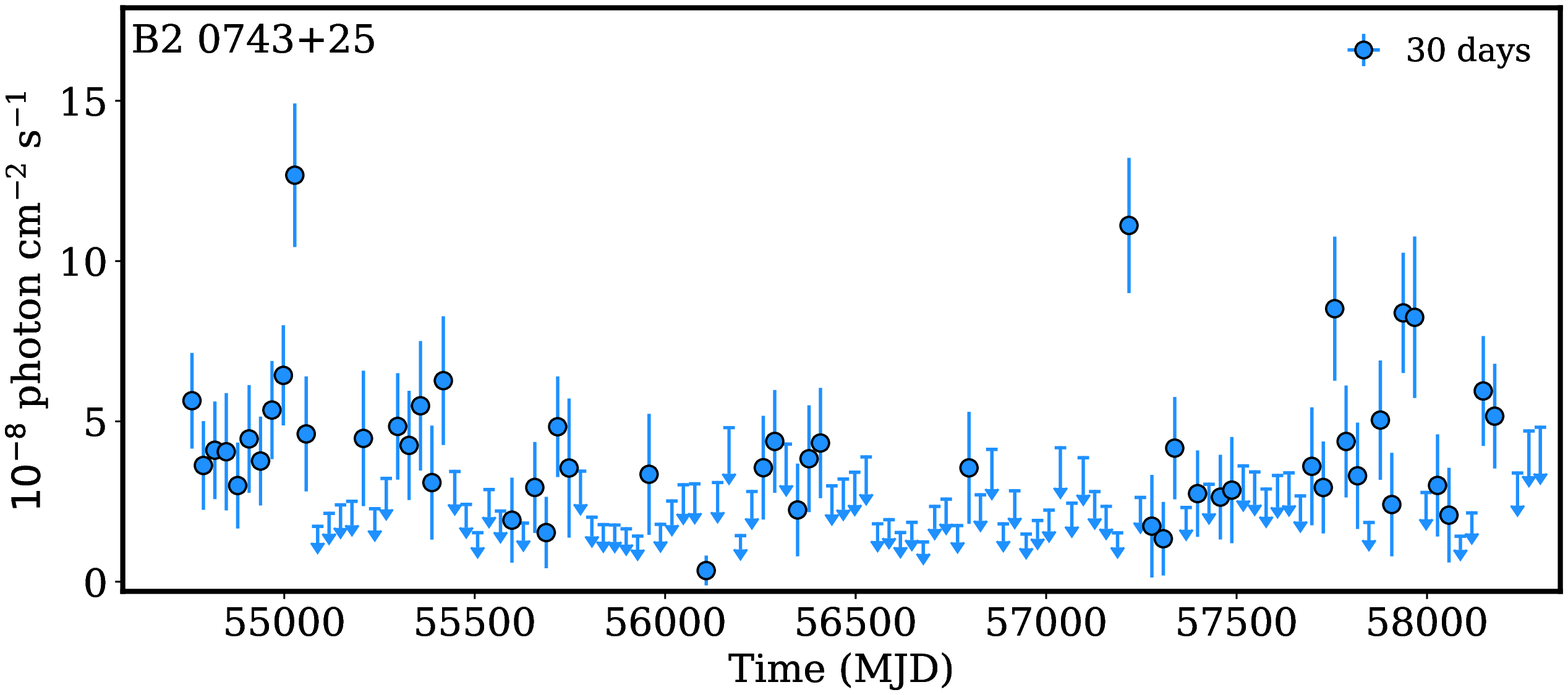}
      \caption{7-day (red) and 30-day (blue) binned light light curves of PKS 0438-43, B3 0908+416B, OD 166, TXS 0907+230, TXS 0800+618, MG3 J163554+3629, GB6 J0733+0456, TXS 1616+517, PMNJ 1441-1523 and B2 0743+25. 
              }
         \label{lightcurvelong}
\end{figure*}
\subsection{X-ray variability}
\begin{figure}
   \includegraphics[width=0.49 \textwidth]{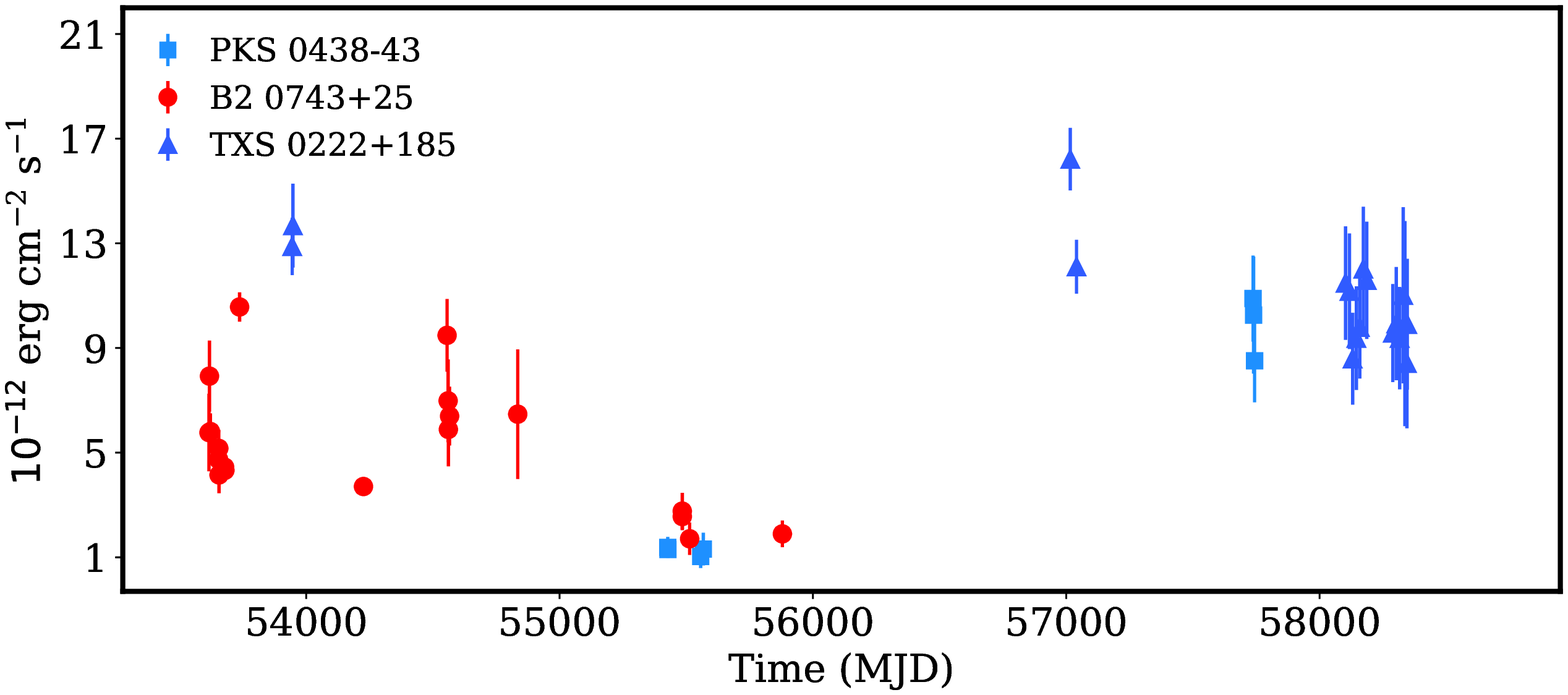}
      \caption{The 0.3-10 keV X-ray fluxes of PKS 0438-43, B2 0743+25 and TXS 0222+185 measured by Swift-XRT.}
         \label{xlightcurve}
\end{figure}

The X-ray flux variability of some sources observed by Swift multiple times have been investigated. The data from each observation were processed and analyzed, but only for eight objects (marked with asterisks in Table \ref{sourcesXray}) the number of counts was enough to constrain the flux and the photon index in a single observation. Except for PKS 0438-43, B2 0743+25 and TXS 0222+185, the X-ray emission appeared to be relatively constant, though the sources were observed in different years. For example, the X-ray flux of PKS 0537-286 was $(4.18\pm0.74)\times10^{-12}\:{\rm erg\:cm^{-2}\:s^{-1}}$ and $(4.53\pm0.89)\times10^{-12}\:{\rm erg\:cm^{-2}\:s^{-1}}$ on 26 October, 2006 and 12 May 2017, respectively. Fitting the flux observed in different years with a constant flux, the $\chi2$ test results in $P(\chi2)=0.38$ and $\chi2/dof=1.07$ which are consistent with no variability. Similarly, the $\chi2$ test shows that the X-ray flux of TXS 0800+618, PKS 0834-20, PKS 0451-28 and B3 1343+451 is constant over different years. There is a marginal evidence ($P(\chi2)=0.038$ and $\chi2/dof=1.68$) that the X-ray flux of TXS 0222+185 is variable; it is mostly around $(0.96-1.1)\times10^{-11}\:{\rm erg\:cm^{-2}\:s^{-1}}$ which increased to $(1.62\pm0.13)\times10^{-11}\:{\rm erg\:cm^{-2}\:s^{-1}}$ and $(1.37\pm0.16)\times10^{-11}\:{\rm erg\:cm^{-2}\:s^{-1}}$ on 24 December 2014 and 31 July 2006, respectively. Instead, the $\chi2$ test shows that the X-ray emission of PKS 0438-43 and B2 0743+25 is variable with $P(\chi2)<5.1\times10^{-6}$. PKS 0438-43 was in a bright X-ray state on 15 December 2016 with a flux of $(1.09\pm0.16)\times10^{-11}\:{\rm erg\:cm^{-2}\:s^{-1}}$ as compared with the flux of $(1.30\pm0.31)\times10^{-11}\:{\rm erg\:cm^{-2}\:s^{-1}}$ in the quiescent state. Similarly, the X-ray flux of B2 0743+25 in the high state on 01 January 2006 was $(1.06\pm0.56)\times10^{-11}\:{\rm erg\:cm^{-2}\:s^{-1}}$. The variation of the 0.3-10 keV X-ray flux of PKS 0438-43, B2 0743+25 and TXS 0222+185 is shown in Fig. \ref{xlightcurve}. No variation of X-ray photon index was found, due to the large uncertainties in its estimation.
\subsection{\gray variability}
The continuous observation of the considered sources by \fermi allows a detailed investigation of their \gray flux variation during the considered ten years. When the source detection significance is $<10\:\sigma$, the data are not enough for variability searches in month scales. Their emission could be variable in longer scales (6 or 12 months) more common for non-blazar AGNs \citep{2018A&A...614A...6S}, which is not investigated here. Initially, the light curves of all sources were calculated with the help of an adaptive binning method. However, the adaptively binned intervals were possible to compute only for the source with a detection significance of $>37\:\sigma$. The adaptively binned light curves of B3 1343+451, PKS 0451-28, PKS 0347-211 and PKS 0537-286 computed above $E_{\rm 0}=163.9$, $163.2$, $187.4$ and $151.0$ MeV, respectively, are shown in Fig. \ref{lightcurve}. B3 1343+451 is the brightest and most variable source in the sample, and in its adaptively binned light curve (Fig. \ref{lightcurve} a) several bright \gray emission periods can be identified: around MJD 55100, MJD 55890, MJD 56170, MJD 56640 and MJD 57060. During the bright \gray flaring periods, the flux changes in sub-day scales; there are $200$ adaptive bins with a high flux and a width shorter than a day; the minimum adaptively binned time width is $6.33$ hours observed on MJD 56176.34 when the source flux was $(7.52\pm1.85)\times10^{-7}\:{\rm photon\:cm^{-2}\:s^{-1}}$ (above 163.9 MeV). Moreover, the width of 83 bins from these $200$ is even shorter than 12 hours, which were mostly observed during the flares around MJD 55890 (24 bins) and MJD 56170 (35 bins). In the quiescent state (e.g., before MJD 55800, expect for the flare on MJD 55100) the \gray flux of the source is $\simeq(1-5)\times10^{-8}\:{\rm photon\:cm^{-2}\:s^{-1}}$ which is lower than that averaged over 10 years (including the bright flaring states), reported in Table \ref{sources}. The highest \gray flux of $(8.77\pm2.16)\times10^{-7}\:{\rm photon\:cm^{-2}\:s^{-1}}$ above $E_{\rm 0}=163.9$ MeV with $\Gamma_{\rm \gamma}=2.48\pm0.29$ was observed on MJD 55891.7 with a detection significance of $9.1\sigma$ within a time bin having a width of 8.2 hours. It corresponds to a flux of $(1.82\pm0.45)\times10^{-6}\:{\rm photon\:cm^{-2}\:s^{-1}}$ above $100$ MeV which is $36.4$ times higher than the \gray flux of the source in the quiescent state (before MJD 55800) but the $\Gamma_{\rm \gamma}$ is within the uncertainties of the value given in Table \ref{sources}. During the hardest \gray emission period, $\Gamma_{\rm \gamma}=1.45\pm0.21$ was detected on MJD 56432 with a significance of $7.2\sigma$, which is unusual for this source. Interestingly, there are twelve periods, mostly during the \gray flares, when its \gray spectrum was hard ($\leq1.70$); one such period had been observed on MJD 55900.53 when, within a time bin having a width of 12.90 hours, $\Gamma_{\rm \gamma}$ was $1.64\pm0.16$ with a detection significance of $10.79\sigma$. In the quiescent state, its \gray luminosity is $(2-4)\times10^{48}\:{\rm erg\: s^{-1}}$ which increased up to $\sim1.5\times10^{50}\:{\rm erg\: s^{-1}}$ during the bright \gray flares (Fig. \ref{lightcurve} (a)).\\
The most distant source showing a substantial \gray flux increase in short periods is PKS 0537-286 at $z=3.10$ (Fig. \ref{lightcurve} b). The flaring activity of this source was first reported in \citep{2017ATel10356....1C} and the rapid (6 and 12 hours) \gray flux variations in \citet{2018ApJ...853..159L}. During the extreme \gray flaring period from MJD 57878.05 to MJD 57881.55, the adaptively binned light curve confirms the intra-day \gray flux increase of PKS 0537-286 ; the adaptive time bin widths are 20.45, 15.96 and 18.90 hours. For comparison, during the flux increase around MJD 55540 and MJD 56390 the minimum time widths are 2.38 and 3.79 days, respectively. The average \gray flux of PKS 0537-286 is $(4.38 \pm 0.18)\times10^{-8}\:{\rm photon\:cm^{-2}\:s^{-1}}$ (Table \ref{sources}), but it significantly increased in MJD 55528-55553, MJD 56264-56400 and MJD 57878.0-57883.4. In the last period, during five consecutive time intervals, the \gray flux above $151.0$ MeV was higher than $3\times10^{-7}\:{\rm photon\:cm^{-2}\:s^{-1}}$, with a maximum of $(6.58\pm 1.35)\times10^{-7}\:{\rm photon\:cm^{-2}\:s^{-1}}$ observed on MJD 57879.2 with a detection significance of $10.57\sigma$. This corresponds to a flux of $(1.29\pm 0.26)\times10^{-6}\:{\rm photon\:cm^{-2}\:s^{-1}}$ above 100 MeV. During these periods, $\Gamma_{\rm \gamma}$ is $2.45\pm0.23$, $2.64\pm0.27$, $2.56\pm0.28$, $2.79\pm0.34$ and $2.91\pm0.35$, not significantly different from the value reported in Table \ref{sources} with no spectral hardening, which shows the emission is dominated by the MeV photons. However, during the flares, the luminosity of the source can be as high as $\simeq4\times10^{49}\:{\rm erg\: s^{-1}}$, putting PKS 0537-286 in the list of the brightest \gray blazars.\\
The adaptively binned light curves of PKS 0347-211 and PKS 0451-28 (Fig. \ref{lightcurve} c and d) show several periods of \gray brightening, when a \gray flux increase within day scales is observed. For example, the shortest time interval when the flux increases is 2.65 days for  PKS 0347-211, and it is 1.56 days for PKS 0451-28. The light curves of both sources reveal several \gray flaring periods when the flux substantially increased. For example, on MJD $54757.04\pm2.71$ the \gray flux of PKS 0347-211 above 187.4 MeV was $(1.57\pm0.41)\times10^{-7}\:{\rm photon\:cm^{-2}\:s^{-1}}$, which corresponds to a flux of $(3.57\pm0.93)\times10^{-7}\:{\rm photon\:cm^{-2}\:s^{-1}}$ above 100 MeV. In the case of PKS 0451-28, the peak \gray flux of $(2.20\pm	0.50)\times10^{-7}\:{\rm photon\:cm^{-2}\:s^{-1}}$ (above $163.2$ MeV) was observed on MJD $56968.60\pm 0.79$ with $9.64\sigma$. This corresponds to a flux of $(3.70\pm0.84)\times10^{-7}\:{\rm photon\:cm^{-2}\:s^{-1}}$ above 100 MeV. During this period, $\Gamma_{\rm \gamma}$ was $2.06\pm0.19$.\\
The light curves generated with the help of the adaptive binning method allowed to identify periods when the flux of some of the sources considered here (Fig. \ref{lightcurve}) increased in sub-day or days scales. It should be mentioned that expect for B3 1343+451, the short time scale variability of the other sources cannot be investigated using the regular time binning method, because in a large number of periods only upper limits are derived. Also, because of low statistics, the adaptively binned light curves were possible to compute only for the source presented in Fig. \ref{lightcurve}. For the other sources included in Table \ref{sources} the variability on week and month scales are investigated.
In order to identify whether the \gray emission is variable or not, a simple $\chi2$ test was performed \citep{2010ApJ...722..520A}; the flux  measured in each interval was fitted by a constant flux and the reduced $\chi2$ and the probability of the flux being constant are computed.\\
The $\chi2$ fitting indicated that the \gray emission of B3 0908+416B, TXS 0800+618, PKS 0438-43, OD 166 and TXS 0907+230 is variability in week scales while that of MG3 J163554+3629, GB6 J0733+0456, B2 0743+25, PMN J1441-1523 and TXS 1616+517 in month scales. For all these sources, $P(\chi2)<2.16\times10^{-4}$ was estimated. The \gray light curves of these sources with an evident increase in the flux are shown in Fig. \ref{lightcurvelong}. For example, during MJD $57729.16\pm3.5$ and MJD $57736.16\pm3.5$, the \gray flux of  PKS 0438-43 above 100 MeV increased $15.2-29.3$ times as compared to its average flux and was $(3.40\pm0.59)\times10^{-7}\:{\rm photon\:cm^{-2}\:s^{-1}}$ and $(6.57\pm	0.86)\times10^{-7}\:{\rm photon\:cm^{-2}\:s^{-1}}$, respectively. Likewise, the 7-day averaged peak values of the \gray flux of B3 0908+416B, OD 166, TXS 0800+618 and TXS 0907+230 were $(3.12\pm 0.37)\times10^{-7}\:{\rm photon\:cm^{-2}\:s^{-1}}$, $(2.50\pm 0.45)\times10^{-7}\:{\rm photon\:cm^{-2}\:s^{-1}}$, $(2.73\pm	0.39)\times10^{-7}\:{\rm photon\:cm^{-2}\:s^{-1}}$ and $(1.32\pm0.34)\times10^{-7}\:{\rm photon\:cm^{-2}\:s^{-1}}$, respectively, which exceed the corresponding values given in Table \ref{sources}. The most distant \gray flaring blazar observed so far is MG3 J163554+3629 at $z=3.65$; this source was reported to be in an active state on July 7 2018, when its daily averaged  peak value of \gray flux was $(6.4\pm	1.15)\times10^{-7}\:{\rm photon\:cm^{-2}\:s^{-1}}$ \citep{2018ATel11847....1P}. The monthly averaged maximum \gray flux of $(1.33\pm	0.28)\times10^{-7}\:{\rm photon\:cm^{-2}\:s^{-1}}$ was observed on MJD $58267.66\pm15$ (May- June 2018). The source was also bright in \grays with a flux of $(1.18\pm0.23)\times10^{-7}\:{\rm photon\:cm^{-2}\:s^{-1}}$ during June-July 2018 when  the maximum daily averaged flux was observed \citep{2018ATel11847....1P}.\\
\begin{figure*}
   \includegraphics[width=0.99 \textwidth]{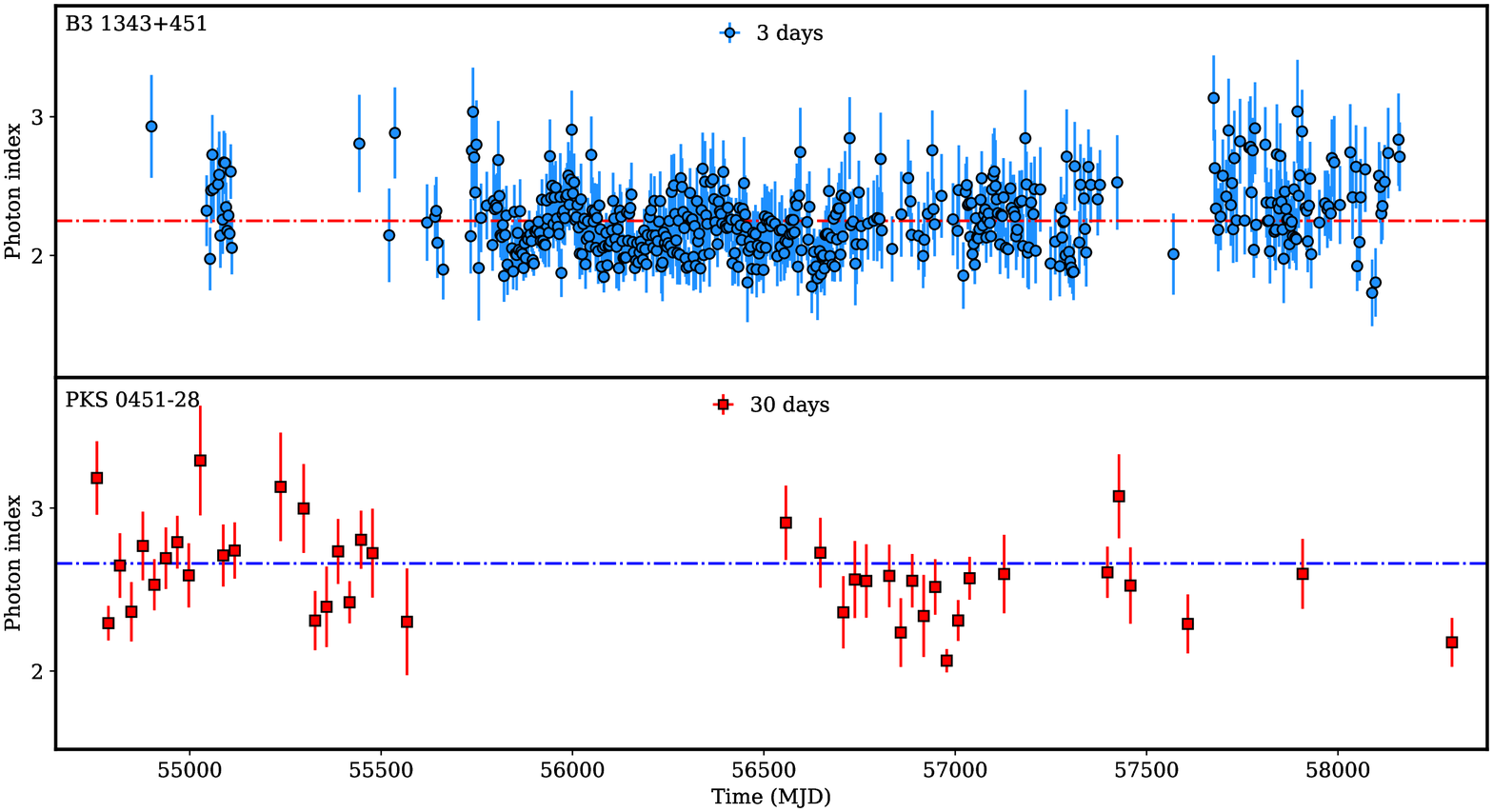}\\
   \includegraphics[width=0.49 \textwidth]{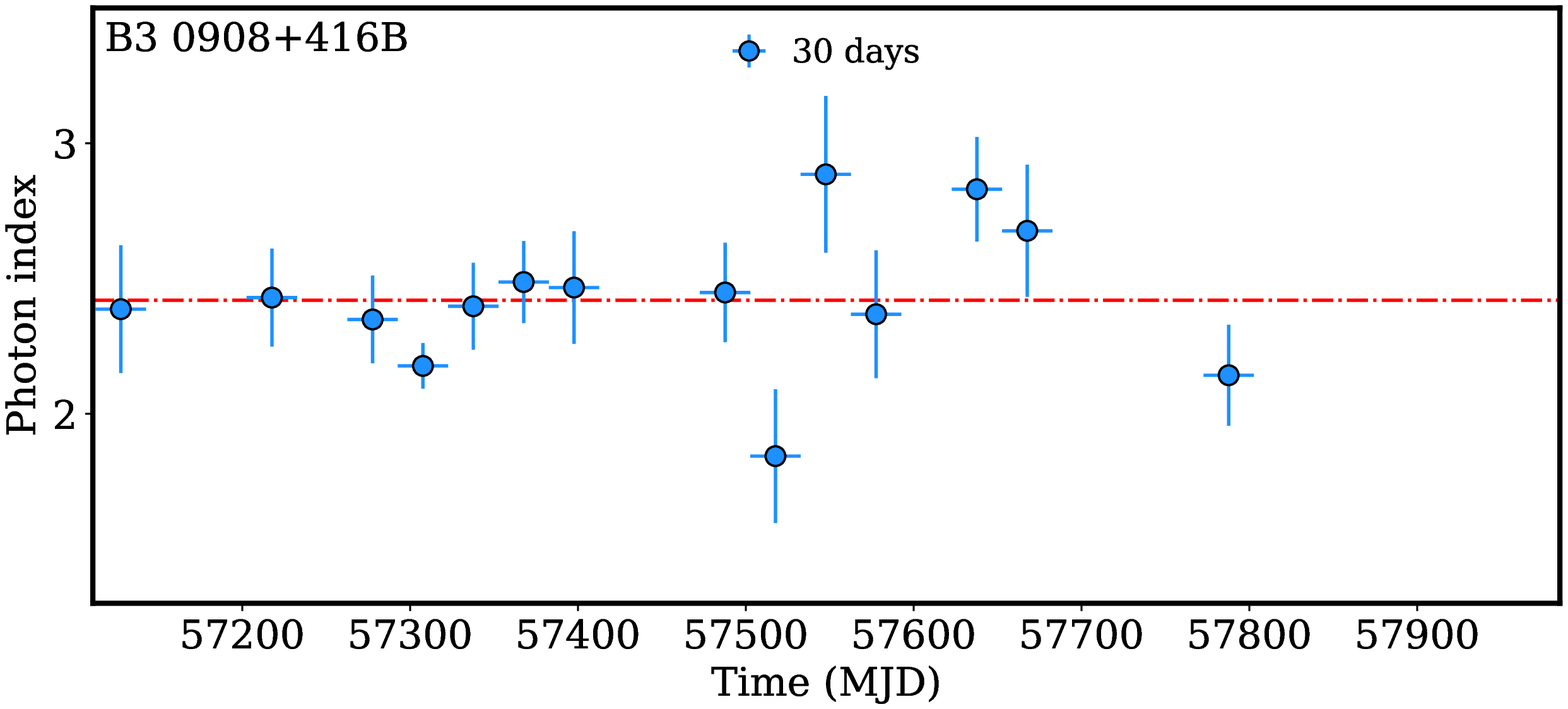}
    \includegraphics[width=0.49 \textwidth]{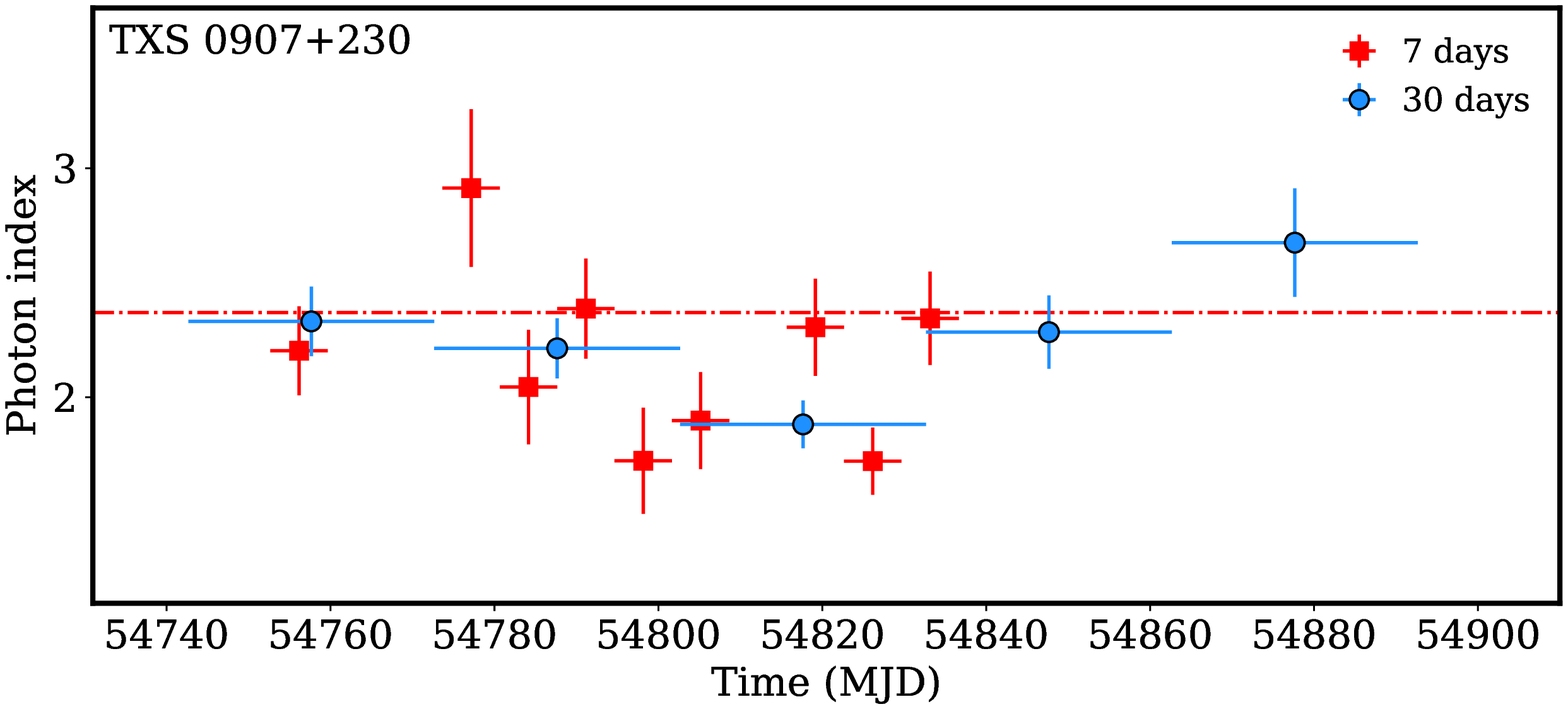}
      \caption{The evolution of the \gray photon indexes of B3 1343+451, PKS 0451-28, B3 0908+416B and TXS 0907+230 in time.
      }
         \label{index}
\end{figure*}
For the considered sources, the \gray photon index evolution in time has also been investigated.The photon index is defined by the processes responsible for particle acceleration and cooling, and its significant changes are directly linked with the processes inside the jet. As the adaptively binned light curves have narrow time bins, the photon indexes are estimated with large uncertainties which introduces difficulties for investigation of their variability. Therefore, the \gray light curves produced in 30 and 7 days (when available) are used to search for photon index variation, except for B3 1343+451 for which a 3-day binned light curve has been used. Moreover, in the light curves only the periods when the source detection was $TS>25$ were considered, otherwise the large uncertainties on the photon index estimation would not allow to make definite conclusions. The simple $\chi2$ test shows that among the considered sources only the photon index of B3 1343+451, PKS 0451-28, B3 0908+416B and TXS 0907+230 varies in time: the variation is highly significant for B3 1343+451 and  PKS 0451-28 with $P(\chi2)=6.9\times10^{-4}$ and $P(\chi2)\leq10^{-5}$, respectively, and $P(\chi2)\simeq0.015$ for B3 0908+416B and TXS 0907+230. The evolution of the photon index of these sources in time is shown in Fig. \ref{index} were the horizontal line corresponds to the averaged photon index estimated in ten years (Table \ref{sources}). The photon index of B3 1343+451 clearly varies, the hardest spectrum being observed on MJD $58089.16\pm1.5$ with $\Gamma_{\rm \gamma}=1.73\pm0.24$; in total there are 60 periods when $\Gamma_{\rm \gamma}<2.0$. Interestingly, the \gray spectrum of the source was hard with $1.95\pm0.07$ when it was in a bright \gray flaring state on MJD $56172.16\pm1.5$. The \gray spectrum of PKS 0451-28 is usually soft but in the 30-day binned light curve the periods when $\Gamma_{\rm \gamma}=2.06\pm0.07$ and $\Gamma_{\rm \gamma}=2.17\pm0.15$ are observed on MJD $56977.66\pm15$ and $58297.66\pm15$, respectively. The first period overlaps with the large \gray flare evident in the adaptively binned light curve (see Fig. \ref{lightcurve} a). The hardening of the \gray spectrum of B3 0908+416B on MJD $57517.66\pm15$ is remarkable when the $\Gamma_{\rm \gamma}$ changed to $1.84\pm0.25$ compared to $2.42\pm0.05$ averaged over 10 years. TXS 0907+230 is the most distant object in our sample ($z=2.66$) with occasional hardening of its \gray spectrum. The 7-day binned light curve of TXS 0907+230 shows that there are three periods (on MJD $54798.16\pm	3.5$, $54805.16\pm3.5$ and $ 54826.16\pm3.5$) when its \gray emission appears with an unusually hard \gray spectrum with $\Gamma_{\rm \gamma}=1.72\pm0.23$, $1.90\pm0.21$ and $1.72\pm0.15$. Yet, in the monthly binned light curve, in the bin covering these periods, $\Gamma_{\rm \gamma}$ is $1.88\pm0.10$ with a detection significance of $12.5\sigma$. Even if the \gray photon index of MG3 J163554+3629 and  PMN J0226+0937 appeared to be constant, hardening of their \gray spectra was occasionally observed. For example, for MG3 J163554+3629 $\Gamma_{\rm \gamma}=2.29\pm0.13$ (with $11.1\sigma$) was observed from MJD 58282 to 58312, likewise for PMN J0226+0937 $\Gamma_{\rm \gamma}$ was $1.80\pm0.16$ (with $9.7\sigma$) on MJD 54892-54922. The hard \gray spectra of the sources discussed above are shown in  Fig. \ref{sed1} in magenta; substantial changes in the \gray spectrum are evident. Such a hard \gray spectrum is more typical for BL Lacs, but it has also been occasionally observed for the FSRQs during the flares \citep[e.g., see][]{2014ApJ...790...45P, 2017MNRAS.470.2861S, 2018ApJ...863..114G, 2020A&A...635A..25S, 2019ApJ...871..211P, 2019A&A...627A.140A}.\\
In addition, \gray light curves with 30-day binning above $1$ GeV are produced to investigate the flux and photon index variation in the GeV band. The periods when the source emission is significant above 1 GeV are also relevant for studying absorption through interaction with EBL photons. There are only a few periods when the considered sources have been detected above 1 GeV with a sufficient significance ($>5\:\sigma$).  The emission, in these periods, is mostly characterized by a soft \gray spectrum, implying these are the same components as those at lower energies. However, the periods when the sources were detected by \fermi correspond to only a small fraction of the total bins (30-day binning), so the poor statistics did not allow us to investigate the possible flux variation or photon index hardening above $1$ GeV.
\begin{figure*}
\includegraphics[width=0.488 \textwidth]{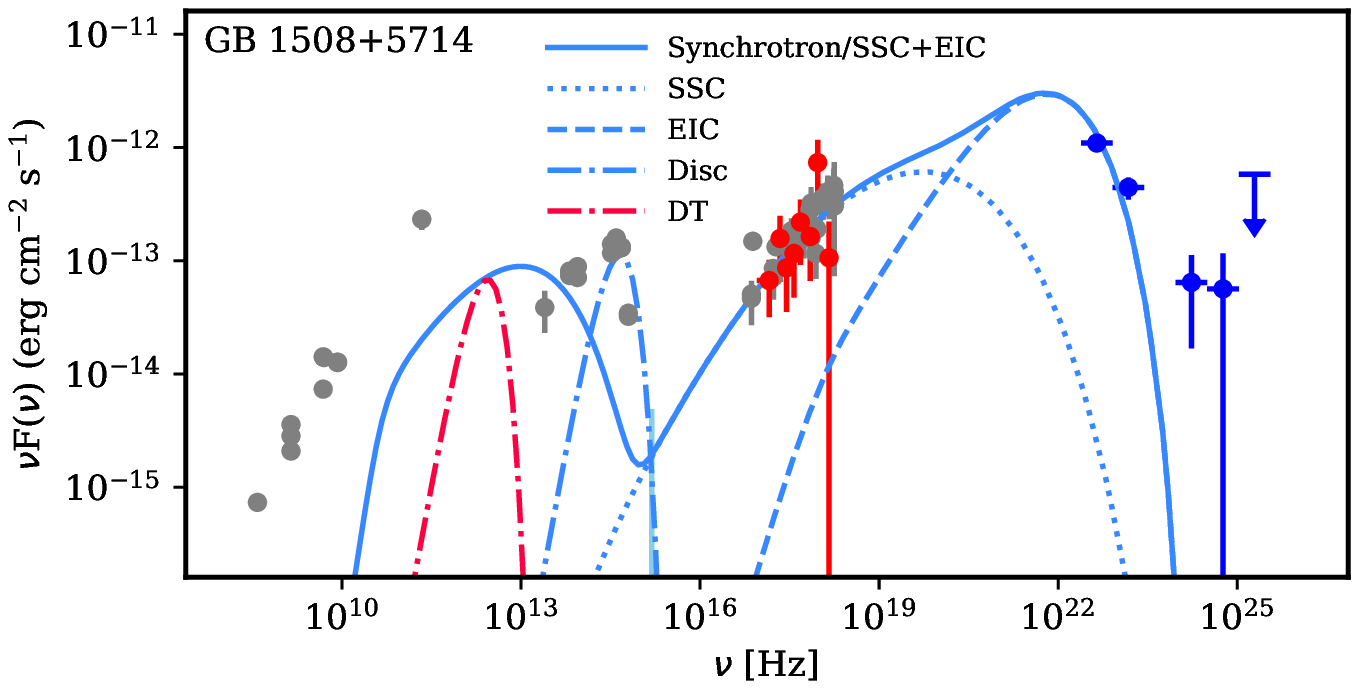}
\includegraphics[width=0.488 \textwidth]{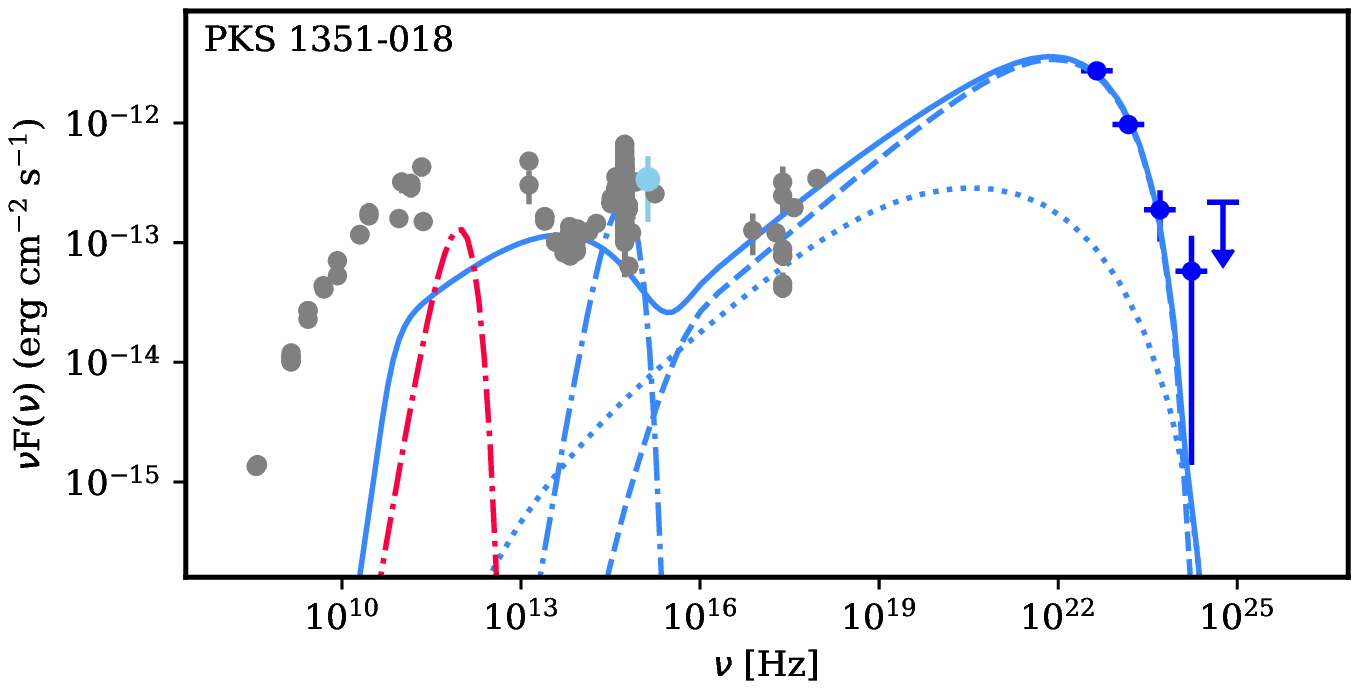}\\
    \includegraphics[width=0.488 \textwidth]{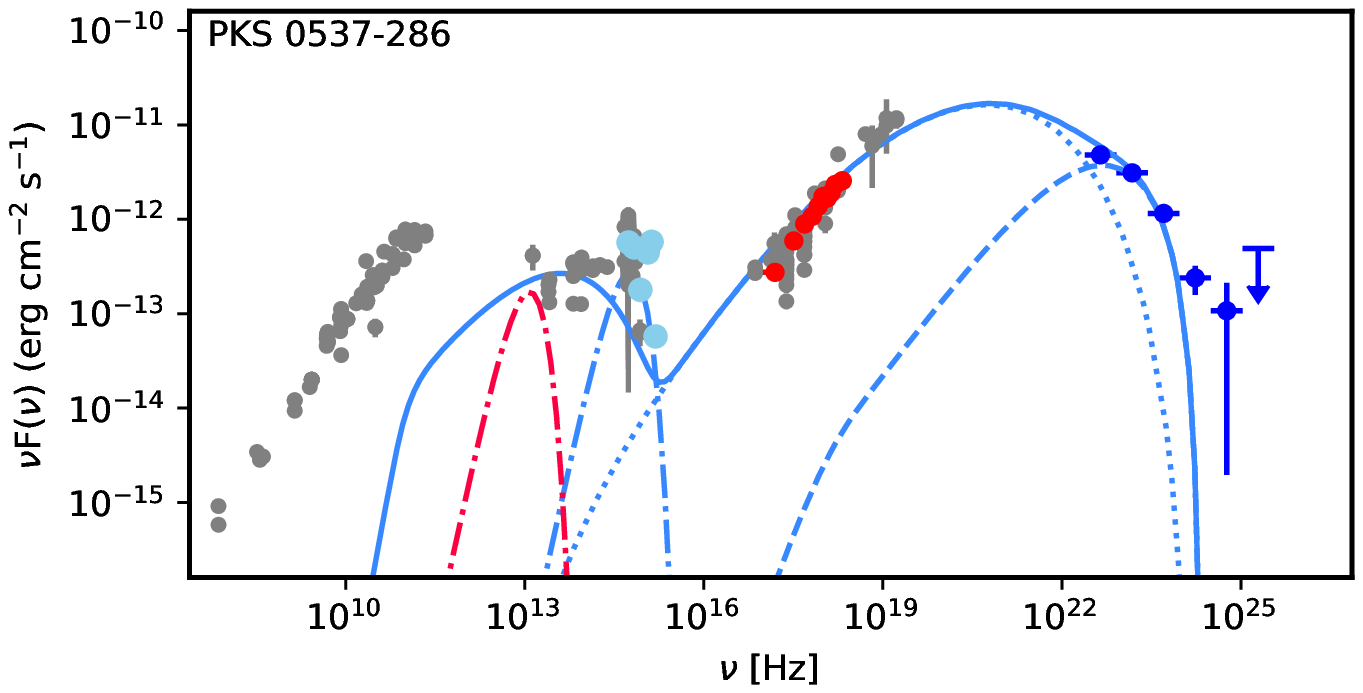}
     \includegraphics[width=0.488 \textwidth]{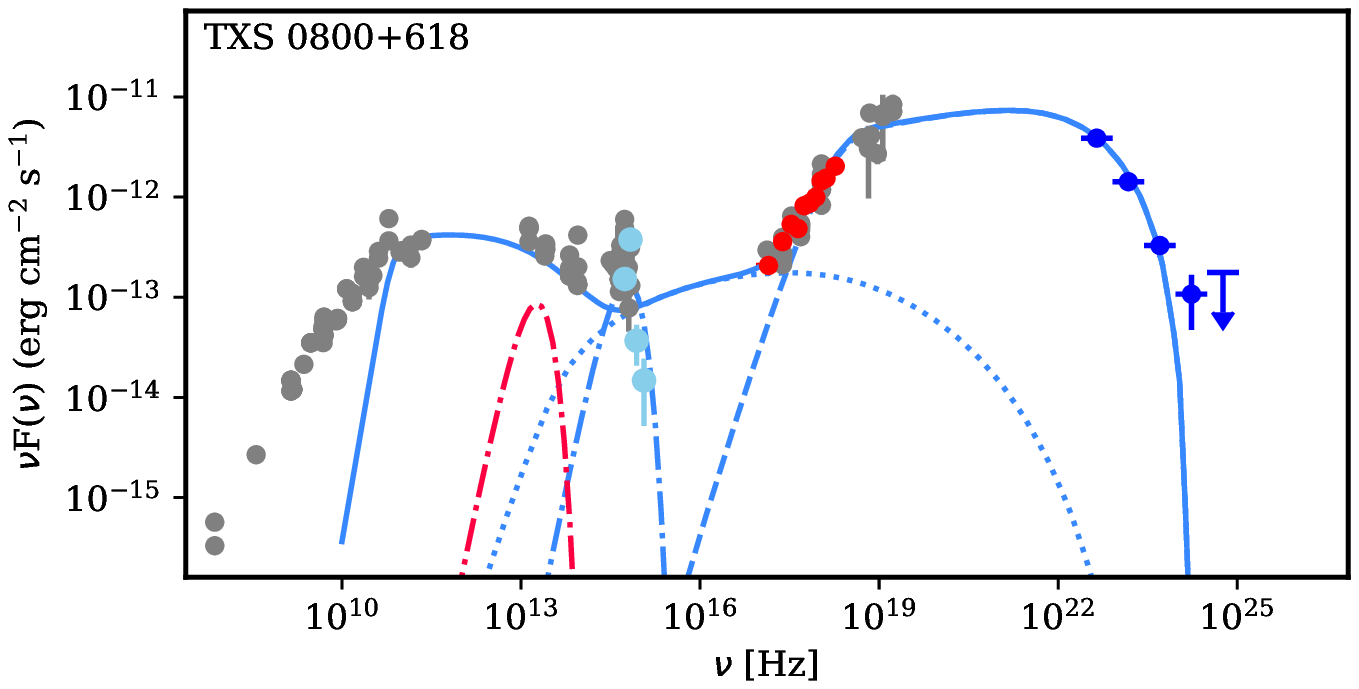}\\
     \includegraphics[width=0.488 \textwidth]{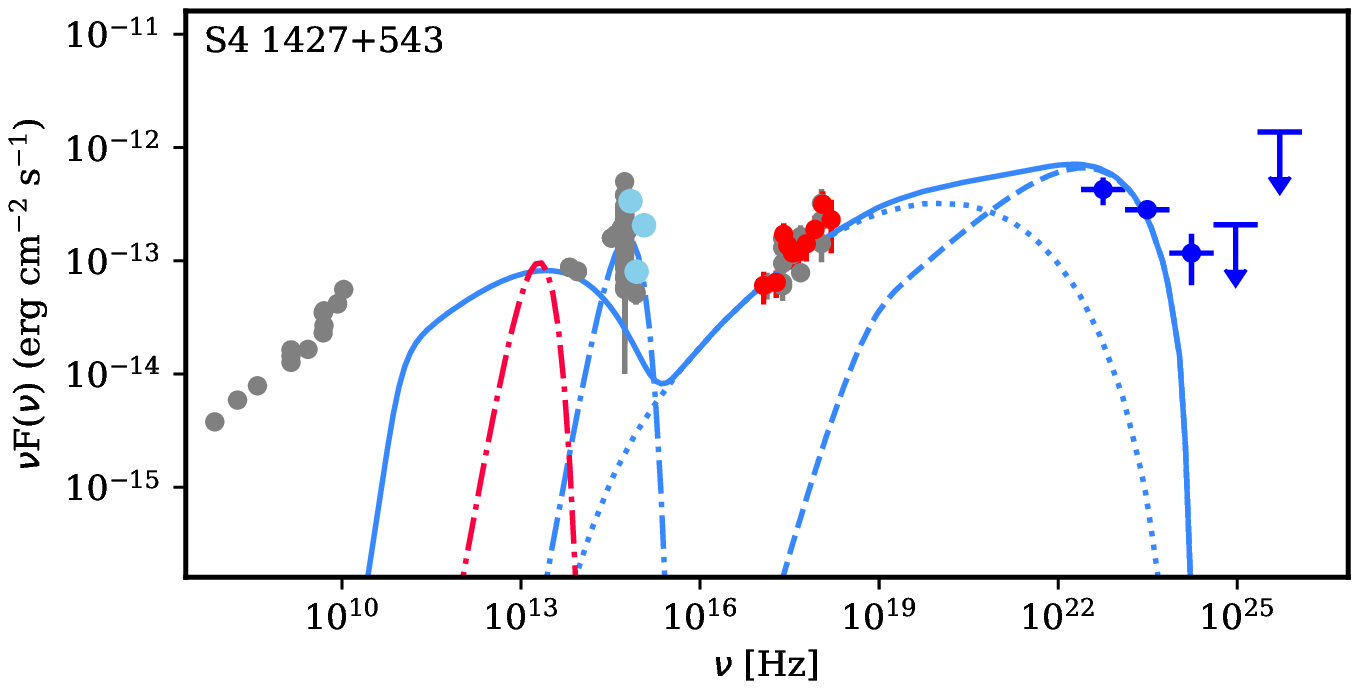}
     \includegraphics[width=0.488 \textwidth]{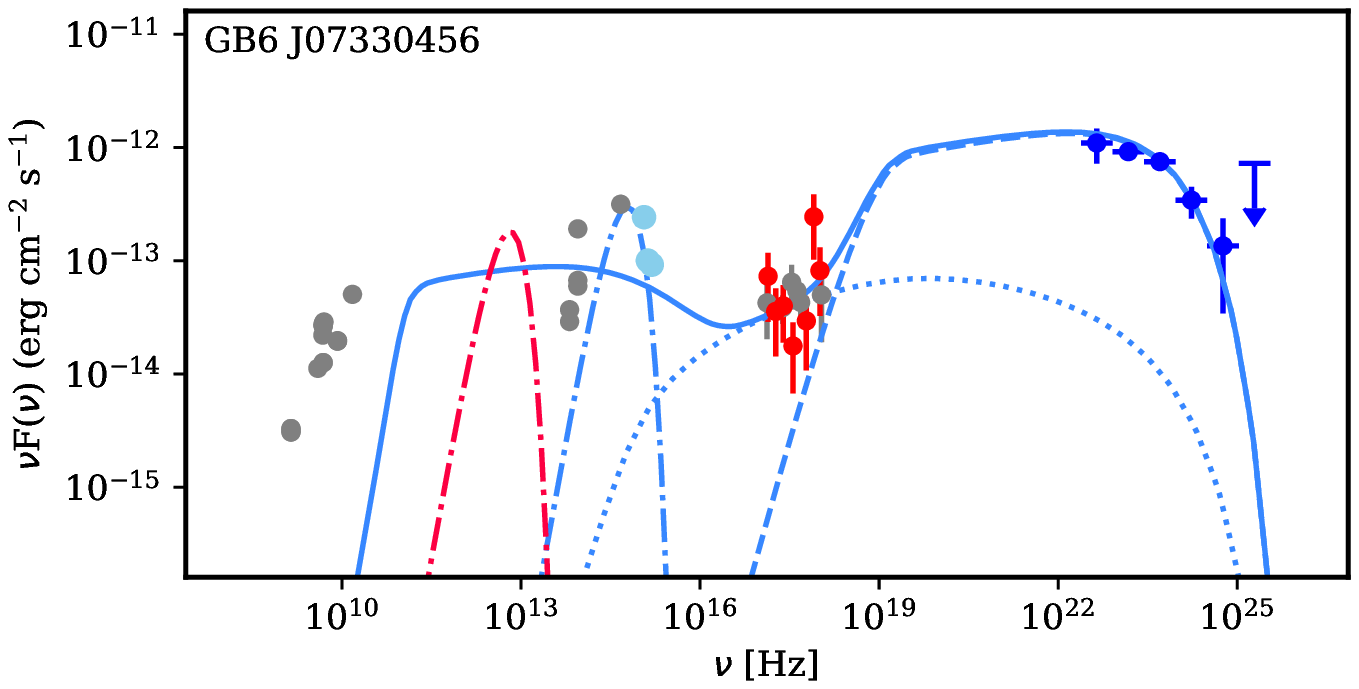}\\
    \includegraphics[width=0.488 \textwidth]{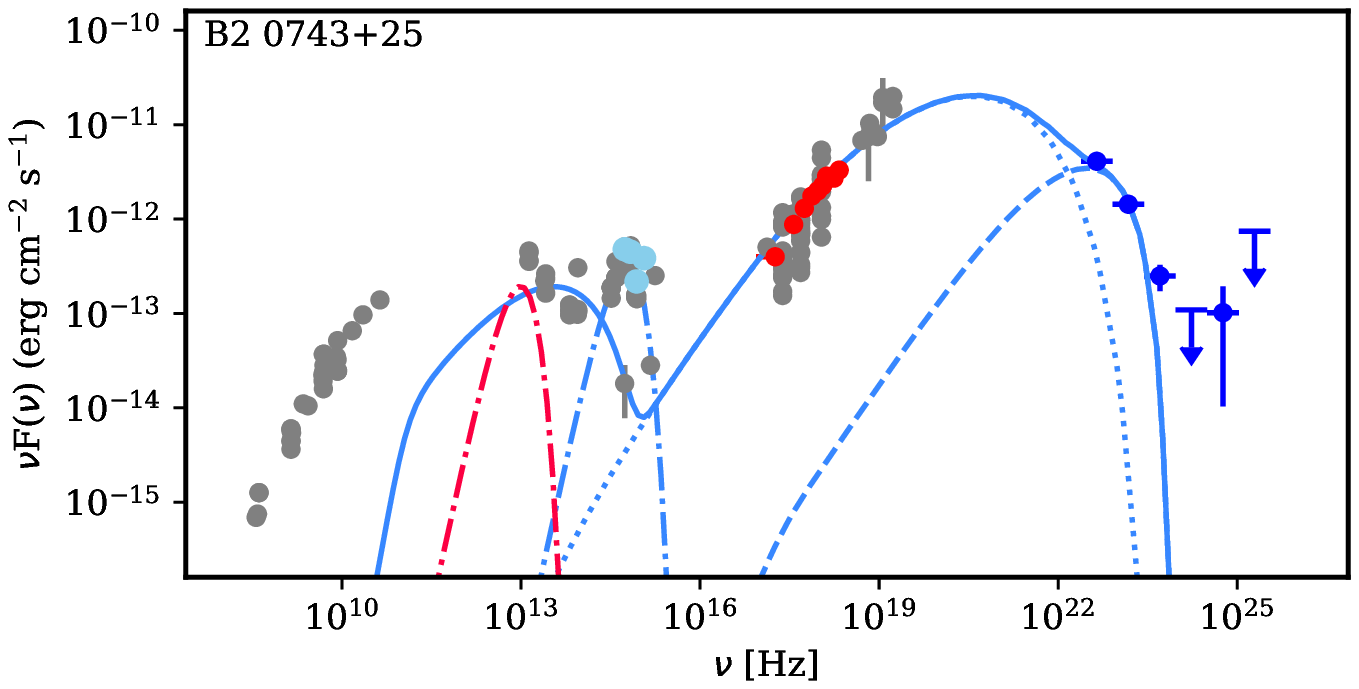}
     \includegraphics[width=0.488 \textwidth]{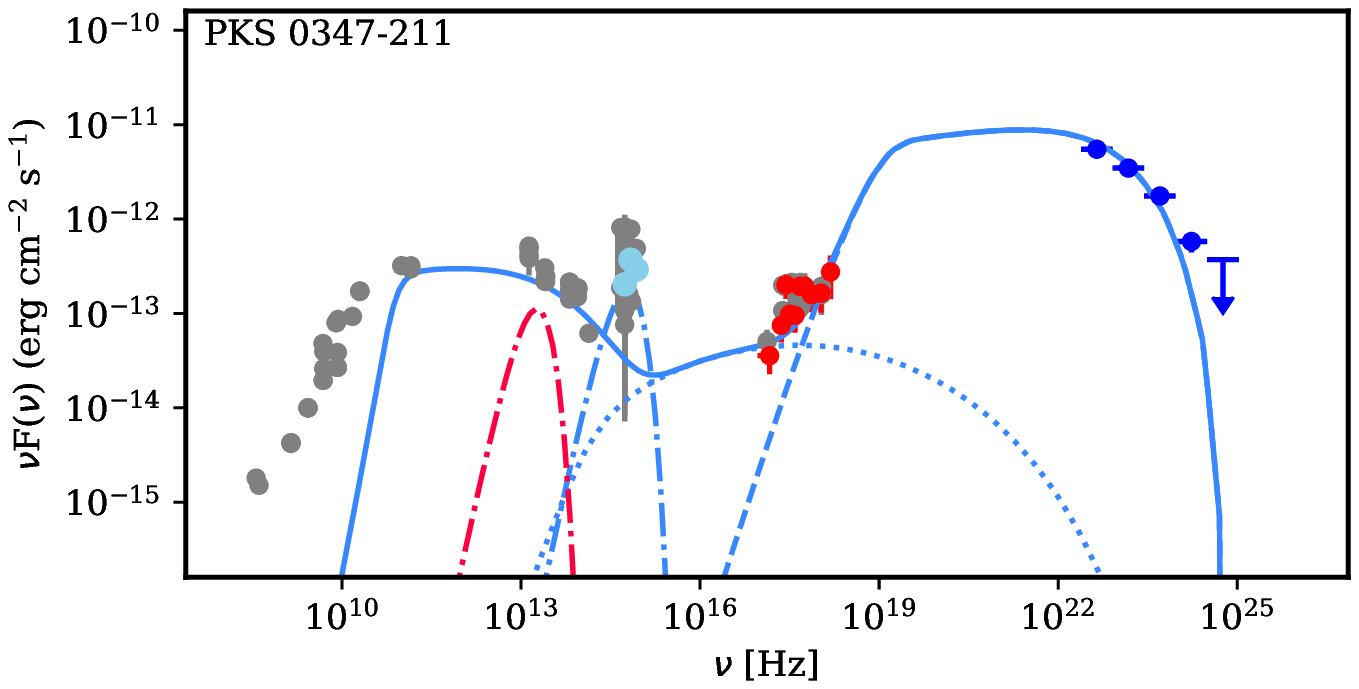}\\
     \includegraphics[width=0.488 \textwidth]{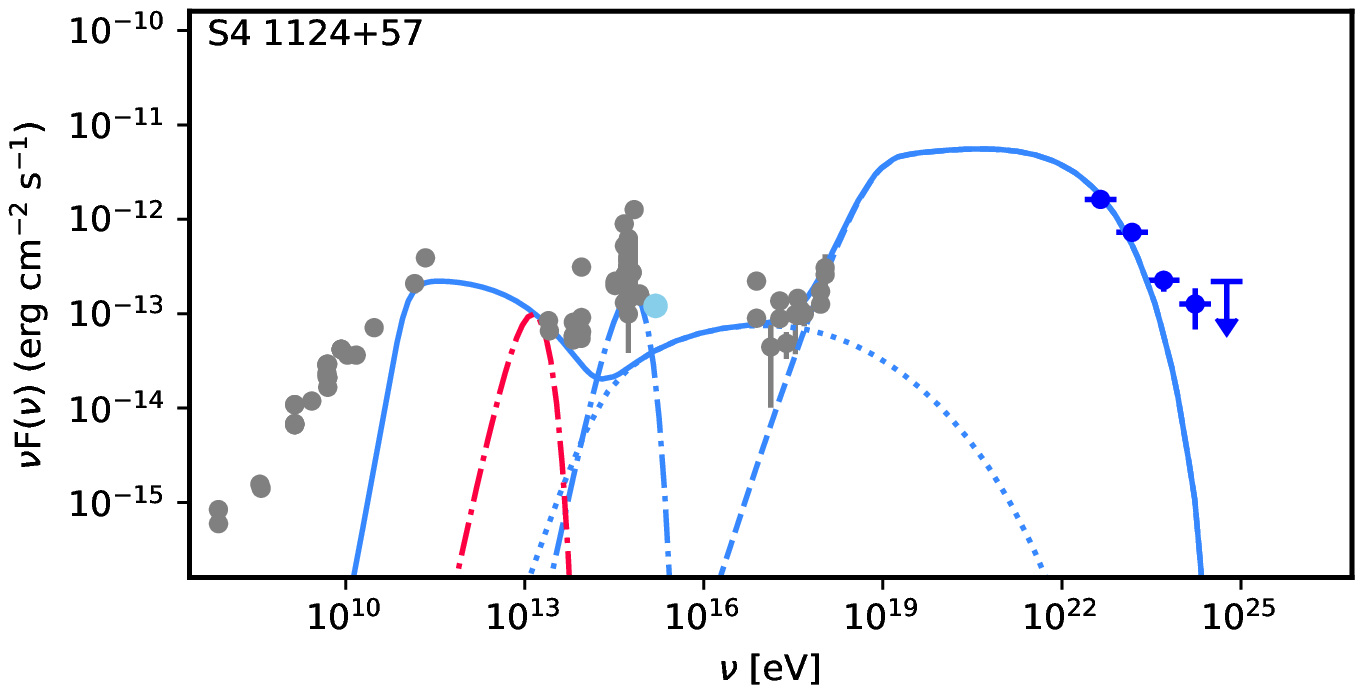}
     \includegraphics[width=0.488 \textwidth]{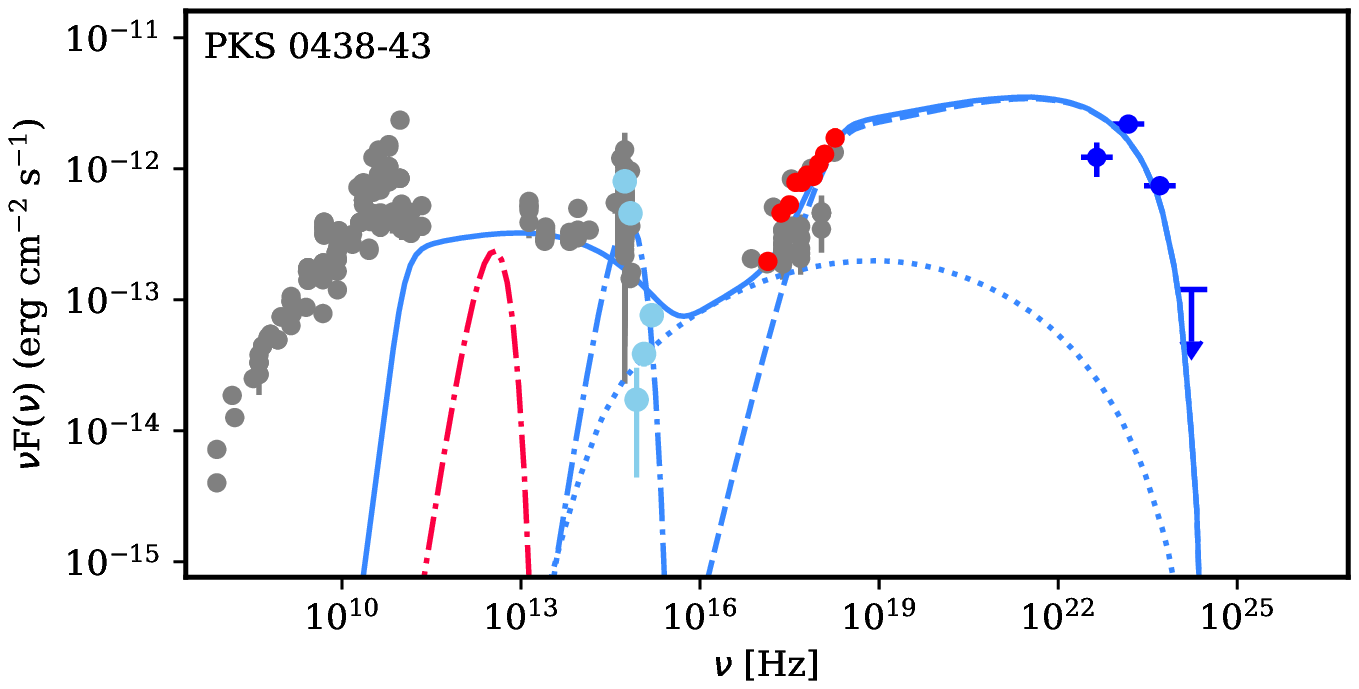}\\
      \caption{Modeling of the broadband SEDs of the considered sources. The Swift UVOT, XRT and \fermi data obtained here are shown with cyan, red and blue, respectively, while the archival data are in gray. 
              }
         \label{sed1}
         \addtocounter{figure}{-1}
\end{figure*}

\begin{figure*}
	\includegraphics[width=0.49 \textwidth]{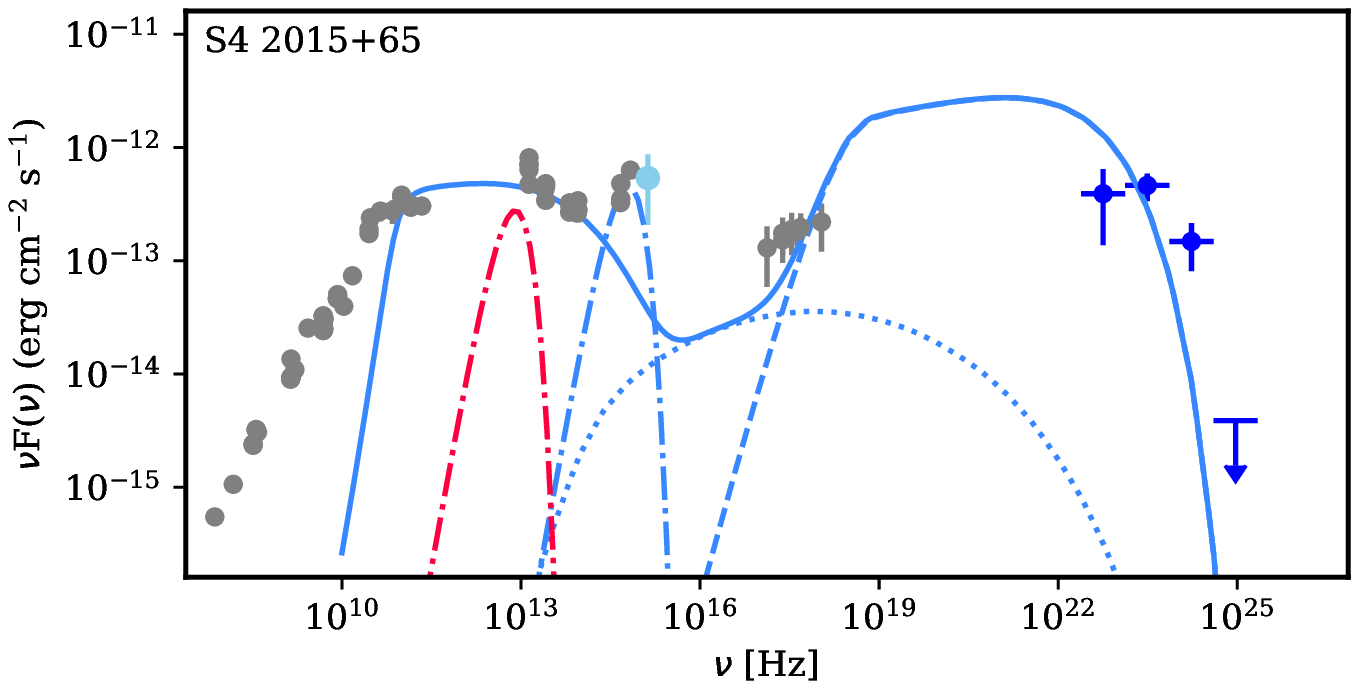}
     \includegraphics[width=0.49 \textwidth]{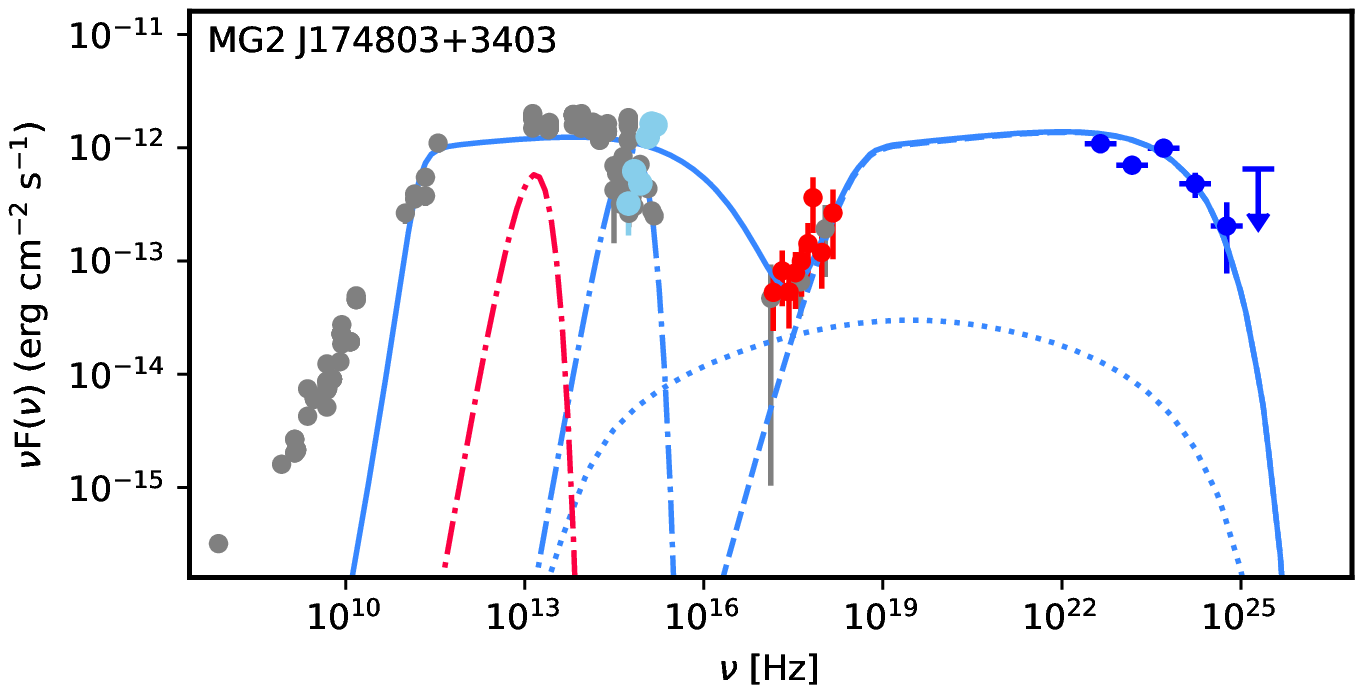}\\
     \includegraphics[width=0.49 \textwidth]{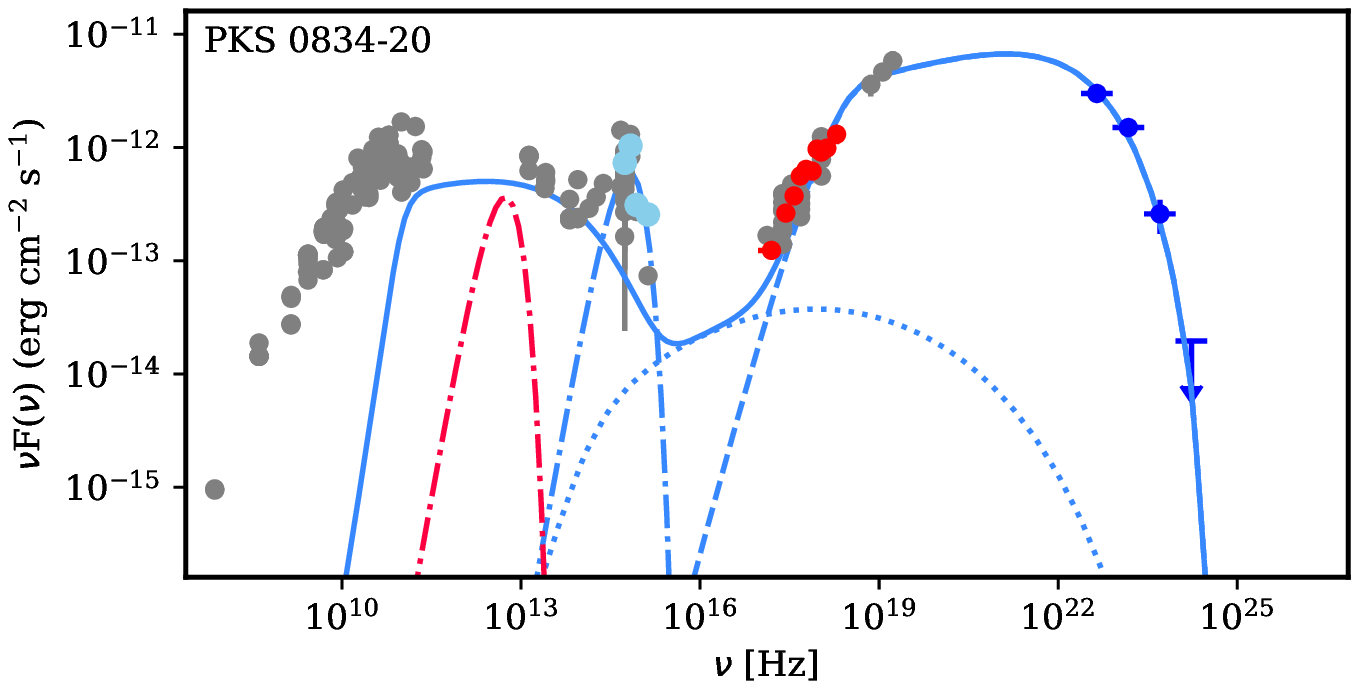}
     \includegraphics[width=0.49 \textwidth]{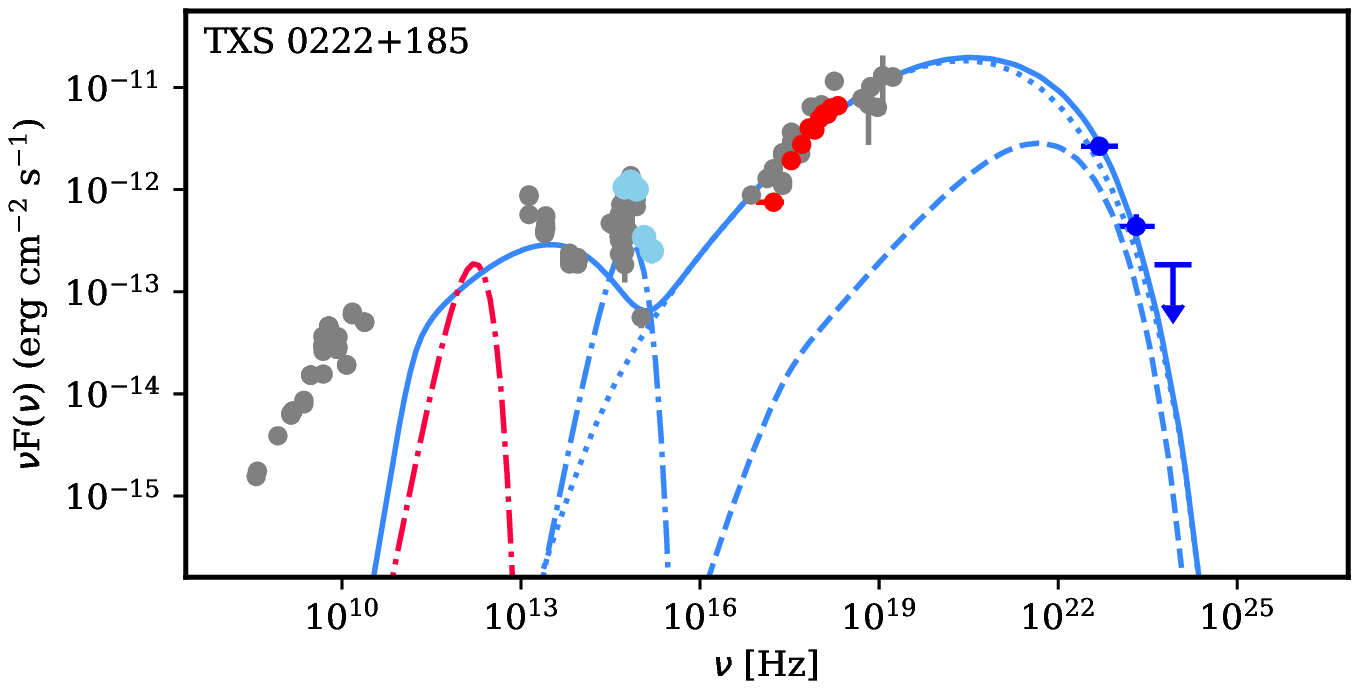}\\
     \includegraphics[width=0.49 \textwidth]{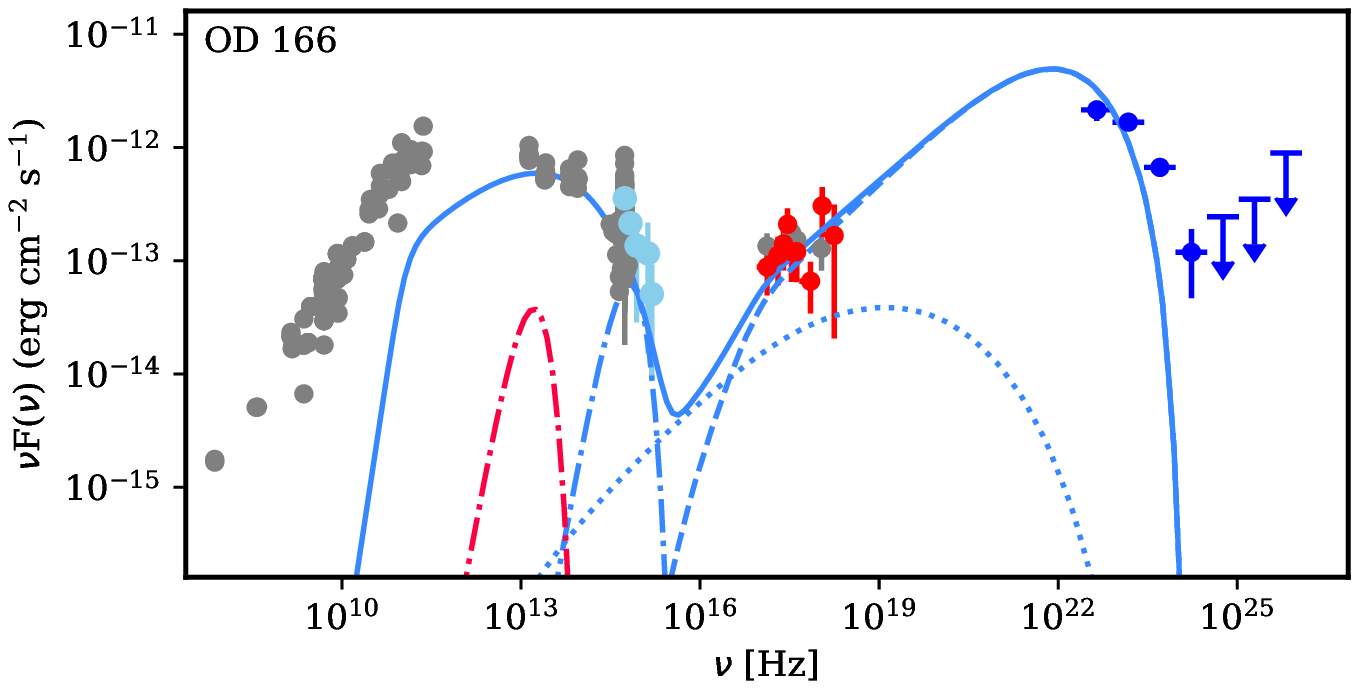}
     \includegraphics[width=0.49 \textwidth]{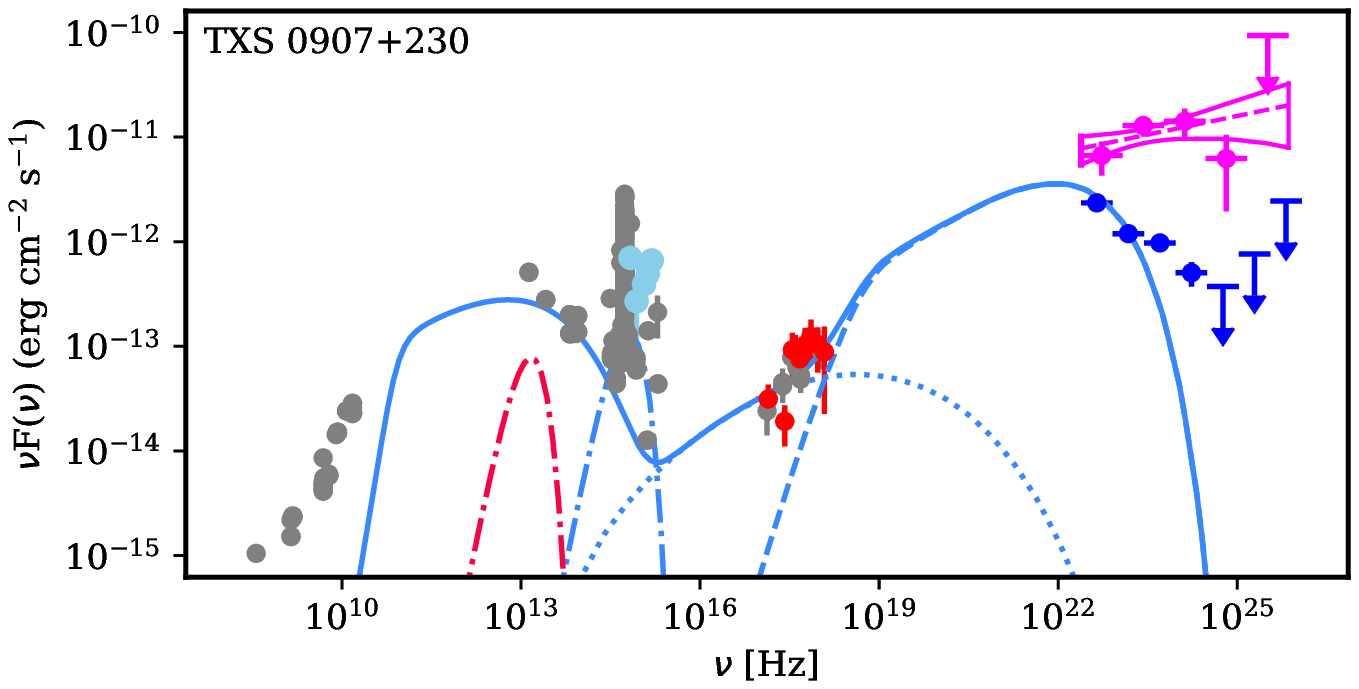}\\
     \includegraphics[width=0.49 \textwidth]{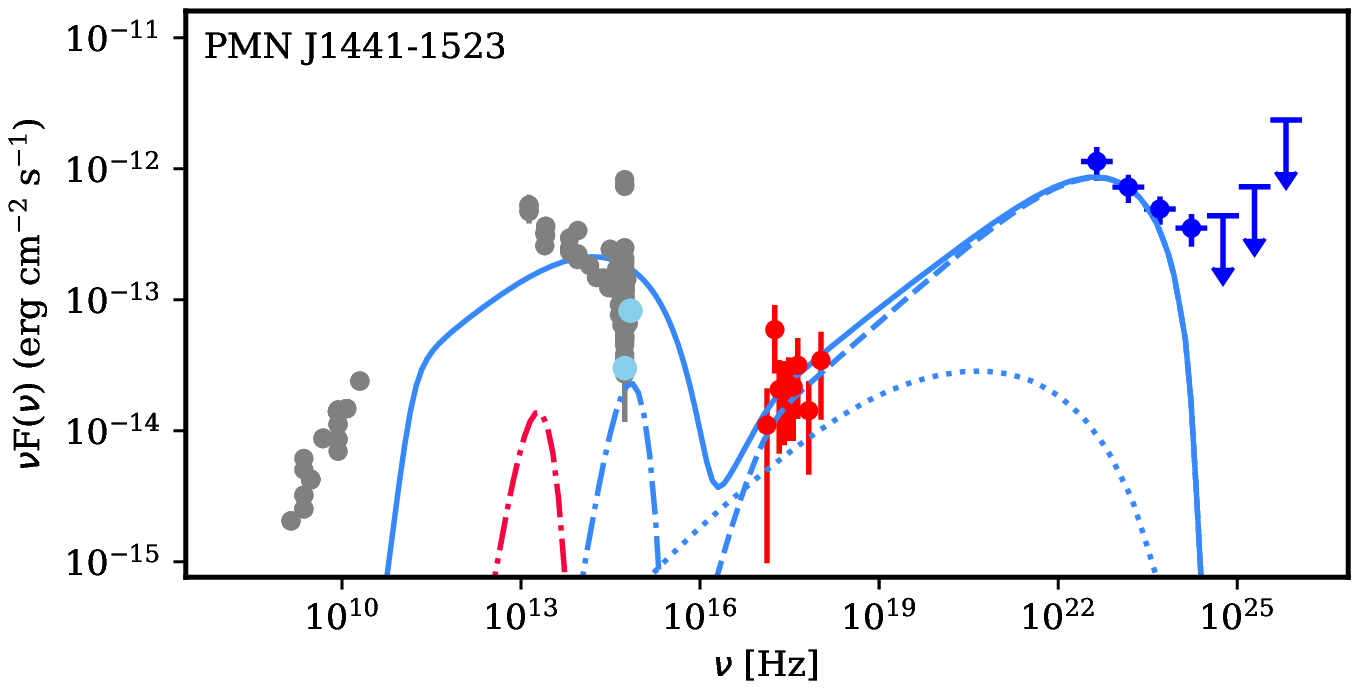}
     \includegraphics[width=0.49 \textwidth]{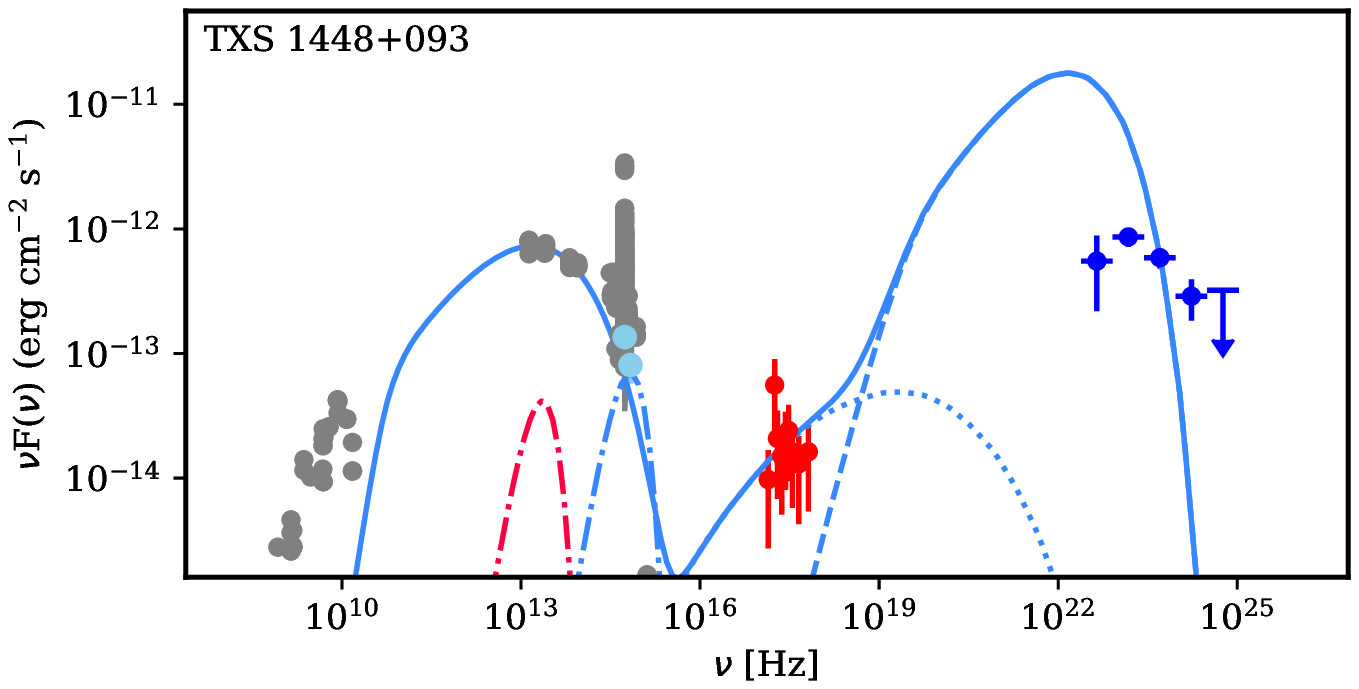}\\
     \includegraphics[width=0.49 \textwidth]{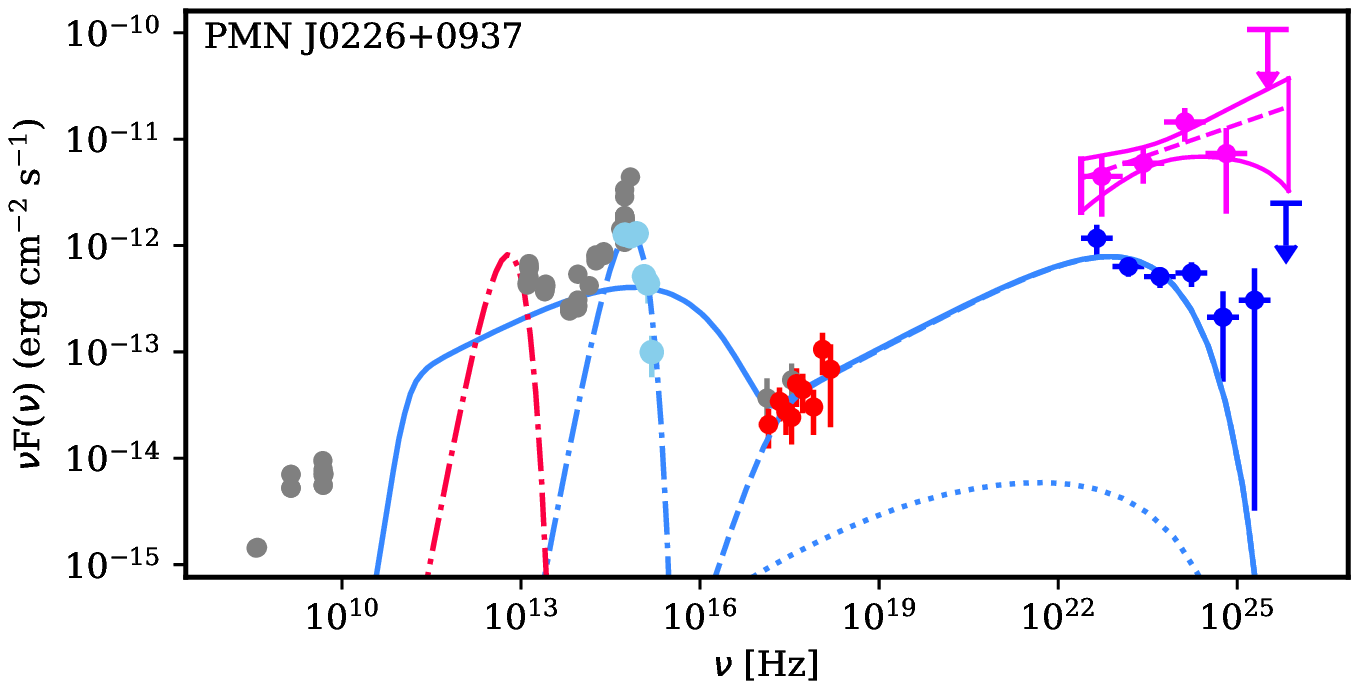}
     \includegraphics[width=0.49 \textwidth]{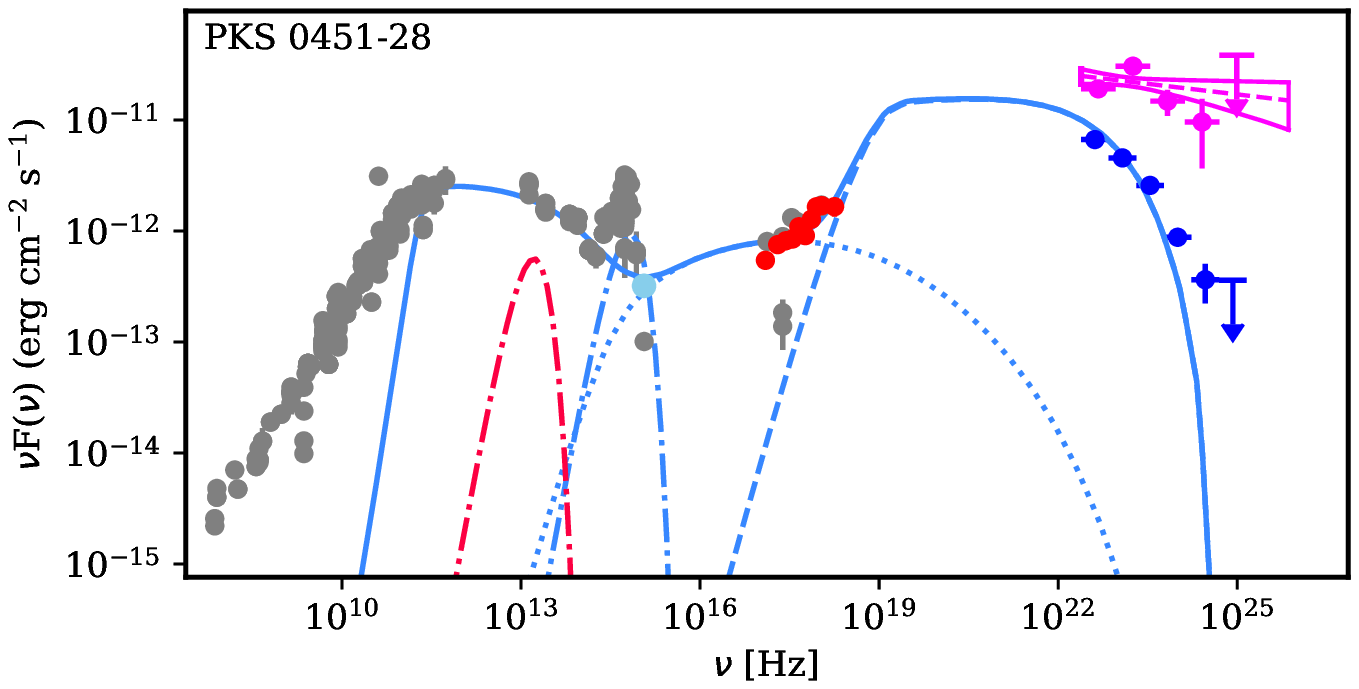}\\
      \caption{(Continued)
              }
\addtocounter{figure}{-1}
\end{figure*}

\begin{figure*}
	 \includegraphics[width=0.49 \textwidth]{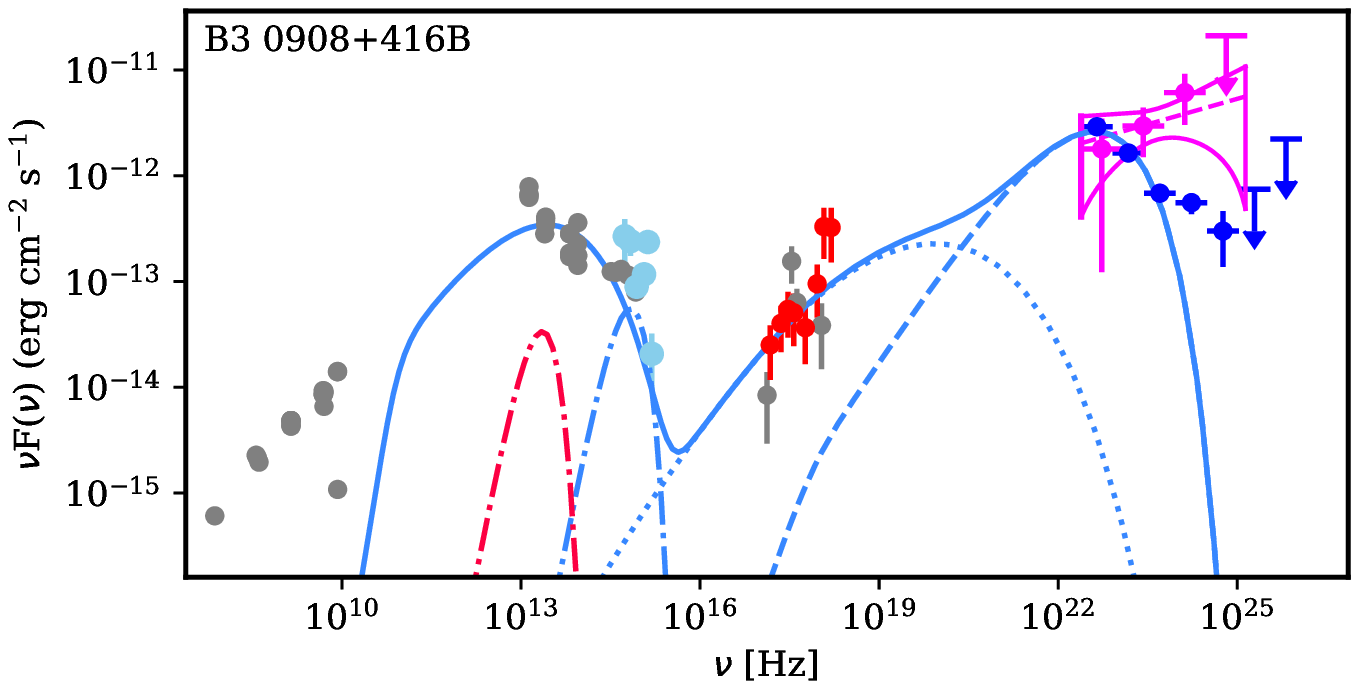}
     \includegraphics[width=0.49 \textwidth]{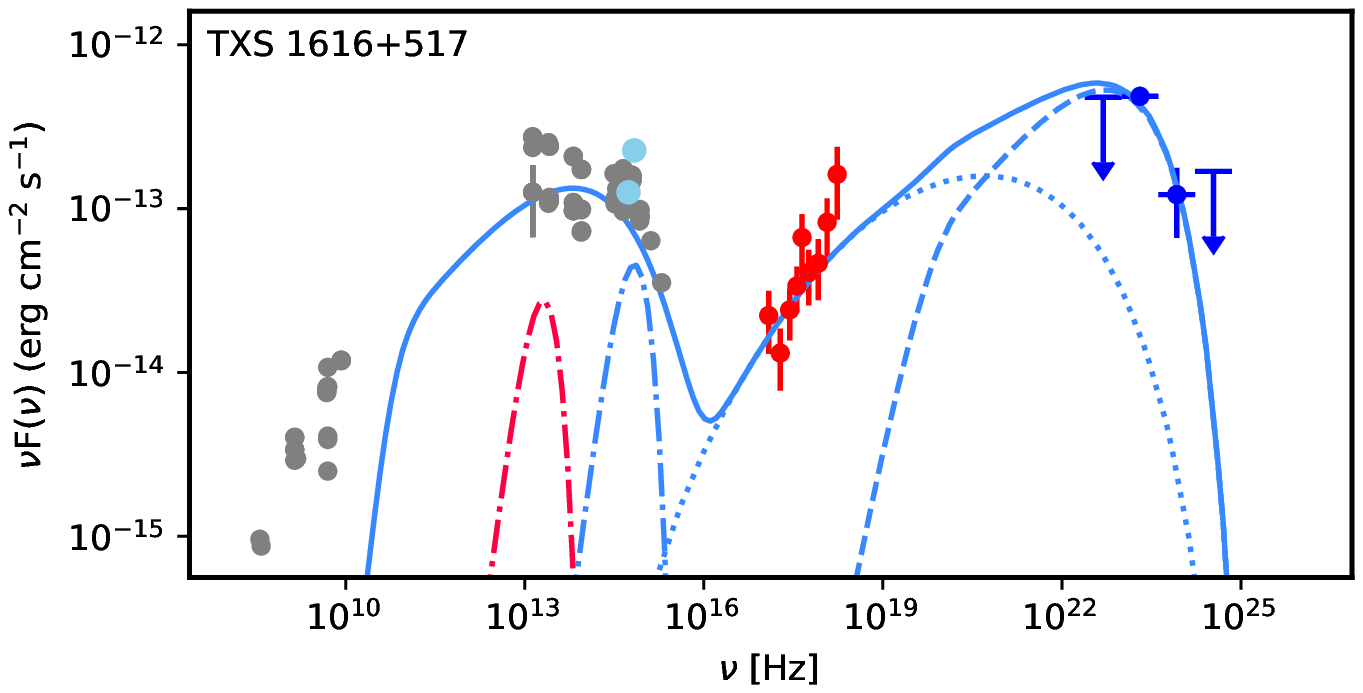}\\
     \includegraphics[width=0.49 \textwidth]{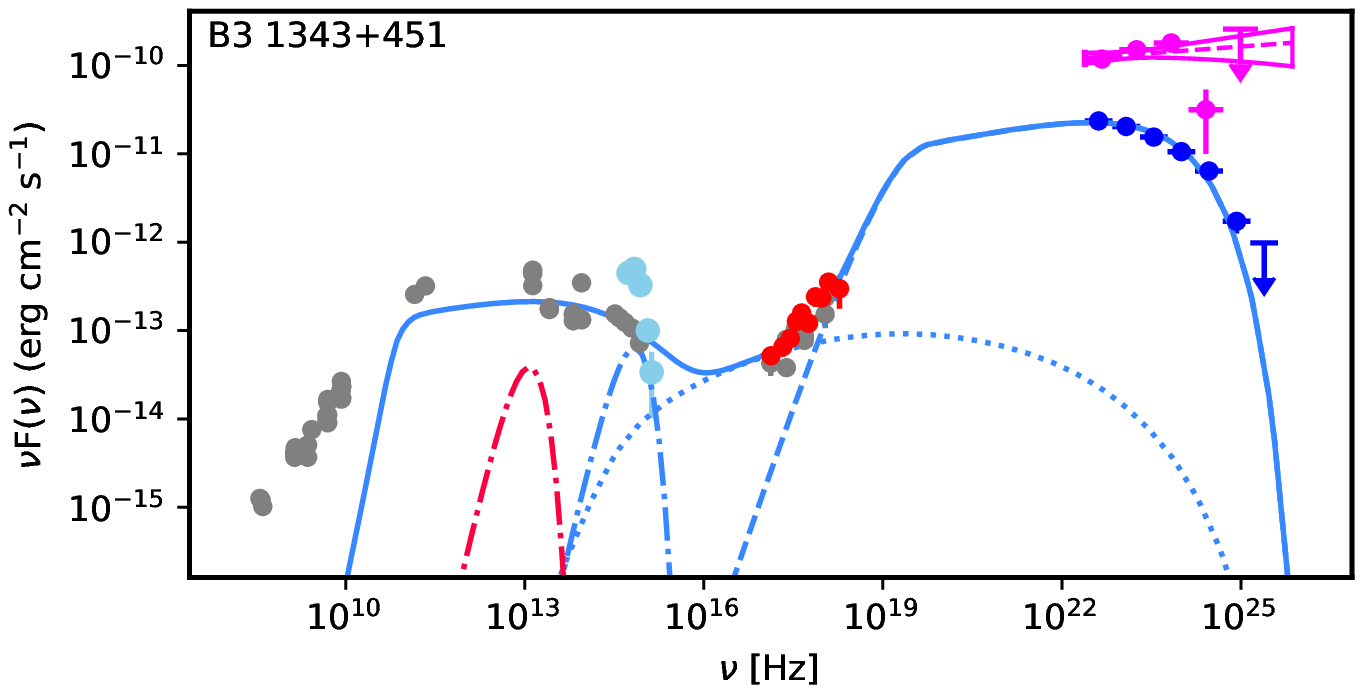}
     \includegraphics[width=0.49 \textwidth]{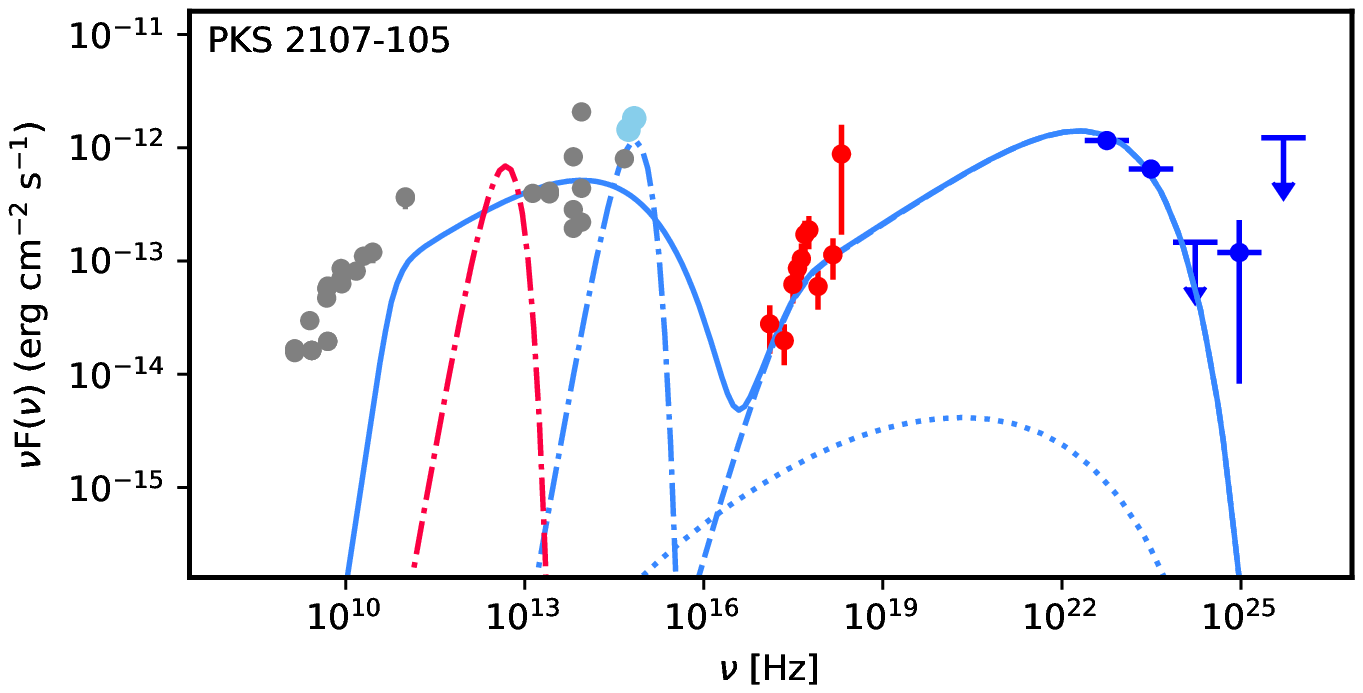}\\
      \caption{(Continued)
              }
         \label{sed3}
         \addtocounter{figure}{-1}
\end{figure*}

\begin{table*}
\caption{\label{t7} Parameters obtained from the modeling of multiwavelength SEDs. [1]: object name. [2]: Doppler factor. [3]: Slope of electron energy distribution. [4] and [5]: the Lorentz factors corresponding to the minimum and cutoff energy of the electron distribution. [6]: Magnetic field in units of $G$. [7] Radius of the emitting region in units of $10^{16}$ cm. [8] and [9]: Electron and magnetic field energy densities. [10]: accretion disk luminosity in units of $10^{46}\:{\rm erg\:s^{-1}}$. [11] and [12]: The power of the jet in the form of the relativistic electrons ($L_{e}$) and magnetic field ($L_{B}$) in units of $10^{45}\:{\rm erg\:s^{-1}}$ and $10^{43}\:{\rm erg\:s^{-1}}$, respectively.}
\centering\renewcommand\cellalign{lc}
\resizebox{\textwidth}{!}{\begin{tabular}{lccccccccccc}
\hline\hline
Sources & $\delta$  &$\alpha$& $\gamma_{\rm min}$& $\gamma_{\rm cut}$ & $B$ & $R$ &$U_{\rm e}$ & $U_{\rm B}$ & $L_{\rm d}$ & $L_{\rm e}$ & $L_{\rm B}$\\
 $[1]$ & [2] & [3] & [4] & [5] & [6] & [7] & [8] & [9] & [10] & [11] & [12]\\
\hline
\hline 
GB 1508+5714 & $15.72\pm1.29$ & $1.17\pm0.07$ & $26.90\pm2.88$ & $1.30\pm0.10$ & $0.19\pm0.02$ & $2.28$ & $0.50$ & $1.49$ & $3.02$ & $2.43$ & $0.73$\\
PKS 1351-018 & $20.47\pm2.49$ & $2.16\pm0.11$ & $2.68\pm0.36$ & $4.99\pm0.71$ & $0.20\pm0.02$ & $2.29$ & $0.54$ & $1.66$ & $4.04$  & $2.68$ & $0.82$\\
PKS 0537-286 & $11.50\pm0.57$ & $1.33\pm0.07$ & $15.70\pm1.49$ & $2.45\pm0.16$ & $0.28\pm0.02$ & $1.14$ & $5.93$ & $3.21$ & $3.44$  & $7.31$ & $0.40$\\
TXS 0800+618 & $14.04\pm0.56$ & $2.75\pm0.04$ & $13.98\pm0.86$ & $2.20\pm0.02$ & $0.26\pm0.01$ & $15.5$ & $0.06$ & $2.73$ & $1.65$ & $13.49$ & $61.84$\\
S4 1427+543 & $10.00\pm0.37$& $2.04\pm0.10$ & $29.00\pm2.55$ & $2.79\pm0.27$ & $0.53\pm0.04$ & $1.90$ & $0.63$ & $10.02$ & $1.83$ & $2.14$ & $3.75$\\
GB6 J0733+0456 & $16.28\pm1.36$& $2.80\pm0.04$ & $47.90\pm3.42$ & $15.73\pm1.60$ & $0.16\pm0.03$ & $2.98$& $0.12$ & $6.22$ & $3.40$ & $0.98$ & $5.20$\\
PKS 0347-211 & $26.00\pm1.02$ & $2.79\pm0.02$ & $23.09\pm1.12$ & $2.62\pm0.16$ & $0.20\pm0.01$ & $8.15$ & $0.03$ & $1.61$ & $1.99$ & $1.72$ & $10.08$\\
B2 0743+25 & $10.02\pm0.45$ & $1.13\pm0.19$ & $7.66\pm0.22$ & $2.03\pm0.08$ & $0.36\pm0.01$ & $0.70$ & $16.90$ & $0.003$ & $3.58$ & $7.80$ & $0.24$\\
S4 1124+57 & $22.17\pm1.37$ & $2.78\pm0.04$ & $22.92\pm1.45$ & $1.28\pm0.10$ & $0.22\pm0.01$ & $5.15$ & $0.14$ & $1.95$ & $1.69$  & $3.39$  & $4.87$\\
PKS 0438-43 & $18.17\pm1.29$ & $2.78\pm0.04$ & $23.13\pm1.54$ & $7.19\pm0.58$ & $0.34\pm0.02$ & $5.42$ & $0.12$ & $4.52$ & $3.91$  & $3.33$  & $12.52$\\
S4 2015+65 & $17.85\pm1.32$ & $2.73\pm0.05$ & $20.63\pm1.64$ & $2.75\pm0.29$ & $0.46\pm0.03$ & $13.50$& $0.07$ & $8.41$ & $4.55$ & $1.26$  & $14.44$ \\
MG2 J174803+3403 & $24.50\pm2.06$ & $2.87\pm0.06$ & $14.67\pm1.58$ & $1.40\pm0.49$ & $1.45\pm0.09$ & $7.41$ & $0.004$ & $83.98$ & $8.87$ & $0.23$ & $434.69$\\
PKS 0834-20 & $27.42\pm0.97$& $2.70\pm0.06$ & $20.57\pm1.51$ & $2.13\pm0.18$ & $0.37\pm0.02$ & $6.83$ & $0.02$ & $5.44$ & $5.51$ & $0.70$ & $23.92$\\
TXS 0222+185 & $10.03\pm0.28$ & $1.62\pm0.05$ & $19.56\pm0.98$ & $2.38\pm0.10$ & $0.35\pm0.02$ & $1.07$ & $10.28$ & $5.04$ & $2.71$ & $11.10$ & $0.54$\\ 
OD 166 & $19.02\pm0.84$ & $1.96\pm0.04$ & $2.58\pm0.15$ & $1.01\pm0.03$ & $1.15\pm0.05$ & $4.32$ & $0.01$ & $0.52$ & $0.53$ & $0.25$ & $92.38$\\
TXS 0907+230 & $21.66\pm1.66$ & $2.23\pm0.11$ & $20.26\pm1.84$ & $1.44\pm0.10$ & $0.31\pm0.02$ & $5.23$ & $0.03$ & $3.72$ & $1.09$  & $0.70$ & $5.96$\\
PMN J1441-1523 & $17.01\pm1.50$ & $2.19\pm0.07$ & $2.86\pm0.29$ & $3.31\pm0.46$ & $1.68\pm0.14$ & $1.47$ & $0.07$ & $112.38$ & $0.17$  & $0.14$ & $22.85$\\
TXS 1448+093 & $17.90\pm1.13$ & $1.52\pm0.15$ & $49.64\pm5.82$ & $0.84\pm0.06$ & $0.70\pm0.05$ & $7.96$ & $0.003$ & $19.44$ & $0.56$ & $0.17$ & $115.93$\\
PMN J0226+0937 & $25.02\pm1.98$ & $2.41\pm0.04$ & $5.37\pm0.59$ & $8.32\pm0.62$ & $1.74\pm0.12$ & $3.09$& $0.03$ & $119.86$ & $10.94$ & $0.03$ & $107.98$\\
PKS 0451-28& $26.14\pm1.27$ & $2.90\pm0.28$ & $21.93\pm1.3`$ & $2.19\pm0.01$ & $0.45\pm0.03$ & $5.90$ & $0.11$ & $8.01$ & $7.20$ & $3.59$ & $26.32$\\
B3 0908+416B & $23.22\pm1.72$  & $1.31\pm0.25$ & $6.41\pm0.72$ & $1.11\pm0.10$ & $0.39\pm0.03$ & $1.76$ & $0.09$ & $6.10$ & $0.43$ & $0.26$ & $1.78$\\
TXS 1616+517 & $10.11\pm0.31$ & $2.09\pm0.09$ & $93.28\pm4.15$ & $4.34\pm0.36$ & $0.52\pm0.02$ & $3.59$ & $0.06$ & $10.79$ & $0.35$ & $0.70$ & $13.11$\\
B3 1343+451& $26.55\pm1.04$ & $2.48\pm0.04$ & $16.49\pm1.31$ & $8.67\pm0.48$ & $0.10\pm0.01$ & $4.16$ & $0.11$ & $0.42$ & $0.48$ & $1.76$ & $0.68$\\
PKS 2107-105 & $27.32\pm1.34$  & $2.30\pm0.06$ & $7.45\pm0.90$ & $3.63\pm0.39$ & $0.67\pm0.05$ & $9.48$ & $0.0006$ & $17.73$ & $8.30$ & $0.05$ & $150.16$\\
\hline
\end{tabular}}
\end{table*}
\section{The origin of multiwavelength emission}\label{sec5}
The multiwavelength emission from blazars produced when the accelerated electrons \citep{ghisellini,bloom, maraschi, blazejowski, sikora, ghisellini009} or protons \citep{dar, beall, bednarek97, 1995APh.....3..295M,1989A&A...221..211M,1993A&A...269...67M,mucke1,mucke2} interact with the magnetic and photon fields, contain valuable information necessary for understanding the physics of their jets. The currently proposed models can explain a single SED snapshot, but they cannot self-consistently explain the dynamical evolution of the radiation and origin of the flares. The flares are most likely caused by changes in the radiating particles or in the emission region \citep[e.g.,][]{2011ApJ...736..128P}, and their origin can be investigated only with contemporaneous multiwavelength data. 
During the \gray flaring of the considered sources (Figs. \ref{lightcurve} and \ref{lightcurvelong}), contemporaneous Swift data are available only for PKS 0438-43 and TXS 0800+618. However, the Swift observations in these periods show that the X-ray photon index and flux did not change, being the same as those given in Table \ref{sourcesXray}, whereas in the optical/UV bands these sources were detected only in one or two UVOT filters with large uncertainties in the flux estimation. So, it is not clear whether or not the low energy component changed or remained the same during the \gray flares. This introduces uncertainties for the theoretical modeling and does not allow to model the SEDs of these two objects during their \gray flaring. Moreover, as there are no multiwavelength data during the \gray flares of the other sources and our goal is to constrain the main physical parameters of distant blazar jets,  only the averaged multiwavelength SEDs have been used in the modeling which represent the typical state of the sources.\\
To understand the origin of the broadband emission of the considered sources, we have used a simple one-zone leptonic model. In this model, the emission region is assumed to be a spherical blob of radius of $R$ moving in the blazar jet with a bulk Lorentz factor of $\Gamma_{\rm j}$ at a viewing angle of $\theta$. The emitting region is filled with a uniformly tangled magnetic field $B$ and with a homogeneous population of relativistic electrons (and positrons), the nonthermal energy distribution of which is described by a power-law with an exponential cut-off at higher energies as:
\begin{equation}
{\rm N(\gamma)\sim \gamma^{-\alpha}exp(-\gamma/\gamma_{cut})\: \: \: \: \: \: \: \: \: \gamma>\gamma_{\rm min}}
\label{EPLC}
\end{equation}
where $\gamma$ is the Lorentz factor of electrons in the blob rest frame, and $\alpha$ is the power-law index. $\gamma_{\rm min}$ and $\gamma_{\rm cut}$ are the Lorentz factors corresponding to the minimum and cutoff energy of the electron distribution in the emission region. The total energy of the electrons in the emitting region is defined as $U_{\rm e}=m_e\:c^2\int \gamma N_e(\gamma)d\gamma$. The energy distribution of electrons given by Eq. \ref{EPLC} is formed when the emitting particles are accelerated with a limiting process at higher energies (e.g., cooling or limited efficiency of the acceleration process). The power-law index ($\alpha$) defines the properties of the acceleration mechanism \citep[e.g.,][]{1989MNRAS.239..995K, 1990ApJ...360..702E, 2012ApJ...745...63S, 1983RPPh...46..973D}, and $\gamma_{\rm cut}$ allows constraining the cooling process of the particles and the state of plasma in the jet \citep{2017MNRAS.464.4875B, 2020A&A...635A..25S}.\\
In this interpretation, the parameters of the emitting region are the radius, Doppler factor ($\Gamma=\delta$, for small viewing angles), and magnetic field. The magnetic field is a free parameter during the fitting (assuming its density $B^2/8 \pi$ scales with $U_{\rm e}$), while in principle a constraint on the other two parameters can be derived from the data. For example, a lower limit on $\delta$ can be imposed using high-resolution radio data or the emitting region size can be constrained from $R\leq \delta\times c\times t/(1+z)$, provided the minimum variability time is known. Unfortunately, both constraints cannot be assessed for all the sources given in Table \ref{sources}, so both parameters have been left free during the fitting. In this case, $R$ is constrained from the observed Compton Dominance (CD- the IC to the synchrotron peak luminosities ratio) and from polynomial fitting of the data (see the documentation of {\rm jetset} \footnote{https://jetset.readthedocs.io/en/latest/}).\\
In the modeling it is assumed that the low-energy peak (from radio to optical/UV) is due to synchrotron emission from ultrarelativistic electrons in the jet with an energy distribution as given by Eq. \ref{EPLC}. Instead, the HE peak is due to the IC scattering of internal \citep[SSC;][]{ghisellini,bloom, maraschi} or external photons \citep[EIC;][]{blazejowski, sikora, ghisellini009}. The IC scattering of external photons is considered, since the SEDs of FSRQS are better explained by EIC, as shown by the previous studies \citep[e.g.,][]{2015ApJ...815L..23A, 2015ApJ...815L..22A, 2018A&A...619A.159M, 2015ApJ...807...79H, 2018ApJ...863..114G}, and the CD is evident in the SEDs of the considered sources (Fig. \ref{sed1}). Localization of the emission region in the jet is an open question and along the jet, depending on the distance from the central black hole, different photon fields can be dominant for the IC scattering \citep{2009ApJ...704...38S}. In this paper we assume that the emitting region is outside the broad-line region (BLR) where the dominant photon field is the IR emission from the dusty torus. \citet{2002ApJ...577...78S} showed that in MeV blazar SEDs the shift of the peak of the HE component to lower energies is most likely due to the comptonization of IR photons from the dusty torus. The IR radiation from the dusty torus is assumed to have a blackbody spectrum with a luminosity of $L_{\rm IR}=0.6\:L_{\rm disc}$ \citep[see][]{2015ApJ...815L..23A} where $L_{\rm disc}$ is the accretion disk luminosity, which fills a volume that for simplicity is approximated as a spherical shell with a radius of $R_{\rm IR} = 2.5\times10^{18}\:(L_{\rm d}/10^{45})^{1/2}\:{\rm cm}$ \citep{2008ApJ...685..160N} with the energy density of $u_{\rm IR}=0.6\:L_{\rm d}/4 \pi R_{\rm IR}^2\:\delta^2$ in the co-moving frame of the jet. In \citet{2011MNRAS.411..901G} and \citet{2017ApJ...839...96M}, the HE component in the SED of distant blazars was modeled by IC scattering of BLR reflected photons, adopting a smooth broken power-law shape of the emitting electrons. We refer the reader to these papers for details on the modeling when BLR reflected photons are considered.\\
The broadband SEDs have been modeled using the {\it jetset version 1.1.2} numerical leptonic code \citep{2011ApJ...739...66T, 2009A&A...501..879T, 2006A&A...448..861M}. The free model parameters (those of the emitting electrons, $\delta$, $R$ and $B$) are constrained by using the Minuit optimizer. The emission directly due to accretion emerges primarily in the UV band, showing a UV excess in the SED, which is modeled by adding a blackbody component \citep{2009MNRAS.399.2041G}. Fitting of this excess allows us to estimate the disc photons temperature and luminosity. When the excess UV component is not distinguished, an upper limit is derived by requiring that the disc emission does not exceed the observed nonthermal emission from the jet. 
\subsection{SED modeling results}
The SEDs modeling results are presented in Fig. \ref{sed1} and the corresponding parameters in Table \ref{t7}. The archival data from the Space Science Data Center \footnote{http://www.ssdc.asi.it} are in gray, while the optical/UV, X-ray, and \gray data, obtained here, are shown in cyan, red, and blue, respectively. The radio data are not included in the SED fits but are considered only as upper limits. The observed radio emission is assumed to originate from a different and extended region. The hard \gray emission spectra of TXS 0907+230, PMN J0226+0937, PKS 0451-28, B3 0908+416B, and B3 1343+451 are in magenta, showing that the \gray flux increases and their spectra extend to higher energies. As high redshift blazars are considered ($z>2.5$), their optical/UV flux could be affected by absorption of neutral hydrogen in intervening Lyman-$\alpha$ absorption systems. Following \citet{2011MNRAS.411..901G} this was corrected using the mean attenuation from \citet{2010MNRAS.405..387G} which was computed for six wavelengths approximately centred in the UVOT filters.\\
The SEDs in Fig. \ref{sed1} contain enough data from radio to HE \gray bands to shape both low and high energy peaks. The applied model reproduces the multiwavelength data relatively well for almost all the sources. The \gray data of TXS 1448+093 are not well explained by the model because in this case the optical/UV data are clearly constraining the HE tail of the electron distribution and if considering $\alpha>2.0$ (softer than the estimated spectrum with $\alpha=1.52\pm0.15$), the flux predicted by the model will be lower than the data. Due to the large uncertainties in the S4 2015+65 \gray data the modeling is difficult, so the best possible one is shown. The electron power-law index is defined by the X-ray data \citep[through $\alpha = 2\Gamma_{\rm X} - 1$ relation][]{1986rpa..book.....R} and depending on whether the SSC or EIC component is dominating in the X-ray band, different values for $\alpha$ are obtained. When the X-ray spectrum is hard and the SSC component is dominating, the energy distribution of the emitting electrons has a hard spectrum as well. For example, for GB 1508+5714, PKS 0537-286, B2 0743+25, TXS 0222+185 and  B3 0908+416B, $\alpha=1.17\pm0.07$, $1.33\pm0.07$, $1.13\pm0.19$, $1.62\pm0.05$ and $1.31\pm0.25$ were estimated which shows that the emission is due to newly accelerated electrons. The X-ray emission from S4 1427+543 and TXS 1616+517 is also dominated by the SSC component but because of large uncertainty in the X-ray photon index and flux estimations, correspondingly $\alpha=2.04\pm0.10$ and $2.09\pm0.09$ have been obtained. On the contrary, when the emission in the X-ray and \gray bands is only defined by the EIC component then $\alpha>2.2$; e.g., $\alpha=2.75\pm0.04$ and $2.70\pm0.06$ are correspondingly estimated for TXS 0800+618 and PKS 0834-20. Correspondingly, $\gamma_{\rm min}$ and $\gamma_{\rm cut}$ are in the range of $2.58-93.28$ and $(1.01-15.73)\times10^{3}$ (excluding TXS 1448+093). The SSC bump in the X-ray band is sensitive to $\gamma_{\rm min}$ values \citep{2009MNRAS.399.2041G, 2000ApJ...544L..23T, 2014ApJ...788..104Z} and those estimated here are well within the range usually estimated for the FSRQs \citep[see Fig. 4 in][]{2014ApJ...788..104Z}. As expected \citep{2008MNRAS.385..283C}, lower values of $\gamma_{\rm min}$ are estimated when the X-rays are produced only by the EIC component: e.g., $\gamma_{\rm min}=2.68\pm0.36$ and $2.58\pm0.15$ are correspondingly estimated for PKS 1351-018 and OD 166. The HE tails of both synchrotron and IC components are well defined by the optical/UV and \gray data, respectively, allowing precise estimation of $\gamma_{\rm cut}=(1.01-15.73)\times10^{3}$. $\gamma_{\rm cut}$ is in a strong dependence on $\alpha$, and its highest value, $(15.73\pm1.60)\times 10^{3}$ was estimated when $\alpha= 2.80\pm0.04$ (the SED of GB6 J0733+0456 in Fig. \ref{sed1}). Meanwhile, when $\alpha=2.2-2.5$, the highest $\gamma_{\rm cut}$ is $(8.67\pm0.48)\times10^3$ for B3 1343+451.\\
The modeling shows that the magnetic field in the emitting region is within $0.10-1.74$ G. The highest values of $1.74\pm0.12$, $1.68\pm0.14$, $1.45\pm0.09$ and $1.15\pm0.05$ G are estimated for PMN J0226+0937, PMN J1441-1523, MG2 J174803+3403, and OD 166, respectively. The estimated Doppler factor is from $\delta=10.00$ to $\delta=27.42$ with a mean of $\delta=19.09$. Although, these are higher than the average values estimated for FSRQs \citep[e.g.,][]{2015MNRAS.448.1060G, 2017ApJ...851...33P}, they are well within the range of physically realistic values \citep[e.g., see][]{2015MNRAS.454.1767L}. The emitting region size is within $R=(0.70-9.48)\times10^{16}$ cm except for TXS 0800+618 and S4 2015+65 for which $R=1.55\times10^{17}$ cm and $1.35\times10^{17}$ cm, respectively. The values estimated for $R$ are consistent with the \gray flux variation in a day or several day scales and suggest that the multiwavelength emission is produced in the sub-parsec scale regions of the jet.\\
The energetics of the considered sources can be estimated using the modeling results. First, the available data allows a straightforward estimation of the disc luminosity of GB 1508+5714, PKS 1351-108, PKS 0537-286, TXS 0800+618, S4 1427+543, GB6 J0733+0456, B2 0743+25, PKS 0347-211, S41124+57, PKS 0438-43, S4 2015+65, PKS 0834-20, TXS 0222+185, TXS 0907+230, PMN J0226+0937, PKS 0451-28 and PKS 2107-105, under the assumption that the disc has a black body spectrum. The estimation shows that $L_{\rm d}\simeq(1.09-10.94)\times10^{46}\:{\rm erg \: s^{-1}}$ with a highest value of $L_{\rm d}\simeq1.09\times10^{47}\:{\rm erg \: s^{-1}}$ estimated for PMN J0226+0937. Such high luminosities are obtained as very powerful blazars are considered here, and they are of the same order as those usually estimated for bright FSRQs \citep{2014ApJ...790...45P} and distant blazars \citep{2011MNRAS.411..901G, 2016ApJ...825...74P}.
The jet power in the form of the magnetic field ($L_{B}$) and relativistic electrons ($L_{e}$) is calculated as $L=\pi R^2 c \:\Gamma^2\:U_{i}$ , where $U_{i}$ is either electron ($U_{e}$) or magnetic field ($U_{B}$) energy density, using the parameters from Table \ref{t7}. The corresponding values are given in Table \ref{t7}, showing that $L_{e}$ is in the range of $(0.03 -13.49)\times10^{45}\:{\rm erg\:s^{-1}}$ while $L_{\rm B}$ in $(0.24-434.69)\times10^{44}\:{\rm erg\:s^{-1}}$. $L_{B}$ and $L_{e}$ are of the same order with $L_{\rm e}/L_{\rm B}=2.92$ and $0.61$ for PKS 0834-20 and PMN J1441-1523 respectively, while for MG2 J174803+3403, OD 166, TXS 1448+093, PMN J0226+0937, PKS 2107-105 $L_{B}/L_{e}=3.7-36.0$. For the other sources of Table \ref{t7}, $L_{\rm e}$ exceeds $L_{\rm B}$, the largest deviation of $L_{\rm e}/L_{\rm B}\geq1800$ being found for PKS 0537-286, B2 0743+25 and TXS 0222+185, which are the only sources when the hard X-ray emission above the Swift XRT band is modeled with a SSC component and the EIC is dominating at higher energies.
However, we note that the luminosities can be higher when the SEDs in the flaring periods are modeled.
\section{Discussion and Conclusion}\label{sec6}
\fermi has detected MeV/GeV emission from $\sim2500$ blazars, which are bright emitters across the whole electromagnetic spectrum. \fermi has sufficient sensitivity to detect blazars farther than $z=2.0$, which are among the most powerful objects in the Universe. The distant objects ($z>2.0$) represent a small fraction of the total observed sources ($3.75$ \%), but their investigation is crucial for the study of the powerful relativistic outflows and measurement of the EBL photon density. We selected the most distant blazars ($z>2.5$) from 4FGL and studied their multiwavelength emission properties by analyzing \fermi, Swift XRT and UVOT data. Also, {the origin of their  multiwavelength emission is investigated} through theoretical modeling of the broadband SEDs.\\
In the X-ray and \gray bands, the spectra of the considered sources have different properties; except for TXS 1448+093, the X-ray spectrum of the other sources is hard with $\Gamma_{\rm X}=1.01-1.86$ while in the \gray band $\Gamma_{\gamma}>2.2$. Thus, the X-ray and \gray data are determining that the second peak in the SED is within $10^6-10^8$ eV. The \gray flux of the considered sources is from $4.84\times10^{-10}$ to $1.50\times10^{-7}\:{\rm photon\:cm^{-2}\:s^{-1}}$, and in the $\Gamma_\gamma-L_{\gamma}$ plane, they occupy the area more typical for bright blazars, which is natural, since the sources at large distances should be powerful enough to be detected. The two BL Lacs included in the sample, 87GB 214302.1+095227 and MG1 J154930+1708, are relatively faint, $\simeq(0.48-1.37)\times10^{-9}{\rm photon\:cm^{-2}\:s^{-1}}$, although it is already unusual to observe BL Lacs at large distances.\\
 In the $F_{\rm X}-\Gamma_{\rm X}$ plane the considered sources occupy the region of hard X-ray spectra ($\Gamma_{\rm X}<2.0$) and a flux from $7.94\times10^{-14}\:{\rm erg\:cm^{-2}\:s^{-1}}$ to $1.17\times10^{-11}\:{\rm erg\:cm^{-2}\:s^{-1}}$. PKS 0451-28, TXS 0222+185, PKS 0834-20, PKS 0537- 286, TXS 0800+618, B2 0743+25 and PKS 0438-43 are separated from the others with a comparably high X-ray flux $F_{\rm X-ray}\geq2.13\times10^{-12}\:{\rm erg\:cm^{-2}\:s^{-1}}$. In the X-ray band, statistically significant flux variation is found for PKS 0438-43 and B2 0743+25 while there is an indication of variability for TXS 0222+185; during the bright X-ray periods, their flux exceeds $10^{-11}\:{\rm erg\:cm^{-2}\:s^{-1}}$.\\
The \gray flux variability can be investigated based on the available data. The most distant flaring blazars are MG3 J163554+3629 and PKS 0537-286 at $z=3.65$ and $z=3.10$, respectively. Though \gray flux amplification is observed in the 30-day bin light curve of MG3 J163554+3629, the \gray flares are more drastic and evident for PKS 0537-286. The adaptively binned light curve PKS 0537-286 shows several bright \gray flaring periods with a maximum flux of $(1.29\pm 0.26)\times10^{-6}\:{\rm photon\:cm^{-2}\:s^{-1}}$ above 100 MeV, observed on MJD 57879.2 in a time bin having a width of $\sim16.0$ hours. For the distance of PKS 0537-286, such flux amplification can be useful for investigation of the $\gamma\gamma$ attenuation, but the photon index is $\Gamma_{\rm \gamma}=2.64\pm0.27$, so the flare is dominated by sub-GeV photons. The \gray flux of B3 1343+451 increases in sub-day scales and that of PKS 0347-211 ($z=2.94$) and PKS 0451-28 ($z=2.56$) in day scales. B3 1343+451 is among the top 30 bright blazars observed by \fermi, showing multiple periods of enhanced \gray emission when the average luminosity of $(2-4)\times10^{48}\:{\rm erg\: s^{-1}}$ increased up to $1.5\times10^{50}\:{\rm erg\: s^{-1}}$. The peak \gray flux of $(1.82\pm0.45)\times10^{-6}\:{\rm photon\:cm^{-2}\:s^{-1}}$ above $100$ MeV was observed on MJD 55891.7 which is $\geq 36.4$ times higher than the \gray flux in quiescent state. The $\chi2$ test showed that the \gray emission of B3 0908+416B, TXS 0800+618, PKS 0438-43, OD 166 and TXS 0907+230 is variable in week scales while that of MG3 J163554+3629, GB6 J0733+0456, B2 0743+25, PMN J1441-1523 and TXS 1616+517 in month scales.\\
The \gray photon index of B3 1343+451, PKS 0451-28, B3 0908+416B and TXS 0907+230 varies in time, and their \gray emission sometimes appears with a hard \gray spectrum. The averaged \gray photon index of these sources determines that the peak of the HE component to be below $10^8$ eV whereas the hard \gray spectrum indicates that the peak is shifted to higher energies. During the flares, different processes can cause the shift of the low-energy or HE peaks. For example, both components will be shifted to HEs when the particles are effectively re-accelerated, resulting in higher electron cutoff energy ($\gamma_{\rm cut}$). It is expected that only the HE component will increase when due to the changes in the location of the emitting region the external photon fields are starting to dominate \citep{2011ApJ...736..128P}. In order to understand the origin of the change in the emission components, extensive multiwavelength observations are required which is not the case for the sources considered here. We note that such changes have already been observed in the previous studies of blazars \citep[e.g.,][]{2014MNRAS.445.4316C, 2015ApJ...815L..23A, 2015ApJ...815L..22A, 2017MNRAS.470.2861S}.\\
The main parameters characterizing the jets of the considered sources are derived by modeling the multiwavelength SEDs. The CD observed in the spectrum of almost all the considered sources implies that the electrons are loosing energy mainly by interacting with the external photons. In the framework of the single-zone scenario, the observed X-ray and \gray data are satisfactorily explained taking into account IC scattering of synchrotron and IR photons which in its turn allows to constrain the  parameters of the emitting electrons. The power-law index of the electrons is from $\sim1.13$ to $2.90$, which is within the range expected from the standard particle acceleration theories. For example, in the diffuse shock particle acceleration, the formed particle spectra can be from very hard ($-1$) to very steep, depending on the shock speed, nature of particle scattering, magnitude of turbulence, shock field obliquity, and other parameters \citep{2012ApJ...745...63S}. On the other hand, the optical/UV and \gray data strongly constrain the cut-off energy of electrons; the particles are effectively accelerated up to $\sim10^{12}$ eV. The cooling of emitting particles defines the cut-off energy formed when the acceleration and cooling time scales are equal. Considering the electrons are mostly cooling by interacting with IR photons, and equating the radiative cooling time $t_{\rm cool}\sim\frac{3\: m_{e}c}{4 \sigma_{\rm T}\:u_{\rm IR}^{\prime}\:\gamma_{\rm e}^{\prime}}$ with the electron acceleration time $t_{\rm acc}\simeq \eta_0 \frac{m_{e} c \gamma_{\rm e}^{\prime}}{e B}$ \citep{2007Ap&SS.309..119R}, one gets that for $10^{12}$ eV electrons these timescales are equal when $\eta^{-1}=5\times10^{-4}$. In other words, the maximum electron energy, if limited solely by the radiative energy losses, might be expected to be much higher than that observed, since often $\eta^{-1}\geq10^{-2}$ is expected \citep[e.g.,][]{1996ApJ...463..555I}. Therefore the cutoff is most likely limited by the physical size of the emitting zone. For a given set of parameters, $R$, $B$, and $\delta$, $\gamma_{\rm min}$ is $<93$ and lower values of $\gamma_{\rm min}$ are estimated when the X-rays are produced only by EIC. This shows that the process responsible for the particle acceleration picks up almost all electrons.\\
The total jet luminosity (defined as $L=L_{e}+L_{B}$) is $\leq1.41\times10^{46}\:{\rm erg\:s^{-1}}$ which is of the same order as that usually calculated for blazars \citep{2014Natur.515..376G}. Thus, the jet power of distant and nearby blazars do not differ substantially. The estimated disc luminosity is within $L_{\rm d}\simeq(1.09-10.94)\times10^{46}\:{\rm erg \: s^{-1}}$, more typical for powerful blazars. The disc and jet luminosities are of the same order for TXS 0800+618 and TXS 0222+185 ($L_{\rm disc}/L_{\rm jet}=1.17-2.44$) while for the others $L_{\rm disc}$ is higher than $L_{\rm jet}$. However, we note that the protons with unknown content in jet are not included in the computation of $L_{\rm jet}$, so in principle higher $L_{\rm jet}$ are expected which could be of the same order as $L_{\rm disc}$ \citep{2003ApJ...593..667M}. Yet, when the accretion disc luminosity and temperature are well measured, an approximate value of the black hole mass can be derived following \citet{2009MNRAS.399L..24G, 2010MNRAS.402..497G}. In general, the black hole mass can be well estimated from the optical spectroscopy \citep{2012ApJ...748...49S} or from fitting the blue bump at the optical/UV band \citep{2013MNRAS.431..210C, 2015MNRAS.448.1060G}. The maximum temperature (and hence the $\nu F\nu$ peak of the disc luminosity) of the standard multi-colour accretion disc temperature profile occurs at $5\:R_{\rm s}$ where $R_{\rm s}$ is Schwarzschild radius. Taking into account that the peak temperature scales as $T_{\rm disc}\sim(L_{disc}/R_{s})^{1/4}$ \citep{2009MNRAS.399L..24G}, $R_s$ can be estimated from which the black hole mass can be derived. The black hole mass estimated for the sources with a clear blue-bump in their SED is in the narrow range of $(1.69-5.35)\times10^{9}\:M_\odot$ where the highest black hole mass of $5.35\times10^{9}\:M_\odot$ is estimated for PMN J0226+0937 which has also the highest disc luminosity. The virial black hole mass of GB 1508+5714, PKS1351-018, S4 1427+543 and B2 0743+25 is also estimated in the quasar catalog of \citet{2011ApJS..194...45S} and is $(3.23\pm0.40)\times10^{8}\:M_\odot$, $(8.91\pm0.64)\times10^{8}\:M_\odot$, $(1.80\pm0.17)\times10^{8}\:M_\odot$ and $(3.89\pm0.26)\times10^{9}M_\odot$, respectively. In the case of B2 0743+25, both methods produce similar results, $3.06\times10^{9}\:M_\odot$ and $(3.89-0.26)\times10^{9}M_\odot$, while for the other sources the masses obtained using optical spectroscopy are slightly lower. Such differences are expected, since both methods rely on a fitting or approximations. For the considered sources, the Eddington luminosity is within $(2.12-5.87)\times10^{48}\:{\rm erg\:s^{-1}}$, and the ratio $L_{\rm disc}/L_{\rm Edd}$ ranges from $0.05$ to $0.16$ \citep[e.g., see][]{2014Natur.515..376G}.
\section{Summary}\label{sec7}
The origin of the multiwavelength emission from distant blazars ($z>2.5$) has been investigated using the \fermi data accumulated in 2008-2018 and Swift XRT/UVOT data observed in the last fifteen years. The main results are summarized as follows: 

\begin{itemize}
 \item[{\it i)}] Twenty-six out of the thirty-three considered sources are FSRQs, five BCUs, and only two are BL Lacs. The two BL Lacs are also the faintest objects in the sample with a flux of $(0.48-1.66)\times10^{-9}\:{\rm photon\:cm^{-2}\:s^{-1}}$) while the others have a flux from $2.73\times10^{-9}\:{\rm photon\:cm^{-2}\:s^{-1}}$ to $\sim1.50\times10^{-7}\:{\rm photon\:cm^{-2}\:s^{-1}}$. 

\item[{\it ii)}]  Except for the two BL Lacs, the photon index of all the considered sources ranges from $2.18$ to $3.05$. Only, the \gray indexes of B3 1343+451, PKS 0451-28, B3 0908+416B and TXS 0907+230 are found to vary in time. The hardest \gray spectra of  B3 1343+451, B3 0908+416B and TXS 0907+230 are with indexes of $1.73\pm0.24$, $1.84 \pm 0.25$, and $1.72\pm0.15$, respectively, while that of PKS 0451-28 - with $2.06 \pm 0.07$.

 \item[{\it iii)}] The Swift XRT observations show a significant X-ray emission only from the FSRQs considered here. Only a few counts have been detected from the other sources, even if some of them have been observed by Swift several times. The X-ray photon index of distant FSRQs is $\Gamma_{\rm X}=1.1-1.8$ and the flux is spanning from $\sim5\times10^{-14}\:{\rm erg\:cm^{-2}\:s^{-1}}$ to $10^{-11}\:{\rm erg\:cm^{-2}\:s^{-1}}$.
 
 \item[{\it iv)}] The brightest X-ray source in the sample is TXS 0222+185 ($z=2.69$) with a flux of $(1.19\pm0.04)\times10^{-11}\:{\rm erg\:cm^{-2}\:s^{-1}}$. The X-ray emission from only PKS 0438-43, B2 0743+25 and TXS 0222+185 showed a substantial flux increase in some observations, whereas the X-ray emission from other sources is relatively constant in different years. 
 
 \item[{\it v)}] The \gray variability of the considered sources has been investigated using the data accumulated for ten years. Fourteen sources from the sample show a variable \gray emission on short and long timescales. The \gray flux of B3 1343+451 ($z=2.53$) and PKS 0537-286 ($z=3.10$) increases in sub-day scales, that of PKS 0347-211 ($z=2.94$) and PKS 0451-28 ($z=2.56$) in day scales. The \gray emission of B3 0908+416B, TXS 0800+618, PKS 0438-43, OD 166 and TXS 0907+230 is variable in a week scale while that of MG3 J163554+3629, GB6 J0733+0456, B2 0743+25, PMN J1441- 1523 and TXS 1616+517 in a month scale. The most distant \gray blazar flaring on short time scales is PKS 0537-286 ($z=3.10$) when its flux increased up to $(1.29 \pm 0.26)\times10^{-6}\:{\rm photon\:cm^{-2}\:s^{-1}}$ above 100 MeV within a time bin having a width of $\sim16.0$ hours. The \gray flux of B3 1343+45 increased significantly from its average level in multiple periods with a maximum flux of $(1.82 \pm 0.45)\times10^{-6}\:{\rm photon\:cm^{-2}\:s^{-1}}$ above 100 MeV, accompanied by moderate hardening of the spectra. 

 \item[{\it vi)}] In the $\Gamma_{\rm \gamma}-L_{\gamma}$ plane, the majority of the considered sources occupy the narrow range of $\Gamma_{\rm \gamma}=2.2-3.1$ and $L_{\gamma}=(0.10-5.54)\times10^{48}\:{\rm erg\:s^{-1}}$, which is more typical for the brightest blazars. However, during \gray flares, the luminosity of variable sources is significantly beyond this boundary, changing within $(10^{49}-10^{50})\:{\rm erg\:s^{-1}}$. For example, the luminosity of B3 1343+451 increases $\sim36.4$ times and corresponds to $L_{\gamma}=1.5 \times 10^{50}\: {\rm erg \:s^{-1}}$, being among the highest values so far observed in the \gray band.
 
  \item[{\it vii)}] The SEDs were modeled within a one-zone leptonic scenario, considering the IC scattering of both synchrotron and IR photons from the dusty torus. The X-ray and \gray data allowed to constrain the peak of the HE component (within $10^6-10^8$ eV) as well as the power-law index and cut-off energy of the radiating electrons; the index and the cutoff energy are within the range of $1.13-2.90$ and $(1.01-15.73) \times 10^3$, respectively. The radius of the emitting region is estimated to be $\leq0.05$ pc while the magnetic field and the Doppler factor are correspondingly within $0.10-1.74$ G and $10.00-27.42$.
 
  \item[{\it viii)}] The jet luminosity is estimated to be $\leq1.41\times10^{46}\:{\rm erg\:s^{-1}}$, which is of the order of the values usually obtained for blazars. For all the sources, expect for TXS 0800+618 and TXS 0222+185 for which $L_{\rm disc}/L_{\rm jet}=1.17-2.44$, the jet luminosity is lower than that of the disc $L_{\rm d}\simeq(1.09-10.94)\times10^{46}\:{\rm erg \: s^{-1}}$ which is estimated by fitting the UV excess. The black hole masses are estimated to be within $(1.69-5.35)\times10^{9}\:M_\odot$ so the disc luminosity is 5-16\% of the Eddington luminosity.
  
\end{itemize}

\section*{Acknowledgements}
We thank the anonymous referee for constructive comments that improved the paper. This work was supported by the RA  MoESCS  Committee of Science, in the frames of the research project No 18T-1C335. This work used resources from the ASNET cloud and the EGI infrastructure with the dedicated support of CESGA (Spain). Part of this work is based on archival data, software or online services provided by the Space Science Data Center - ASI. This research has made use of the Swift XRT Data Analysis Software (XRTDAS). This work made use of data supplied by the UK Swift Science Data Centre at the University of Leicester.
\section*{Data availability}
The data underlying this article will be shared on reasonable request to the corresponding author.



\bibliographystyle{mnras}
\bibliography{biblio} 




\bsp	
\label{lastpage}
\end{document}